\documentclass[12pt, a4paper]{article}

\usepackage[english]{babel}
\usepackage[toc,page]{appendix}
\usepackage[utf8]{inputenc}
\usepackage{amsmath}
\usepackage{amssymb}
\usepackage{array}
\usepackage{graphicx}
\usepackage{placeins}
\usepackage{caption}
\usepackage{subcaption}
\usepackage[colorinlistoftodos]{todonotes}
\usepackage{tensor}
\usepackage{fullpage}
\usepackage{braket}
\usepackage{framed}
\usepackage{feynmf}
\usepackage{cancel}
\usepackage{empheq}
\usepackage[affil-it]{authblk}
\usepackage{dsfont}
\usepackage{bbm}
\usepackage{sectsty}
\usepackage{sectsty}
\usepackage{lipsum}
\usepackage{hyperref}
\usepackage{framed}
\usepackage{tikz}
\usetikzlibrary{cd}
\usetikzlibrary{positioning}
\usepackage{pgfplots}
\usetikzlibrary{decorations.markings}
\usetikzlibrary{decorations.pathmorphing}
\hypersetup{
    colorlinks,
    citecolor=blue,
    filecolor=blue,
    linkcolor=blue,
    urlcolor=blue
}
\definecolor{shadecolor}{rgb}{1,0.9,1}

\usepackage{titlesec}

\titleformat{\section}
  {\normalfont\fontsize{13}{10}\bfseries}{\thesection}{1em}{}

\titleformat{\subsection}
  {\normalfont\fontsize{10.5}{10}\bfseries}{\thesubsection}{1em}{}

\titleformat{\subsubsection}
  {\normalfont\fontsize{10.5}{10} \itshape}{\thesubsubsection}{1em}{}

\titleformat{\paragraph}[runin]{\bfseries}{}{}{}

\newcommand{\La}{{\cal L}}

\newcommand{\Ham}{{\cal H}}
\newcommand{\cW}{{\cal W}}

\newcommand{\Op}{{\cal O}}

\newcommand{\N}{{\cal N}}

\newcommand{\I}{{\cal I}}

\newcommand{\cR}{{\cal R}}

\newcommand{\half}{\frac{1}{2}}

\newcommand{\Real}{{\mathbb{R}}}

\newcommand{\pardev}[2]{\frac{\partial #1}{\partial #2}}

\newcommand{\eq}[1]{(\ref{eq:#1})}
\numberwithin{equation}{section} 

\tikzset{->-/.style={decoration={
  markings,
  mark=at position .5 with {\arrow{>}}},postaction={decorate}}}

\tikzset{-->-/.style={decoration={
  markings,
  mark=at position .75 with {\arrow{>}}},postaction={decorate}}}
 
\tikzset{->--/.style={decoration={
  markings,
  mark=at position .25 with {\arrow{>}}},postaction={decorate}}}

\usepackage[font=small, margin=2.5cm]{caption} 

\usepackage{fancyhdr}
\fancypagestyle{plain}{%
	\fancyhead[R]{DMUS–MP–20/07 \\ LTH 1241}
	
	}

\usepackage[
            marginparsep=-8mm,
            marginparwidth=35mm
            ]{geometry}

\begin{document}

\title{Cosmological Solutions, a New Wick-Rotation, and the First Law of Thermodynamics}

\author{J. Gutowski\footnote{J.Gutowski@surrey.ac.uk}${\ }^1$,  T. Mohaupt\footnote{Thomas.Mohaupt@liv.ac.uk}${\ }^2$ and G. Pope\footnote{Giacomo@liv.ac.uk}${\ }^2$ \vspace{0.5cm}}

\affil{\normalsize $^1$Department of Mathematics \\ University of Surrey \\Guildford, GU2 7XH, UK}
\affil{$^2$Department of Mathematical Sciences \\ University of Liverpool \\ Liverpool, L69 7ZL, UK}
\date{}
\maketitle
\abstract{
We present a modified implementation of the Euclidean action formalism suitable for studying the thermodynamics of a class of cosmological solutions containing Killing horizons. To obtain a real metric of definite signature, we perform a ``triple Wick-rotation" by analytically continuing all spacelike directions. The resulting Euclidean geometry is used to calculate the Euclidean on-shell action, which defines a thermodynamic potential. We show that for the vacuum de Sitter solution, planar solutions of Einstein-Maxwell theory and a previously found class of cosmological solutions of $\mathcal{N} = 2$ supergravity, this thermodynamic potential can be used to define an internal energy which obeys the first law of thermodynamics. Our approach is complementary to, but consistent with
the isolated horizon formalism. For planar Einstein-Maxwell solutions, we find dual solutions in Einstein-anti-Maxwell theory where the sign of the Maxwell term is reversed. These solutions are planar black holes, rather than cosmological solutions, but give rise, upon a standard Wick-rotation to the same Euclidean action and thermodynamic relations.
}
\pagenumbering{gobble}
\sectionfont{\large}
\subsectionfont{\normalsize}

\newpage
\tableofcontents

%
%
\newpage
\pagenumbering{arabic}

\section{Introduction and motivation}

The laws of black hole mechanics \cite{Bardeen:1973gs} were initially thought of as
formal analogies of the laws of thermodynamics, but subsequent work has 
shown that they have a genuine thermodynamical interpretation. This suggests that 
the relation between classical and quantum gravity
may be analogous to the relation between thermodynamics and statistical mechanics \cite{Bekenstein:1973ur, hawking1975, Gibbons:1976ue, Gibbons:1977mu}. 
This is among the most compelling clues that we have about the nature of quantum gravity.  
It is therefore important to identify ever-larger classes of solutions to classical gravity
which obey the laws of black hole mechanics, or variant versions thereof. 

One setting in which the laws of black hole mechanics can be derived is static, asymptotically
flat spacetimes containing a Killing horizon.\footnote{This generalizes to the larger class of stationary spacetimes, where the first law also contains a term involving the angular momentum and
rotation velocity of the spacetime. But since the solutions which we will consider in this paper are of the
more restricted, static type, we will neglect this term from the beginning.} 
Then the first law of black hole mechanics takes the form
\begin{equation*}
    dM = \frac{\kappa}{8 \pi G} dA + \mu_i d\mathcal{Q}^i,
\end{equation*}
where $M$ is the mass, $\kappa$ is the surface gravity, $A$ is the area of the black hole horizon and $\mathcal{Q}^i, \; \mu_i$ are a set of conserved charges and their associated potentials. This statement
does not involve any thermodynamics and is derived using geometrical reasoning. Yet, 
the seminal work by Bekenstein \cite{Bekenstein:1973ur} and Hawking \cite{hawking1975} has
demonstrated that this relation
can be interpreted as the first law of thermodynamics through identifying
\begin{equation*}
    S = \frac{A}{4 G}, \qquad T_H = \frac{\kappa}{2 \pi},
\end{equation*} 
where $S$ is the entropy and $T_H$ is the Hawking temperature. With these identifications,
the first law of black hole mechanics becomes the first law of thermodynamics:
\begin{equation*}
        dE = T_H dS + \mu_i d\mathcal{Q}^i.
\end{equation*}
Here the internal energy $E$ is understood to be equal to the mass of the black hole solution,
while the charges $\mathcal{Q}^i$ replace the particle numbers of a grand ensemble, as usual in relativistic thermodynamics. 

One obstruction in generalising this statement to Killing horizons in more general spacetimes --- in particular, those which are not asymptotically flat and not static ---
is the definition of the mass $M$, which takes the role of energy $E$. Diffeomorphism invariance prevents one from assigning a total momentum four-vector, and hence a mass
to regions of spacetime in general. 
For asymptotically flat spacetimes the ADM construction can be used to define a total mass 
 \cite{arnowitt1959}. For Killing horizons with an asymptotically flat static region, this is equivalent to the Komar construction \cite{Ashtekar:1979cc}, where the mass is a conserved charge associated with a timelike Killing vector which becomes null on the horizon.
In these constructions, the normalisation of the mass is implied by the `natural' normalisation of this Killing vector field, which is that the Killing vector field has unit norm at infinity. 
Mass-like quantities can be defined in more general situations.
For example, the quasi-local mass of  Brown and York \cite{Brown:1992br} 
 which is well-defined when the region is stationary, though the resulting 
 mass parameter is necessarily position-dependent and thus does not have a natural normalisation.
Wald's formalism \cite{Wald:1993nt,Iyer:1994ys} allows one to associate conserved
charges to closed surfaces in general diffeomorphism invariant theories of 
gravity, and provides a setting for deriving the first law.  For space-times which are locally asymptotic
to $AdS$ spaces, conserved charges and thermodynamics can be defined
using the variational principle and holographic renormalization see \cite{hep-th-0505190}
and references therein.
In \cite{An:2016fzu} it was stressed that the essential requirement 
to formulate black hole thermodynamics is to have a consistent variational principle, 
which then automatically takes care of the finiteness of conserved charges. This
approach is not limited to boundary conditions which are
locally asymptotically $AdS$. In particular, it was applied
\cite{An:2016fzu}  for the STU-model of four-dimensional $\mathcal{N}=2$
supergravity with conical boundary conditions. 
Further work which applies variational principles to spacetimes with $AdS$ asymptotics
includes  \cite{Anabalon2016,Cardenas2016,Cassani:2019mms}.

There also are approaches which formulate black hole thermodynamics
strictly in terms of near horizon data. Where a comparison to approaches
with asymptotic boundary conditions is possible, they agree up to an
ambiguity associated with the normalization of the horizontal Killing vector
field. One such approach is Asthekar's isolated horizon formalism \cite{Ashtekar:2000hw},
which recasts the first law using only quantities defined locally on the horizon. This includes
a mass-like parameter whose definition and normalization is fixed by 
imposing that the first law takes its standard form. More recently, there has
been much interest in the near horizon behaviour of near-extremal black holes,
in the context of studying non-integrable conserved charges and scalar hair.
In these approaches one uses the existence of an asymptotic $AdS_2$ factor
to perform a reduction to two dimensions. The reduced system is largely
determined by its symmetries, and related to the JT and SYK models, 
see for example
\cite{Maldacena:2016upp, Almheiri:2016fws,Cvetic:2016eiv,Gaikwad:2018dfc,
Nayak2018, Li2020, Ruzziconi2020}.

In the present paper we study a class of spacetimes with Killing horizons, 
which has a causal structure complementary to black holes in the sense
that the exterior region is non-static and asymptotic to a Kasner cosmology
in the infinite past and future, while the interior region is static and terminates
in a timelike curvature singularity. 
  These spacetimes, which are solutions to $\N=2$ supergravity with vector 
multiplets, have planar symmetry and were found in  \cite{Gutowski:2019iyo}, 
when attempting to generalise the black brane type solutions of \cite{Errington:2014bta}.
The simplest member of this family, which is obtained
by imposing that all scalar fields are constant, is the planar version of the 
Reissner-Nordstr\"om solution of Einstein-Maxwell theory. As explained in  \cite{Gutowski:2019iyo},
the planar symmetry has the effect of preventing the existence of the static asymptotically flat regions
familiar from the spherically symmetric Reissner-Nordstr\"om solution. One 
is left with a dynamical region which is now the outer part, and a static inner region around
the singularity. The resulting conformal diagram is that of a maximally extended 
Schwarzschild spacetime, rotated by 90 degrees, or equivalently, the conformal 
diagram of a spherical Reissner-Nordstr\"om solution with the assymptotically 
flat  regions (called `type I' in most references) removed, see Figure  \ref{Fig:pRN}.

While this situation is complementary to the usual set-up of black hole thermodynamics,
we will show that it is nevertheless possible to define a mass-like quantity $E$ and
to derive a relation which takes the form of the first law of thermodynamics, as 
well as a Smarr relation, with $E$
playing the role of energy.  Since the static region terminates in a singularity, there 
are two options we will explore:
the first is to work with the exterior, non-static region, the second is to use
the isolated horizon
formalism which only requires local near-horizon data. We find that both approaches
lead to mutually consistent results. Most of the paper is devoted to adapting the
Euclidean action formalism to the dynamic patch (non-static region). While applying
the isolated horizon formalism is more straightforward, we will see that the 
Euclidean action formalism provides us with additional insights. In particular
we obtain a thermal partition function, and we will discover an interesting `duality' between the thermodynamics
of cosmological solutions and the thermodynamics of planar black holes in 
theories where the sign of the Maxwell term has been flipped.\footnote{Such sign flips
appear in type-II$^*$ string theories, as we will discuss in Section \ref{Sect:Disc+Outlook}.}

Our adaptation of the Euclidean action approach works as follows.
In the standard setting based on a static exterior region with a well 
behaved asymptotic boundary, one Wick-rotates the time coordinate and 
obtains a smooth, positive definite metric on a real slice of the complexification 
of the original solution. 
One then substitutes the Euclideanised solution into the action, thus obtaining
a function which depends on the parameters of the solution. In this step of the
procedure, boundary terms play a central role. The exponential of the resulting
Euclidean action can be viewed as the saddle point approximation to the full 
Euclidean functional integral. Following Gibbons and Hawking \cite{Gibbons:1976ue}
we can interpret this expression as a thermal
partition function. Given this, the energy $E$ can be computed by taking suitable derivatives
with respect to combinations of parameters, of the solution which correspond to
thermodynamic variables.
By then computing its variation $\delta E$, one can check whether the first law is satisfied. 
For the type  of solution we are interested in, the singularities in the
static region prevent us from computing the Euclidean action in the static patch,
as its boundary is a singularity.
Therefore, we work instead in the dynamic outer region, 
as it is well behaved at its asymptotic boundary, located at  
past  timelike infinity.\footnote{The global solutions have a second cosmological region which is related to this region by time-reversal, see Figure \ref{Fig:pRN}. For thermodynamics we choose pairs of patches such that the exterior and interior are related by future-pointing null rays, that is regions III/IV or III/I.} 
Since the horizontal 
Killing vector field is spacelike in this region, we cannot apply a Wick-rotation in time,
which would make spacetime complex. Instead, we perform a triple Wick-rotation 
in all spacelike coordinates, which provides a real slice of the complexified spacetime
with a (negative) definite metric. Using this slice, we obtain a well behaved Euclidean action. 
We note that the standard argument for 
identifying the resulting Euclidean action with a thermodynamic potential 
depends on the Killing vector being timelike, and thus being related to time translations 
and energy. In the dynamic outer patch, the Killing vector is spacelike and thus 
corresponds to spatial translations and momentum. We proceed formally and
 relate our Euclidean action  to a thermodynamic potential, leaving 
 questions about the underlying microscopic theory aside.
The `energy' $E$ is defined as 
a derivative of this potential, and we prove that  its variation $\delta E$
satisfies a relation which takes the exact form of the first law. As a further consistency
check, we also apply the isolated horizon formalism,
which imposes the first law and this way obtains an expression for the energy, 
and we find that the results of both formalisms agree. 


The structure of the paper is as follows. We begin by introducing the Euclidean action formalism and 
reviewing the relevant ingredients. After an overview of standard techniques, the procedure of the triple Wick-rotation is defined. Following this, three examples of the triple Wick-rotation are given. First, the de Sitter solution is discussed in Section \ref{sec:desitter}, serving as a simple example of the first law for a vacuum solution, where we can compare against a standard Wick rotation in the static region as a consistency check. Then, in Section \ref{sec:pem}, the planar Reissner-Nordstr\"om solution 
to the Einstein-Maxwell system is studied using the triple Wick-rotation. A thermodynamic potential is derived, and from this, the first law is verified.  This solution can be regarded as a limit
of a family of solutions to the STU model of $\N=2$ supergravity, for which we verify 
the first law in Section \ref{sec:pstu}. The method is applied again 
for the full STU model, allowing the definition of a mass-like parameter which, when varied, gives the first law. In the following Section \ref{Sect:Isolated_horionzs}, these results are supported through an alternative calculation using the isolated horizon technique of \cite{Ashtekar:2000hw}. In Section \ref{sec:antimax} we find a dual planar Reissner-Nordstr\"om solution in Einstein-anti-Maxwell theory, that is, in a theory where the sign of the Maxwell term is flipped. In this solution, the roles of the interior and exterior region are exchanged, and so the solution is a planar black hole with a static exterior region. This allows us to apply a standard Wick rotation, and we find that this solution has the same Euclidean action and thermodynamic relations as the planar 
Reissner-Nordstr\"om solution. While we do not discuss the embedding of these solutions into string theory, we point out 
that this duality between solutions, as well as their connection through a common
Euclidean section, is related to the existence of  a `twisted' version of the 
$\mathcal{N}=2$ supersymmetry algebra, and to timelike T-duality \cite{Hull:1998vg,Cortes:2019mfa}. 
We conclude with a physical interpretation and discussion of the work completed in this paper in Section \ref{sec:discussion}. Some of the calculational details are relegated to the appendices, together with a summary of the conventions used in this work. Specifically, Appendix \ref{sec:Conventions} summarises our conventions, while Appendix \ref{sec:extrinsiccurvature}
reviews extrinsic curvature, to the extent that is needed to compute boundary terms for the
gravitational action. Appendix \ref{sec:cad} reviews the definition and normalization of charges,
and gives details of the dualization of magnetic to electric charges that we use in the main part.
The quite substantial Appendix \ref{app:classification} presents details of the maximal analytical
extensions for all solutions considered plus the Schwarzschild solution for reference. This 
includes the definition of Kruskal and of advanced and retarded Eddington-Finkelstein coordinates, the computation of the expansion of null congruences, and the classification of horizons. We also
show how type A-III vacuum Einstein solutions arise as asymptotic limits, and we show that the
maximally extended planar Reissner-Nordstr\"om solution is a bouncing cosmology, which 
interpolates between, and regularizes, two Kasner cosmological solutions. Appendix \ref{app_thermo}
collects some thermodynamic relations for reference.

\section{Euclidean action formalism}
In this section, we first review the standard Euclidean action formalism, which interprets the
saddle point approximation of the partition function for a gravitational theory as a 
thermodynamic partition function  \cite{Gibbons:1976ue, Gibbons:1977mu}. Then
we present a modification which assigns a Euclidean action to a dynamic\footnote{Here 
and in the following `dynamic' means `non-stationary', that is a spacetime without a 
timelike Killing vector field.} spacetime
by using a triple Wick-rotation. 

\subsection{Gravitational and thermodynamic partition functions}

The thermodynamic canonical partition function $Z(\beta)$ for a system with a Hamiltonian $\hat{H}$ is defined by
\begin{equation*}
    Z(\beta) := e^{-\beta F}  = \textrm{Tr} e^{- \beta \hat{H}}\;,
\end{equation*}
where $F$ is the free energy and $\beta$ is the inverse temperature. For a system with 
a conserved charge $\mathcal{Q}$, the thermodynamic potential depends on the conserved charge
in addition to its dependence on temperature, 
$F=F(\beta, \mathcal{Q})$. The grand canonical ensemble is defined by keeping the charge constant
and letting the corresponding intensive thermodynamic variable, the chemical potential $\mu$,
fluctuate. The corresponding thermodynamic partition function is the 
grand canonical partition function:
\begin{equation*}
    \mathcal{Z}(\beta, \mu) := e^{-\beta \Omega}  = \textrm{Tr} e^{- \beta \hat{H}}\;,
\end{equation*}
where $\Omega(\beta, \mu)$ is the grand potential. Note that we are suppressing the contribution of a pressure/volume term usually seen in the thermodynamic potentials. From a gravitational perspective, these arise from rotations and angular momentum, or, in the case of planar solutions, translations
and linear momentum, which are not present in the 
solutions we consider in this paper. The thermodynamic relations for such an ensemble
are summarized in Appendix \ref{app_thermo} for convenience.

To illustrate the correspondence between partition functions of quantum (field) theories and thermodynamic partition functions, we consider the case of a quantum particle. The
time-evolution operator admits a path integral representation involving the classical action
\begin{equation*}
    \bra{x} e^{-i t H} \ket{x'} = \int \mathcal{D} x e^{i S[x]},
\end{equation*}
where we have set $\hbar = 1$. 
By Wick-rotating the time coordinate $t \rightarrow -i\beta$ and taking the trace, which 
in the path integral corresponds to integrating over paths periodic in time, one obtains
\begin{equation*}
    \textrm{Tr} e^{ -\beta H}  = \int \mathcal{D} x e^{-S_E[x]} = e^{-\beta F},
\end{equation*}
where $\beta$ is interpreted as inverse temperature, and where $F$ is the free energy.

It is straightforward, at least at a formal level, to extend this prescription to quantum field theories.
In a quantum theory including gravity, the path integral is performed 
over the space of all metrics $g$, as well as over the matter fields $\varphi$,
\begin{equation*}
    Z = \int \mathcal{D} g \mathcal{D} \varphi e^{-S_E[g, \varphi]} \;.
\end{equation*}
While it is challenging to give a precise meaning to the full path integral, one 
can proceed formally and attempt to make sense of it in a saddle point approximation. 
This leads to the expression 
$Z \simeq e^{- S_E}$ , where the Euclidean action $S_E$ is evaluated on an
on-shell field configuration satisfying suitable boundary conditions \cite{Hawking:1995ap}. 

 Employing this, we obtain a relation between the Euclidean on-shell action and the
 free energy:
\begin{equation*}
    \log (Z) \simeq  - S_E \simeq - \beta F \quad  \Rightarrow \quad F \simeq \frac{S_E}{\beta}.
\end{equation*}
When gauge fields are present, the boundary conditions are chosen such that 
the total charge $\mathcal{Q}$ is fixed. Then the Euclidean action depends 
on the associated chemical potential $\mu$, so that  
$S_E  = S_E(\beta, \mu)$, and one obtains the following
relation between the Euclidean on-shell action and the grand potential $\Omega(\beta, \mu)$:
\begin{equation*}
    \log (\mathcal{Z}) \simeq  - S_E \simeq - \beta \Omega, \quad \Rightarrow \quad \Omega \simeq \frac{S_E}{\beta}.
\end{equation*}

\subsection{Simple Wick-rotation}

We now use the Einstein-Maxwell theory with a cosmological constant to 
review the standard Wick-rotation.
Our conventions for actions are explained in Appendix \ref{sec:Conventions}.
We follow \cite{York1986} for the gravitational action, and generalise this by 
including the cosmological constant and 
the Maxwell action:
\begin{equation}
\label{eq:totact1}    
\begin{aligned}
        S &= S_{\text{bulk}} + S_{\text{GHY}}
        \\ 
        &= -\frac{1}{16 \pi} \int_M \sqrt{|g|} (R - 2\Lambda) d^4 x  - \frac{1}{16 \pi} \int_M \sqrt{|g|} F_{\mu \nu} F^{\mu \nu} d^4 x
        \\
        &+ \frac{\epsilon}{8 \pi} \int_{\partial M} \sqrt{|\gamma|} (K - K_0) d^3x \;.
\end{aligned}
\end{equation}
The middle line is the bulk term, containing the Einstein-Hilbert action with the Ricci scalar
$R$, a cosmological constant $\Lambda$ and the Maxwell term. The second line is the Gibbons-Hawking-York boundary 
term  $S_{\text{GHY}}$ \cite{York:1972, Gibbons:1976ue}, 
which is needed to cancel
boundary terms arising from the variation of the Einstein-Hilbert action if spacetime is not
closed (compact without boundary). The spacetime metric $g$ induces a metric
$\gamma$ on the boundary $\partial M$.
$K$ is 
trace of the 
extrinsic curvature of $\partial M$ as an embedded submanifold of spacetime $M$, see
the Appendix \ref{app:extrinsic_curvature} for details. 
The constant $\epsilon$ takes the values $\epsilon = \pm 1$ for boundaries with unit normals which are either spacelike $(+)$ or timelike $(-)$. 
To obtain a finite value for the on-shell action, we include a background term
$K_0$. For an asymptotically flat spacetime $K_0$ is the extrinsic curvature
of the boundary embedded into a flat spacetime, which ensures that the action 
of Minkowski space, which is a solution for $\Lambda=0$,  is zero rather than divergent.

We now apply the Wick-rotation $t \rightarrow -i t$ to (\ref{eq:totact1}) to map
$\exp(i S) \rightarrow \exp(- S_E)$. Following \cite{York1986} we first consider
the gravitational terms. The bulk gravitational term receives a factor of $-i$ from the measure:
\begin{equation*}
    -\frac{1}{16 \pi} \int_M \sqrt{g} (R - 2\Lambda) d^4 x 
 \rightarrow i \frac{1}{16 \pi} \int_M \sqrt{g} (R - 2\Lambda) d^4 x.
\end{equation*} 
For the transformation of the GHY-term we need to distinguish two cases.
\begin{enumerate}
\item
For surfaces with a timelike unit normal: 
\begin{equation*}
\epsilon = -1, \qquad K \rightarrow i K, \qquad \sqrt{\gamma} d^3x \rightarrow \sqrt{\gamma} d^3x	\;.
\end{equation*}
\item
For surfaces with a spacelike unit normal:
\begin{equation*}
\epsilon = 1, \qquad K \rightarrow K, \qquad \sqrt{\gamma} d^3x \rightarrow -i \sqrt{\gamma} d^3x \;.
\end{equation*}
\end{enumerate} 
The resulting Euclidean Gibbons-Hawking-York term is the same for both types of hypersurfaces and transforms as
\begin{equation*}
    + \frac{\epsilon}{8 \pi} \int_{\partial M} \sqrt{|\gamma|} (K - K_0) d^3x \rightarrow  -i \frac{1}{8 \pi} \int_{\partial M} \sqrt{|\gamma|} (K - K_0) d^3x .
\end{equation*}
We now consider the Maxwell field. Before Wick-rotation, we use that the Maxwell action is evaluated on-shell, allowing us to rewrite its contribution as a total derivative\footnote{In
terms of differential forms, $F\wedge \star F = dA \wedge \star F = d(A \wedge \star F)$, if
$d \star F=0$.}
\begin{equation*}
	F^{\mu \nu} F_{\mu \nu} = 2 \nabla_\mu (A_\nu F^{\mu \nu}).
\end{equation*}
Applying Stoke's theorem, we can write the bulk contribution as an integral over the boundary
\begin{equation*}
	- \frac{1}{8 \pi} \int_M \sqrt{|g|} \nabla_\mu \left( A_{\nu} F^{\mu \nu} \right) d^4 x = \frac{1}{8 \pi} \int_{\partial M} F^{\mu \nu} A_{\mu} d\Sigma_{\nu},
\end{equation*}
where the volume element on the boundary is defined as $d\Sigma_{\mu} = 
n_\mu \sqrt{|\gamma|} d^3x$ and $n^\mu$ is the outward-pointing unit normal vector. Applying a Wick-rotation, we find the Maxwell action transforms as
\begin{equation*}
     \frac{1}{8 \pi} \int_{\partial M} F^{\mu \nu} A_{\mu} d\Sigma_{\nu} \rightarrow  - i \frac{1}{8 \pi} \int_{\partial M} F^{\mu \nu} A_{\mu} d\Sigma_{\nu} \;,
\end{equation*}
where note explicitly that each pieces transforms as: $d\Sigma_{\mu} \rightarrow -id\Sigma_{\mu} $, $A_{\mu} \rightarrow -iA_{\mu}$, and $F^{\mu \nu} \rightarrow iF^{\mu \nu}$.

Taking all contributions together, the Euclidean action is 
\begin{equation}
\label{eq:euclact}
\begin{aligned}
            S_E = -i S_{\mathrm{Wick-rotated}}
        &= \frac{1}{16 \pi} \int_M \sqrt{g} (R - 2\Lambda) d^4 x  \\
        &- \frac{1}{8 \pi} \int_{\partial M} \sqrt{|\gamma|} (K - K_0) d^3x - \frac{1}{8 \pi} \int_{\partial M} F^{\mu \nu} A_{\mu} d\Sigma_{\nu} \;.
\end{aligned}
\end{equation}

\subsection{Triple Wick-rotation}
The standard simple Wick-rotation can be applied for static spacetimes
which upon continuation 
remain real, so that the Euclidean on-shell action can be 
interpreted as a thermal partition function. The static patches of the planar solutions
found in  \cite{Gutowski:2019iyo} take the form
\begin{equation}
\label{static_patch}
ds^2 = - f(r) dt^2 + \frac{dr^2}{f(r)} + r^2 (dx^2 + dy^2) \;,
\end{equation}
which at first appears suitable for this procedure. However, we also need smooth field
configurations to obtain a well-defined and finite Euclidean on-shell action. 
For the solutions of  \cite{Gutowski:2019iyo} the static patches have a curvature singularity for some finite value $r=r_{\text{sing}}$ of the transverse coordinate $r$, which makes the Euclidean on-shell action ill-defined. 

However, these static patches have a horizon at another finite value $r=r_h>r_{\text{sing}}$
of the coordinate $r$, and 
by analytic continuation one obtains a dynamic
patch, where the metric, after relabelling $r\leftrightarrow t$, takes the form
\begin{equation}
\label{dynamic_patch}
ds^2 = - \frac{dt^2}{\tilde{f}(t)} + \tilde{f}(t) dr^2 + t^2 (dx^2 + dy^2) \;.
\end{equation}
Note that the function $\tilde{f}(t)$ has been modified with an additional sign: $\tilde{f}(x) = -f(x)$. This ensures that $\tilde{f}(t)$ is positive definite within the domain of $t \in (t_h, \infty)$. For the remainder of the discussion, the tilde will be dropped and it is understood that functions $f$ appearing in the line element are positive definite for each patch, and that the coordinate denoted $t$ is timelike while 
the coordinate denoted $r$ is spacelike.

It was shown in  \cite{Gutowski:2019iyo} that these solutions have
a well behaved asymptotic behaviour for $t\rightarrow \infty$.\footnote{To be precise, 
there are two extensions of the static patch, and depending on the extension, $t\rightarrow
\infty$ either corresponds to future or to past timelike infinity.  We refer to the Appendix \ref{app:classification} for a discussion of the global structure of the solution.}
In the dynamic patch,
the horizontal Killing vector field is spacelike rather than timelike, 
and the application of the simple Wick-rotation leads to a complex line element and
action. To work with this dynamic patch, we will need to modify the standard Euclidean method. There are
some examples where complex line elements are used in the literature, the
canonical example being the Kerr metric \cite{Gibbons:1976ue}. In this
case, the generalisation is to admit timelike Killing vector fields which are not
hypersurface orthogonal, and the complexification
arises from cross terms in the line element. This is different from our case,
where the Killing vector field is still hypersurface orthogonal, but spacelike. 

We therefore explore an alternative procedure, which in principle 
can be applied to any metric which has no timelike-spacelike cross-terms, and
depends explicitly on time but not on the spatial coordinates.
We choose to Wick-rotate all three spacelike coordinates of the line element. Since the examples
for which we will obtain a well defined Euclidean action 
are of the form (\ref{dynamic_patch}), we denote the spatial coordinates $r,x,y$ so that
the triple Wick-rotation takes the form
\begin{equation*}
    (r,x,y) \rightarrow \pm i (r,x,y) \;,
\end{equation*}
where we admit either choice of sign.
As we work with the mostly plus conventions, the resulting Euclidean line element will be \emph{negative-definite}.

Applying this transformation to \eq{totact1} the Euclidean action associated with the triple Wick-rotation is calculated. The bulk contribution transforms as
\begin{equation*}
\begin{aligned}
    - \frac{1}{16\pi} \int_{M} (R - 2\Lambda) \sqrt{-g} d^4 x 
    \rightarrow 
     &- (\pm i)^{3}  \frac{1}{16\pi}  \int_{M} (R - 2\Lambda) \sqrt{-g} d^4 x \\
     = &\pm i \frac{1}{16\pi}  \int_{M} (R - 2\Lambda) \sqrt{-g} d^4x \;.
\end{aligned}
\end{equation*}
The GHY-term, as with the single Wick-rotation, transforms with the same sign for $\epsilon = \pm 1$.

\begin{enumerate}
    \item  
    For surfaces with a timelike unit normal
    \begin{equation*}
    	\epsilon = -1,  \qquad K \rightarrow K, \qquad \sqrt{\gamma} d^3x \rightarrow (\pm i)^{3} \sqrt{\gamma} d^3x.
    \end{equation*}
\item 
For surfaces with a spacelike unit normal, 
\begin{equation*}
	\epsilon = 1, \qquad K \rightarrow \mp i K, \qquad \sqrt{\gamma} d^3x \rightarrow (\pm i)^2 \sqrt{\gamma} d^3x,
\end{equation*}
\end{enumerate}
and we see that for either hypersurface, the GHY term transforms under a triple Wick-rotation as
\begin{equation*}
    + \frac{\epsilon}{8 \pi} \int_{\partial M} \sqrt{|\gamma|} (K - K_0) d^3x \rightarrow  \pm i \frac{1}{8 \pi} \int_{\partial M} \sqrt{|\gamma|} (K - K_0) d^3x.
\end{equation*}

As with the standard Wick-rotation, we can write the gauge field contribution as a boundary term as we evaluate the action on shell. Performing the triple Wick-rotation, we find
\begin{equation*}
     \frac{1}{8 \pi} \int_{\partial M} F^{\mu \nu} A_{\mu} d\Sigma_{\nu} \rightarrow  \mp i \frac{1}{8 \pi} \int_{\partial M} F^{\mu \nu} A_{\mu} d\Sigma_{\nu},
\end{equation*}
where we have used that $d\Sigma_\mu \rightarrow (\pm i)^3 d\Sigma_\mu$, $A_\mu \rightarrow \pm iA_\mu$ and $F^{\mu \nu} \rightarrow \mp iF^{\mu \nu}$. Piecing this all together, the triple Wick-rotated Euclidean action is given by
\begin{equation}
    \label{eq:3EucAct}
\begin{aligned}
        S_E = &\pm \frac{1}{16 \pi} \int_{M} \sqrt{g} (R - 2\Lambda) d^4 x 
        \\& \pm \frac{1}{8\pi} \int_{\partial M} \sqrt{|\gamma|}(K-K_0) d^3x 
        \mp \frac{1}{8\pi} \int_{\partial M} F^{\mu \nu } A_\mu d\Sigma_\nu.
\end{aligned}
\end{equation}
 We then identify the thermodynamic potential as we do in the standard formulation, evaluating the partition function $\mathcal{Z}$ in a saddle point approximation to obtain
\begin{equation}
\label{eq:thermalguess}
    \log \mathcal{Z} = - S_E(\beta, \mu) = - \beta \Omega\;,
\end{equation} 
where the inverse temperature $\beta$ and chemical potential $\mu$ can be expressed in terms of parameters of the triple-Wick rotated solution. 


\subsection{Surface gravity and temperature}
\label{sec:surfacetemp}
When working with the Euclidean action formalism, the temperature associated with a Killing horizon is usually determined by the periodicity of Euclidean time, which in turn is fixed by imposing the
absence of a conical singularity after Wick-rotation. The near horizon approximation of the line element 
has the Rindler-like metric 
\begin{equation*}
    ds^2 = - \kappa^2 r'^2 dt^2 + dr'^2 + (A + \cdots)  d\vec{X}^2,
\end{equation*}    
where $\kappa$ is the surface gravity, $r' = r-r_h$ is a shifted coordinate which vanishes
at the horizon, and $A$ is independent on $r'$. The term $d\vec{X}^2$ is the standard
line element on the unit-sphere or on the Euclidean plane, depending on whether we
impose spherical or planar symmetry. 
While the term proportional to $d\vec{X}^2$ in the line element is manifestly
regular for $r'=0$, there is a conical singularity in $(t,r')$-plane unless the Euclidean
time coordinate $t$ satisfies $\kappa t \sim \kappa t + 2\pi$ and thus is periodic with 
period $2\pi \kappa^{-1}$. This determines the temperature $T_H$ associated with the
horizon, $\beta = T_H^{-1} = 2\pi \kappa^{-1}$. We observe that since the surface gravity
$\kappa$ enters the line element quadratically, this procedure does not actually determine
whether $T_H$ is positive or negative. However, the sign can be set through computing
the Hawking temperature using curved spacetime quantum field \cite{hawking1975}, 
or the tunneling effect for a quantum particle \cite{Parikh:1999mf}.
As an aside, we note that by removing the conical singularity, the horizon becomes the origin in the Wick-rotated spacetime. As a result, the Wick-rotated spacetime has only one boundary, located at $r \rightarrow \infty$. This means that when we calculate boundary terms for the Euclidean action, there will only be asymptotic contributions.

The surface gravity $\kappa$ of a Killing horizon 
is defined by
\begin{equation}
\label{kappa}
\xi^\nu \nabla_\nu \xi^\mu \big|_{r = r_h} = \kappa \xi^\mu \;,
\end{equation}
evaluated on the horizon, where $\xi$ is a Killing vector field which is null on the horizon and non-null outside the
horizon. We observe that $\kappa$ changes sign under $\xi \rightarrow -\xi$.
For static, asymptotically flat spacetimes, the sign can be fixed by defining $\kappa$ 
to be the acceleration of a test mass at the horizon, multiplied by the redshift factor
\cite{Wald:106274}. This also fixes the magnitude of $\kappa$, which changes 
under rescalings of $\xi$. The standard formula \eqref{kappa} applies to the case where $\xi$ 
has unit norm at spatial infinity. 

Since we will investigate a  non-standard situation, we will not
assume that $\kappa$ and $T_H$ are positive. Furthermore, as the asymptotic region is not flat, we have to specify how we normalize the horizontal Killing vector field. For this purpose we will follow the work of  \cite{Hayward:1997jp} and \cite{Binetruy:2014ela, Helou:2015zma}, which 
provides a method for computing the surface gravity and temperature of 
trapping horizons. While the definition of an event horizon requires the knowledge
of the global causal structure of a spacetime, trapping horizons are defined 
quasi-locally by the existence of marginally trapped surfaces.
That is, on a trapping horizon the expansion of one of the two future-directed null congruences 
defined by ingoing and outgoing light rays changes sign, so that the horizon
separates a non-trapping region where one congruence expands and the other contracts
from a trapping region where both congruences either expand or contract. The Killing horizons
of the solutions  \cite{Gutowski:2019iyo} are event horizons, and thus in particular 
trapping horizons, so that the formalism can be applied.\footnote{While \cite{Hayward:1997jp} and \cite{Helou:2015zma} assume spherical symmetry, their formalism extends straightforwardly 
to situations with planar symmetry.} In the literature the term `trapping horizon' is used 
for hypersurfaces where the expansion of one null congruence vanishes, while 
spatial cross sections of a trapping horizon are called `apparent horizons.' We will use
both terms interchangeably.

In the so-called Kodama-Hayward approach \cite{Kodama:1980, Hayward:1997jp}, the surface gravity is obtained as follows. 
The metric is required to have the structure
\begin{equation*}
    ds^2 = \gamma_{ij}(x) dx^i dx^j + C^2(x) d\vec{X}^2 \;,\quad i=0,1,
\end{equation*}
where $\gamma_{ij}$\footnote{Note that here $\gamma$ is the metric for the transverse coordinates and is distinct from the boundary metric $\gamma$ used in our Euclidean action calculations. As it is 
unlikely to cause confusion, we allow this duplicity in our notation in order that this section uses the same conventions as the cited papers \cite{Binetruy:2014ela, Helou:2015zma}.} 
 and $C$ only depend on the coordinates $(x^0,x^1) = (t,r)$. For spherically symmetric spacetimes, $d\vec{X}^2 = d\Omega_2$ is the standard metric on the two-sphere. In our calculations, we allow planar symmetry and hence $d\vec{X}^2$ is the standard
metric on $\mathbb{R}^2$. The surface gravity in the Kodama-Hayward formalism is
\begin{equation*}
    \kappa = \frac{1}{2 \sqrt{-\gamma}} \partial_i \left(\sqrt{-\gamma} \gamma^{ij} \partial_i C 
    \right) = \frac{1}{2}
    \Delta_\gamma C \;.
\end{equation*}
For later reference, we compute the Kodama-Hayward surface gravity for line elements
of the form 
\begin{equation*}
    ds^2 = - \frac{dt^2}{f(t)} + f(t) dr^2 + t^2 d\vec{X}^2,
\end{equation*}
which include the dynamic patches of  \cite{Gutowski:2019iyo}, and obtain 
\begin{equation}
\label{eq:THK}
\kappa = -\frac{1}{2} \partial_t f(t).
\end{equation}
Following \cite{Binetruy:2014ela, Helou:2015zma},  trapping horizons and their 
Kodama-Hayward surface gravity subdivide into four cases, as follows. One chooses a local frame containing two future-directed 
null vectors, $N_\pm^\mu$ which are `outgoing' ($+$) and `ingoing' ($-$). Using the definition for the expansion
\begin{equation}
\theta_\pm = \nabla_\mu N^\mu_\pm,
\end{equation}
the four types of horizons are determined by calculating their expansions $\theta_\pm$ and their Lie derivative, evaluated on the horizon. 

Non-trapping regions in spacetime are those where $\theta_+ \theta_- <0$. The convention
taken in  \cite{Binetruy:2014ela, Helou:2015zma} is that $\theta_- < 0$ and $\theta_+ > 0$ in all 
non-trapping regions so that the outgoing congruence $N_+^\mu$ is diverging (or expanding) while 
the ingoing  congruence $N_-^\mu$ is converging (or contracting). Trapping regions are those where
$\theta_+ \theta_->0$, so that either both congruences expand, or both congruences
contract. Apparent horizons occur at boundaries where $\theta_\pm =0, \theta_\mp \not=0$. 

In Appendix \ref{app:classification}, we construct the maximal extensions of 
spacetimes with a line element of the form \eqref{static_patch}, using only
qualitative properties of the function $f$, namely its zeros and asymptotic behaviour. 
We compute the expansions of null congruences, identify the types of the trapping horizons and 
identify the global causal structure.  For comparison, we also include 
spacetimes like the Schwarzschild spacetime where the static region
is $r>r_h$ rather than $r<r_h$. 
In all cases the maximally extended spacetime consists of four regions, 
which are separated by trapping horizons which happen to be Killing horizons.
We identify one of the two non-trapping regions with the region where the line element 
\eqref{static_patch} is static, and use this `standard static patch' (or standard non-trapping
patch) to fix the direction of physical time for the extended spacetime. In this patch
we identify
two future-pointing null geodesic congruences $N^\mu_\pm$, such that their expansions satisfy 
$\theta_+>0$ and $\theta_-<0$. Then we choose Kruskal-like 
coordinates $\{X_+, X_-, \ldots \}$ in such a way that they are 
adapted to the standard patch, that is, such that in the standard patch 
ingoing future-pointing null congruences have constant $X_+$ and propagate towards increasing $X_-$, 
while outgoing future-pointing null congruences have constant $X_-$ and 
propagate towards increasing $X_+$. When extending the metric, the vector fields $N^\mu_\pm$ and the scalars $\theta_\pm$ to the full maximally extended spacetime, we observe that there always
is a second static, non-trapping patch. In this second static patch the roles of ingoing
and outgoing congruences are reversed so that we have the
non-standard assignments $\theta_+<0$ and $\theta_->0$. In addition, there always
are two trapping regions, one where both congruences expand, one where both contract. 
The transverse coordinate $r$ used in \eqref{static_patch} can always be 
extended beyond the trapping horizon and covers two regions of the extended spacetime,
depending on the choice of the static patch and the way in which we continue. 
We call regions interior regions when they contain a curvature singularity, and exterior regions
if null geodesics can be extended to infinite affine parameter in one direction. 
The coordinate $r$ takes values $r<r_h$ in the interior and $r>r_h$ in the
exterior regions. For our main examples, the de Sitter, planar Einstein-Maxwell and planar
STU solutions, the inner region $r<r_h$ is static while the exterior region $r>r_h$
is dynamic. We will therefore relabel $r\rightarrow t$ in exterior regions to emphasize
that this cooordinate is timelike. 
For more details, we refer the reader to Appendix \ref{app:classification}.

The variation of the expansions are given by their Lie derivatives $\La_\pm \theta_\mp$ 
with respect to the lightcone coordinates $X_\pm$. Since $\La_\pm \theta_\mp$ does not change
sign across horizons, there are four types of horizons \cite{Binetruy:2014ela, Helou:2015zma}.
\begin{enumerate}
\item
Future outer horizons: $\theta_+ =0, \; \theta_- <0, \; \La_{N_-}\theta_+ <0$. The sign of $\theta_+$ 
changes from positive to negative with growing $X_-$. The situation is analogous
to regions I and II of the extended Schwarzschild solution, see the left diagram in Figure
\ref{fig:comparison}. For sufficiently small $X_-$
(`outside the horizon')
the outgoing congruence is expanding, while for sufficiently large $X_-$ (`inside the
horizon') both
congruences contract. Therefore future outer horizons can be taken as local definitions of black holes. 
\item
Past outer horizons: $\theta_-=0, \; \theta_+ >0, \; \La_{N_+} \theta_-<0$. The sign of $\theta_-$ 
changes from positive to negative with growing $X_+$. For sufficiently large $X_+$
(`outside the horizon') 
the ingoing congruences are converging, while for sufficiently small $X_+$  (`inside the
horizon') 
both congruences expand. The small $X_+$ region is analogous to the
time-reflected region IV of the extended Schwarzschild solution, while the
large $X_+$ region is analogous to region I. 
Therefore past outer horizons can be taken as local
definitions of white holes. 
\item
Future inner horizons: $\theta_+= 0, \; \theta_-<0, \; \La_{N_-}\theta_+ >0$. The sign of $\theta_+$ 
changes from negative to positive with increasing $X_-$. The inside region (large $X_-$)
is non-trapping while in the outside region (small $X_-$) both congruences contract. 
Therefore future inner horizons can be taken as local definitions of contracting cosmologies, where all
null congruences become converging for large enough distances from the observer.
\item
Past inner horizons: $\theta_-=0, \; \theta_+ >0, \; \La_{N_+} \theta_->0$. The sign of $\theta_-$ changes
from negative on the inside (small $X_+$) to positive on the outside (large $X_+$). The interior
region is non-trapping while in the exterior region both congruences expand. 
Therefore
past inner horizons are local definitions of expanding cosmologies, where all null 
congruences become expanding for large enough distances from the observer. 
\end{enumerate}
The surface gravities of these horizons are related to variations of the expansions by 
$\kappa \propto -  \La_\pm \theta_\mp$. Thus outer horizons have 
positive surface gravity, while inner horizons have negative surface gravity. A further 
sign has been argued for in the relation between surface gravity and temperature. In \cite{Binetruy:2014ela, Helou:2015zma}
the Hawking temperature of an apparent horizon was computed using the Parikh-Wilczek
tunnelling method. It was found that $T_H \propto \pm \kappa$, with the upper sign for future
horizons and the lower sign for past horizons. The net effect is that future outer 
horizons (black holes), and past inner horizons (expanding cosmologies) have positive
temperature, while future inner horizons (contracting cosmologies) and past outer
horizons (white holes) have negative temperature. 

Negative temperature was 
argued to indicate the absence of Hawking radiation, since future inner and past 
outer horizons cannot separate virtual particle pairs created by vacuum fluctuations,
thus not enabling the Hawking effect \cite{Binetruy:2014ela, Helou:2015zma}. In thermodynamics,
the inverse temperature is related to the entropy $S$ and internal energy $E$ by
\begin{equation*}
    \beta = \pardev{S}{E}.
\end{equation*} 
Therefore, negative temperature can occur if one drops the usual assumption that
the entropy increases monotonically with the energy. A toy model for negative
temperature is provided by a system with finite maximum energy \cite{Ramsey:1956}.
Taking a system with two energy eigenstates $E_1< E_2$ as the simplest example, this will be in a maximally
ordered state, $S=0$, if all particles are either in the lower or in the higher state, while
a maximally disordered state is realized when half of the particles are in either state. 
Upon heating up such a system, entropy and temperature first increase,
with the temperature reaching $+\infty$  when entropy becomes maximal.
Upon further heating, the entropy decreases and the temperature jumps at the 
turning point from $+\infty$ to $-\infty$. After this point, it increases, approaching $0$ 
from below when reaching a situation where all particles are in the higher state. 
Thus negative temperatures are `higher' than positive temperatures and
correspond to `population inversion.'
We will see later that some of the horizons we are interested in have negative
surface gravity and negative temperature, and that this is necessary in order to
for the first law to take its standard form when using our triple Wick rotated Euclidean formalism.
 

\section{Thermodynamics of the de Sitter solution}

\label{sec:desitter}

As an introductory example of the implementation of the triple Wick-rotation in spacetimes with dynamic asymptotic regions, we study the de Sitter solution of Einstein's equations with a cosmological constant. This example is somewhat simpler than the solutions of  \cite{Gutowski:2019iyo} since
it is a vacuum solution. However, it allows us to demonstrate that the results we obtain 
using a triple Wick-rotation in the dynamic patch agree with those obtained previously 
using a single Wick-rotation in the static patch. 

\subsection{Static patch, single Wick-rotation}
The de Sitter spacetime line element in static coordinates is given by
\begin{equation}
\label{eq:desitterline}
    ds^2 = -\left(1 - \frac{r^2}{L^2} \right) dt^2 + \left(1 - \frac{r^2}{L^2} \right)^{-1} dr^2 + r^2 d\Omega^2_2,
\end{equation}
with the cosmological horizon located at $r_h = L$, where $L$ is the de Sitter radius and the domain of our radial coordinate is $r \in [0,r_h)$. At $r=r_h$ there is a Killing horizon for the Killing vector field
$\xi = \partial_t$, which becomes spacelike when we continue to $r>r_h$. 
The thermodynamics of de Sitter space can be 
calculated within the static patch $0<r<r_h$ using standard methods.
The cosmological constant $\Lambda$ can be written generally as a function of the de Sitter radius
\begin{equation*}
    \Lambda = -\frac{(d-1)(d-2)}{2 L^2} = -\frac{3}{L^2},
\end{equation*}
where for reference, we first give the relation for general dimension $d$ before setting 
$d=4$. Note the minus sign, which is due to 
our sign conventions where the cosmological constant is proportional to the 
Ricci scalar, while the Ricci scalar is negative for de Sitter space.
We expand on our conventions in Appendix \ref{sec:Conventions}.

Under the Wick-rotation $t \rightarrow -i \tau$, the line element \eq{desitterline} maps to the positive definite line element 
\begin{equation}
\label{eq:eucds}
    ds^2 = \left(1 - \frac{r^2}{L^2} \right) d\tau^2 + \left(1 - \frac{r^2}{L^2} \right)^{-1} dr^2 + r^2 d\Omega^2_2.
\end{equation}

\paragraph{Entropy:}
Using the Bekenstein-Hawking area law, the entropy is determined by
\begin{equation}
\label{eq:dsBH}
    S_{dS} = \frac{A}{4} = \pi^2 L.
\end{equation}

\paragraph{Temperature:}

\begin{figure}[!h]
\centering
\begin{tikzpicture}
 	\draw[thick] (0,0) -- (6,0) node[midway,below=0.2cm] {$\mathcal{J}^-$}
   -- (6,6) node[rotate=90,midway,below] {South Pole}
   -- (0,6) node[midway,above=0.2cm] {$\mathcal{J}^+$}
   -- (0,0) node[rotate=90,midway,above] {North Pole};

	\draw[dashed] (0,0) -- (6,6);
	\draw[dashed] (0,6) -- (6,0);
	
	\draw[thick, blue, ->] (5.5,5) to [bend right] (5.5,1);
	\draw[thick, blue, ->] (0.5,1) to [bend right] (0.5,5);
	\draw[thick, blue, ->] (1,0.5) to [bend left] (5,0.5);
	\draw[thick, blue, ->] (5,5.5) to [bend left] (1,5.5);
	
	\node (a) at (4,3) {I};
	\node (a) at (2,3) {IV};
	\node (a) at (3,4) {II};
	\node (a) at (3,2) {III};
\end{tikzpicture}
\caption{Penrose-Carter diagram for the global de Sitter solution. Dashed lines denote the cosmological horizon located for $r = L$ and the North/South poles are identified for $r = 0$. Blue curved arrows denote the flow of the Killing vector field. \label{de_Sitter}}
\end{figure}
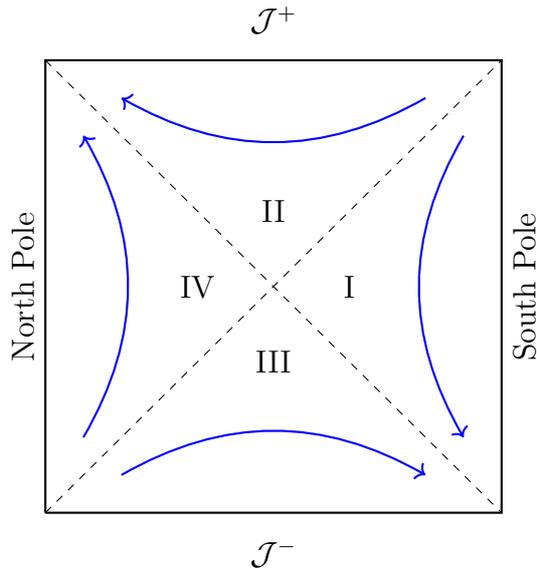

The temperature associated with the horizon is proportional to the Kodama-Hayward surface gravity, which is found to be $\kappa = -L^{-1}$, thus yielding the Hawking temperature
\begin{equation*}
    T_H = \frac{\kappa}{2 \pi} = - \frac{1}{2\pi L}.
\end{equation*}
Note that the Hawking temperature is negative. We employ the definitions of 
\cite{Binetruy:2014ela, Helou:2015zma} and consider horizons which can be crossed
by future directed null rays and future directed time-like curves (`observers') from 
the outside to the inside. These are the regions III and IV in Figure \ref{de_Sitter}, where
the global time orientation is chosen such that the Killing vector field is future-pointing
in region III, that is, globally time flows `upwards'.
This choice of regions is natural because it has the same
causal structure as the part of the extended Schwarzschild spacetime which 
describes a black hole (regions I and II in the left diagram of Figure \ref{fig:comparison}). 
As we show in Appendix  \ref{app:classification}, the horizon between regions III and IV
in the global de Sitter spacetime is a \emph{future inner horizon}, which therefore has
negative surface gravity and temperature. This is different from the assignments
made in other references, including \cite{Gibbons:1977mu, spradlin2001les},
where the Hawking temperature is positive: $T_H>0$. However according to 
\cite{spradlin2001les} this implies that the entropy is negative. A positive temperature for de Sitter horizons is consistent with considering the \emph{past inner} horizon separating regions IV and II. 
In contrast, in this paper 
the sign of the temperature is determined by the type of apparent horizon, but the
entropy is always defined by the area law and therefore positive. 
Note that the expression $TdS$ entering into the first law is the same in both approaches.

\paragraph{Euclidean action:}
Global de Sitter space is a maximally symmetric space of constant positive curvature 
with topology $\mathbb{R} \times S^3$. Its Kruskal diagram decomposes into 
four regions, two of which have a timelike Killing vector field and do not intersect the
boundary, which is spacelike with topology $S^3$. If we evaluate the Euclidean action on a static patch, the 
boundary terms do 
not contribute and the de Sitter action is completely determined by the bulk terms: 
\begin{equation*}
    S_E = \frac{1}{16\pi} \int_{M} \sqrt{g}(R - 2\Lambda),
\end{equation*}
where the Ricci curvature is constant: $R = -12 L^{-2}$ and the integral over the four-manifold gives
\begin{equation*}
    S_E = \frac{1}{16\pi} \left(-\frac{6}{L^2}\right) \int_0^\beta d\tau \int_{S^2} \sin\theta d\theta d\phi \int^0_{r_h} r^2 dr = -\pi L^2.
\end{equation*}
Note the limits on the integration of the radial coordinate $r$, which have been chosen to run from $r_h$, the origin of the Euclidean manifold, to the North pole for $r = 0$.\footnote{As this might be confusing, let us justify the integration bounds. Although we interpret $r = 0$ as the coordinate origin for static coordinates of the de Sitter solution, this is not the origin for the Wick-rotated Euclidean manifold. When we Wick-rotate, the location of the horizon: $r = r_h$, becomes the origin with the identification $\tau \simeq \tau + \beta$ made to avoid a conical singularity. Our integration limits are then chosen to match the conventions from the origin of the Euclidean space to the boundary and as such we integrate from $r = r_h$ to $r = 0$.}
As there are no charges in the solution, we work in the canonical ensemble
and we have the following relations:
\begin{equation*}
    \log(Z) = -S_E = -\beta F, \qquad F = E - T S .
\end{equation*}
The de Sitter solution is a maximally symmetric vacuum solution and thus interpreted as 
a ground state.  We therefore choose the natural normalisation $E=0$. Following from this we obtain
\begin{equation*}
    S_E = \beta F = - S \qquad \Rightarrow \qquad S = \pi L^2.
\end{equation*}
We see that the thermodynamic entropy matches with \eq{dsBH} and the first law is satisfied
though in a `degenerate way',  as the entropy is constant: $dS = 0 = TdS = dE = d(0) = 0$.

\subsection{Dynamic patch, triple Wick-rotation}

The static patch is not complete and by analytical extension of the coordinate $r$
through the Killing horizon to values $r > r_h$ we obtain a second, dynamical patch, 
with asymptotic region $r \rightarrow \infty$.
When crossing the horizon, the function $f(r)$ becomes strictly negative, and we find that the 
coordinates $t,r$ exchange their roles. The timelike coordinate $t$ becomes spacelike, while
the spacelike coordinate $r$ becomes timelike. We adopt the convention to relabel coordinates
in the dynamic patch, so that $t$ is always timelike and $r$ always spacelike. 

Then the  line element in the dynamic patch is
\begin{equation}
\label{eq:desitterasy}
    ds^2 = -\left( \frac{t^2}{L^2} -1\right)^{-1} dt^2 +\left(\frac{t^2}{L^2} - 1\right) dr^2 + t^2 d\Omega^2_2\;.
\end{equation}
The coordinate domain is $t\in (t_h, \infty)$ where $t_h$ is the Killing horizon located at $t_h=L$.
Note that while this cannot be read of from the local form of the line element, we have 
chosen the continuation from region IV to region III, so that $t\rightarrow \infty$ corresponds
to past timelike infinity, see Appendix \ref{app:classification} for details. This is relevant
because it determines the sign of the temperature. 

\paragraph{Triple Wick-rotation:}
We now perform a triple Wick-rotation where $r \rightarrow \pm i r$ and where 
the sphere $S^2$ is analytically continued to the hyperbolic plane
$\Ham_2$ by  $(\theta, \phi) \rightarrow \pm i (\theta, \phi)$. The line element \eq{desitterasy} is mapped to the \emph{negative-definite} line element
\begin{equation*}
    ds^2 = -\left( \frac{t^2}{L^2} -1\right)^{-1} dt^2 -\left(\frac{t^2}{L^2} - 1\right) dr^2 - t^2 d\Ham^2_2\;,\;\;\;
    d\Ham^2_2 = d\theta^2 + \sinh^2 \theta d\phi^2 \;.
\end{equation*}

\paragraph{Temperature \& entropy:} The temperature associated to the Killing horizon 
is the same as in the previous calculation. Using the Kodama-Hayward expression \eq{THK}, the surface gravity is found to be $\kappa = -L^{-1}$ and for a future inner horizon the Hawking temperature is
\begin{equation*}
T_H = \frac{\kappa}{2\pi} =  - \frac{1}{2\pi L}.    
\end{equation*}
The entropy is identical to the static solution and is given by
\begin{equation*}
    S_{dS} = \frac{A}{4} = \pi^2 L.
\end{equation*}

\paragraph{Euclidean action:}
The dynamical patches of global de Sitter space intersect the boundary, which 
is spacelike with topology $S^3$. 
Therefore we need to take boundary terms 
into account. After our triple Wick-rotation, the boundary has topology ${S}^1 \times \Ham_2$, 
where the radius of the $S^1$ is fixed by imposing the absence of a conical singularity. 

The Euclidean action for the triple-Wick-rotated system is 
\begin{equation*}
    S_E = \pm \frac{1}{16 \pi} \int_M \sqrt{g} (R-2\Lambda) d^4 x \pm \frac{1}{8 \pi} \int_{\partial M} \sqrt{\gamma} K d^3x  +  \int_{\partial M}  L_{ct}[\gamma] d^3x, 
\end{equation*}
where a counter term $L_{ct}$ has been included to remove divergences from the action. 
The boundary is at $t \rightarrow \infty$, and we  first  integrate $t$ in the domain $t \in [t_h, \epsilon^{-1})$ and then take the limit of $\epsilon \rightarrow 0$. The volume
\begin{equation*}
    \omega = \int_{\Ham_2} \sinh \theta d\theta d\phi,
\end{equation*}
of the hyperbolic plane is divergent. While one option in this situation is to work 
with densities, we keep $\omega$ as a formal constant which corresponds to
the parametric volume $\omega_{S^2} = 4 \pi$ of the two-sphere in the static patch.

The bulk term of the Euclidean action is 
\begin{equation*}
    S_{\text{Bulk}} = \pm \frac{1}{16 \pi} \int_M \sqrt{g} (R-2\Lambda) d^4 x \;,
\end{equation*}
where
\begin{equation*}
    R = -\frac{12}{L^2}, \qquad \Lambda = -\frac{3}{L^2}, \qquad \sqrt{g} = t^2 \sinh \theta.
\end{equation*}
Putting these into the action and integrating over the manifold we find:
\begin{equation*}
    \begin{aligned}
        S_{\text{Bulk}} &= \pm \frac{1}{16 \pi} \int_M \sqrt{g} (R-2\Lambda) d^4 x \\
        &= \pm \frac{1}{16 \pi} \left( -\frac{6}{L^2} \right) \int_0^\beta dr \int_{\Ham_2} \sinh \theta d\theta d\phi \int_L^{\epsilon^{-1}} dt \; t^2,\\
        &= \mp \frac{\beta \omega}{16 \pi} \frac{2}{L^2} \left(\frac{1}{\epsilon^3} - L^3\right) \;.
    \end{aligned} 
\end{equation*}
The Gibbons-Hawking-York term 
\begin{equation*}
    S_{\text{GHY}} =  \pm \frac{1}{8 \pi} \int_{\partial M} \sqrt{-\gamma} K d^3x,
\end{equation*}
can be calculated in the following way:
the normal vector to the boundary for constant $t$ is
\begin{equation*}
    n^\mu = \left(- \sqrt{f}, 0, 0, 0 \right) \quad \Rightarrow \quad n_\mu n^\mu = -1 \;.
\end{equation*}
The trace $K$ of the extrinsic curvature, evaluated on a surface of constant $t = t_0$,
can then be computed using \eqref{K_as_divergence}:
\begin{equation*}
    K = \nabla_\mu n^\mu = \frac{3t_0^2 - 2L^2}{t_0 L^2 \sqrt{f}}, \; \; \sqrt{- \gamma} = \sqrt{f(t_0)} t_0^2 \sinh \theta
     \;,
\end{equation*}
\begin{equation*}
    K \sqrt{-\gamma} = \frac{3 t_0^3}{L^2} - 2t_0 \; .
\end{equation*}
Combining these, we find that the boundary contribution at $t_0 = \epsilon^{-1}$ is:
\begin{equation*}
    \begin{aligned}
        S_{GHY} &=  \pm \frac{1}{8 \pi} \int_{\partial M} \sqrt{-\gamma} K d^3x \\
        &= \pm \frac{1}{8 \pi} \left(\frac{3}{L^2 \epsilon^3} - \frac{2}{\epsilon} \right) \int_0^\beta dr \int_{\Ham_2} \sinh \theta d\theta d\phi \\
        &= \pm \frac{\beta \omega}{8 \pi} \left(- \frac{2}{\epsilon} + \frac{3}{L^2 \epsilon^3}\right)  \;.
    \end{aligned}
\end{equation*}
The counter term is constructed from the geometric data of the boundary metric:
\begin{equation*}
    \int_{\partial M}  L_{ct}[\gamma] d^3x  = \int_{\partial M}  d^3x \sqrt{|\gamma|} (c_1 + c_2 R[\gamma]),
\end{equation*}
where $R[\gamma]$ is the Ricci curvature associated to the boundary manifold, and $c_{1,2}$ are renormalisation constants. We can expand out the counter terms in orders of $\epsilon$ and find:
\begin{equation*}
    \sqrt{|\gamma|} = \left(\frac{1}{L \epsilon^3} - \frac{L}{2 \epsilon} + \Op(\epsilon^1)  \right) \sinh \theta,
\end{equation*}
\begin{equation*}
    R[\gamma] \sqrt{|\gamma|} = \left(- \frac{2}{L \epsilon}  + \Op(\epsilon^1)  \right)\sinh \theta.
\end{equation*}
Comparing terms of order $\epsilon$ we find that the counter term is:
\begin{equation*}
    \int_{\partial M}  L_{ct}[\gamma] d^3x  = \mp \frac{1}{4 \pi L} \int_{\partial M} d^3x \sqrt{|\gamma|} \left(1 +  \frac{L^2}{4} R[\gamma] \right).
\end{equation*}
By construction, our action is now finite at the boundary $\epsilon\rightarrow 0$ and is of the form:
\begin{equation*}
    S_E = \pm \frac{\beta \omega}{8 \pi} \left(\frac{t_h^3}{L^2} \right) = \mp \frac{2 \pi L \omega}{8 \pi} \frac{L^3}{L^2} = \mp \frac{\omega L^2}{4}.
\end{equation*}
Picking the sign
\begin{equation*}
    (r,\theta,\phi) \rightarrow + i (r,\theta,\phi), 
\end{equation*}
for  the triple Wick-rotation, the signs of the Euclidean actions agree for both patches, 
and the actions only differ by the numerical factors $\omega, \omega_{S^2}=4\pi $. 
As these are numbers,
which we could eliminate by taking the Euclidean action per coordinate area,
the resulting thermodynamics is the same.

\section{Planar solutions to the Einstein-Maxwell theory}
\label{sec:pem}

Our next examples are vacuum solutions of the Einstein-Maxwell equations with
planar symmetry, or  `planar Reissner-Nordstr\"om solutions.' These solutions
are the simplest examples of a class of planar solutions to the STU model
of $\N=2$ supergravity  \cite{Gutowski:2019iyo}, and correspond to the limit
where all scalar fields are taken to be constant. It was shown in  \cite{Gutowski:2019iyo}
that planar Einstein-Maxwell solutions already show all the qualitative features
of the global causal structure of the full class of solutions. Similarly, we will see
in the next section that the thermodynamics of planar Einstein-Maxwell solution
is simpler than, but representative of, the thermodynamics of planar solutions
of the STU model. 

Following  our conventions, which are summarized in  Appendix \ref{sec:Conventions}, 
the Lorentzian bulk action for Einstein-Maxwell theory is
\begin{equation*}    
        S  = \frac{1}{16\pi} \int d^4x \, e \,  \left( - R - F^{\mu \nu} F_{\mu \nu} \right) .
\end{equation*}
In particular, we work in a convention where Newton's constant is set to unity, $G=1$, so that
the gravitational coupling $\kappa_4$ satisfies $\kappa_4^2 = 8\pi$. 
For later use we observe that the
Maxwell term has the coefficient $(16 \pi)^{-1} = (4 g^2)^{-1}$, 
where $g=\sqrt{4\pi} $ is interpreted as a coupling constant, see Appendix 
\ref{sec:cad}.

\subsection{Static patch}

Solving the Einstein-Maxwell equations while imposing planar symmetry and staticity
leads to a Ricci flat solution with the line element
\begin{equation}
\label{eq:planarEM}
    ds^2 = -f(r) dt^2 + \frac{dr^2}{f(r)} + r^2 (dx^2 + dy^2), \qquad f(r) = -\frac{2M}{r} + \frac{q^2}{r^2},
\end{equation}
where we must choose $M > 0$ in order to ensure the presence of a horizon.\footnote{Solutions with $M<0$ have naked singularities.} 
The transverse coordinate $r$ takes values in the interval $0 < r < r_h$, where
$r=0$ is the location of a curvature singularity, while $r_h$ is the location of a 
Killing horizon, where  $f(r_h) = 0$. 
Since we assume that the solution only carries electric charge,
 the gauge field is given by
\begin{equation}
\label{eq:gaugefield}
    F = \left( - \frac{q}{r} \right) dt \wedge dr\;.
\end{equation}
The gauge potential $A$ is found through integration of \eq{gaugefield} together with the 
standard boundary condition $A(r_h) = 0$:\footnote{See \cite{Dempster:2016} Appendix F for an explanation.}
\begin{equation}
\label{eq:gaugepotential}
    A = \left(- \frac{q}{r} + \frac{q}{r_h} \right) dt.
\end{equation}

\paragraph{Charge \& chemical potential:}
The chemical potential is given by the asymptotic value of the gauge potential \cite{Hartnoll:2009sz}; taking this limit for  \eq{gaugepotential} gives 
\begin{equation*}
    \mu :=  \lim_{r \rightarrow \infty} A_t = \frac{q}{r_h} = \frac{2 M}{q} \;.
\end{equation*}
Note that while $r \rightarrow\infty$ is outside the static patch $0<r<r_h$, 
we will see below that we can analytically extend spacetime to $0<r<\infty$, 
so that this limit makes sense. 
The conserved electric charge is computed using Gauss' law, which for planar symmetric gives
\begin{equation*}
    \mathcal{Q} = \frac{1}{4\pi} \int_{\Real^2} \star F = \frac{q \omega}{4\pi}, \qquad     \omega = \int_{\Real^2} dx \wedge dy \;.
\end{equation*}
Here $\omega$ is the divergent parametric area of the horizon, which we keep as a formal
constant to allow comparison to the spherically symmetric case. 
The factor of $4\pi$ is due to the normalization we have chosen for the gauge field.  In our conventions the volume form is defined using the conventional choice $\epsilon_{trxy} = 1$.

\subsection{Dynamic patch}

\begin{figure}[h!]
\centering
\begin{tikzpicture}[scale=0.8, every node/.style={scale=1}]

\node (I) at (0,6) {II};
\node (II) at (0,0) {III};
\node (III) at (-2,3) {IV};
\node (IV) at (2,3) {I};

\node (i) at (0,9.5) {$i^+$};
\node (i) at (0,-3.5) {$i^-$};

\path 
 (I) +(90:3) coordinate[label=90:] (IItop)
 +(-90:3) coordinate(IIbot)
 +(0:3) coordinate[label=360:] (IIright)
 +(180:3) coordinate[label=180:] (IIleft)
 ;
\draw[thick] (IIleft) -- (IItop) node[midway, above, sloped] {$\mathcal{J}^+$} -- (IIright)  -- (IIbot) -- (IIleft) -- cycle;

\path 
 (II) +(90:3) coordinate (Itop)
 +(-90:3) coordinate (Ibot)
 +(180:3) coordinate (Ileft)
 +(0:3) coordinate (Iright)
 ;
\draw[thick] (Ileft) -- (Itop) -- (Iright) -- (Ibot) node[midway, below, sloped] {$\mathcal{J}^-$} -- (Ileft) -- cycle;

\draw[decorate,decoration=snake,draw=black, thick] (Ileft) -- (IIleft);

\draw[decorate,decoration=snake, draw=black, thick] (Iright) -- (IIright);
\end{tikzpicture}
\caption{Conformal diagram of the planar 
Reissner-Nordstr\"om solution and planar solution of the STU model. 
The standard static region, where the Killing vector field $\partial_t$ is timelike and 
future-pointing is Region IV. For thermodynamics we consider the future inner horizon 
between Regions III and IV, which can be crossed by causal geodesics from the
outside to the inside.
\label{Fig:pRN}}
\end{figure}
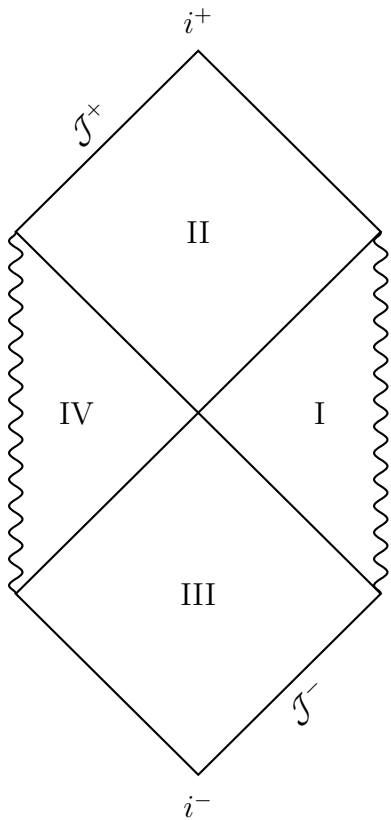

Due to the presence of a curvature singularity at $r=0$ we cannot apply the standard 
thermodynamic formalism in the static patch. By analytic continuation, using advanced
Eddington-Finkelstein coordinates at an intermediate step, see 
Appendix \ref{app:classification}, 
we can extend space time to the dynamical region $r_h < r < \infty$, 
where the horizontal Killing vector field becomes spacelike. Since $r$ becomes
a timelike coordinate in the dynamic patch, we apply the same convention as 
in the de Sitter example, and as in \cite{Gutowski:2019iyo}, we relabel the 
coordinates $(t,r) \rightarrow (r,t)$, and redefine $f$ by a minus sign. Then the line element of the dynamic 
patch takes the form
\begin{equation}
\label{eq:planarEM2}
    ds^2 = -\frac{dt^{2}}{f(t)} +  f(t) dr^{ 2}  + t^{2} (dx^2 + dy^2) \qquad f(t) = \frac{2M}{t} - \frac{q^2}{t^2} \qquad t_h = \frac{q^2}{2M},
\end{equation}
which covers region III of Figure \ref{Fig:pRN}, with $t\rightarrow \infty$ corresponding
to past timelike infinity. 
It has been shown in  \cite{Gutowski:2019iyo} that this line element becomes 
asymptotic to a Kasner cosmological solution in the limit $t \rightarrow \infty$.
Using advanced Eddington-Finkelstein coordinates, one can show that the Killing 
horizon between regions III and IV (and I)
is an apparent horizon of future inner type, consistent with the interpretation 
as a contracting cosmological solution, see Appendix \ref{app:classification}.

%
%

\paragraph{Temperature \& entropy:}

To compute the surface gravity and temperature of the future inner horizon, we use 
the Kodama-Hayward formalism.  Applying \eq{THK} to the line element \eq{planarEM2},
and taking into account that for future horizons the surface gravity and temperature have the same sign we 
obtain
\begin{equation}
\label{eq:EMtemp}
    \kappa = - \frac{4M^3}{q^4} \qquad \Rightarrow \qquad T_H = \frac{\kappa}{2\pi} = - \frac{2 M^3}{\pi q^4}.
\end{equation}
The Bekenstein-Hawking area law gives a relation for the entropy of the horizon in terms of the 
horizon area: 
\begin{equation*}
    S_{BH} = \frac{A}{4} = \frac{\omega t_h^2}{4} = \frac{q^4 \omega}{16M^2}.
\end{equation*}
As with other extensive quantities, we keep the divergent volume $\omega$ as a formal
constant rather than using densities. 

\subsection{Euclidean action}

Our main goal is to show that the future inner horizon satisfies the first law of horizon mechanics, which
takes the same form as the first law of thermodynamics. This requires to identify geometrically
defined quantities of the solution with thermodynamic quantities. In standard black hole 
thermodynamics the mass $M$ of the black hole is identified with the internal energy 
of a canonical or grand canonical ensemble. Due to the planar symmetry, and since
we are not working in a static patch,  we do not have a 
natural candidate for a mass-like quantity. Since there is no asymptotically flat region which we
could use to normalize the mass or the horizontal timelike Killing vector field, 
we cannot apply the ADM or Komar approach. We will trade this problem for the one
of obtaining a well behaved Euclidean action which we interpret as a grand canonical
partition function.  The mass-like quantity we identify with the internal energy is then obtained
using standard thermodynamic relations. The remaining problem in defining the Euclidean
action is its normalisation. For solutions which are asymptotic to a `vacuum', 
that is to a maximally symmetric spacetime, the normalisation is fixed by adding 
a boundary term such  
that the Euclidean action is zero when evaluated on the vacuum solution. We do not
have this option since our solution is not asymptotic to a maximally symmetric spacetime. 
Moreover, the GHY-boundary term will turn out to be finite, so there is no need to
add counterterms. However, the integral over the two planar directions is divergent, 
and while we can formally absorb this in a constant $\omega$, we will allow for 
a finite multiplicative factor $\mathcal{N}$ between the Euclidean action $S_E$ and the 
grand potential $\Omega$:
\begin{equation}
\beta \Omega = \mathcal{N} S_E \;.
\end{equation}
The constant $\mathcal{N}$ parametrizes the relative normalisation between
thermodynamic and geometric quantities. To fix it we can impose one relation,
which we choose to be Gauss' law. That is, we identify the charge $\mathcal{Q}$ defined
by the gauge field of our field configuration with the negative derivative of $\Omega$ with 
respect to the chemical potential
\begin{equation}
\label{eq:EMBoundary}
    \pardev{\Omega}{\mu} \overset{!}{=} - \mathcal{Q} \;.
\end{equation} 
Once $\mathcal{N}$ has been fixed by this condition, all thermodynamic relations 
must take their standard form, if our interpretation of $\mathcal{Z} = \exp( -\mathcal{N} S_E)$
as a thermodynamic partition function is correct.

Performing the triple Wick-rotation
\begin{equation*}
    (r,x,y) \rightarrow \pm i ({r}, {x},{y}),
\end{equation*}
we obtain the negative definite line element
\begin{equation}
    ds^2 = -f(t)^{-1} dt^2 - f(t) d{r}^2 - t^2 (d{x}^2 + d{y}^2).
\end{equation}
The Euclidean action is given by
\begin{equation*}
\begin{aligned}
        S_E = &\pm \frac{1}{16 \pi} \int_{M} \sqrt{g} R d^4 x \pm \frac{1}{8\pi} \int_{\partial M} \sqrt{|\gamma|}Kd^3x \\
        &\mp \frac{1}{8\pi} \int_{\partial M} F^{\mu \nu } A_\mu d\Sigma_\nu.
\end{aligned}
\end{equation*}
The bulk gauge field term has been transformed into a boundary term.
Since  the planar Einstein-Maxwell solution has a vanishing Ricci scalar, $R = 0$, the action is completely determined by the boundary terms, which are evaluated in the limit where $t \rightarrow \infty$.
We do not include a background boundary term, because $S_E$ will turn out to be finite. 

The hypersurface $\Sigma = \partial M$ is obtained as the limit of a sequence of slices
of the spacetime $M$ for constant time $t_0$, and has an extrinsic curvature with trace $K$ when 
considered as an embedded submanifold of $M$. The GHY-term is determined by 
$K$ and by the induced boundary metric $\gamma$ \cite{Brown:1992br}. 
It can be computed using the formulas reviewed in Appendix \ref{app:extrinsic_curvature}
with the result
\begin{equation*}
    K = \frac{3Mt_0 -q^2}{t_0^2 \sqrt{2 M t_0 - q^2}} ,\qquad \sqrt{|\gamma|} = t_0^2 \; \sqrt{\frac{2M}{t_0} - \frac{q^2}{t_0^2}}.
\end{equation*}
Evaluating this in the limit $t_0 \rightarrow \infty$ gives
\begin{equation*}
    S_{GHY} = \pm \frac{1}{8\pi} \int_{\partial M} \sqrt{|\gamma|} K  = \pm 
    \frac{3 M \beta \omega}{8\pi} \;.
    \end{equation*}
The factor $\beta \omega$ is the parametric volume of the boundary.     
After Wick-rotation the coordinate $r$ becomes periodic with period $\beta$, in order to
avoid a conical singularity at $t=t_h$.\footnote{To be precise, the conical method
determines the period up to sign, and we choose $\beta$ to have the sign 
determined  by the Kodama-Hayward method.}     
    
As the gauge potential has only one non zero component, the boundary term is simply calculated
\begin{equation*}
        \mp \frac{1}{8\pi} \int_{\partial M} F^{\mu \nu } A_\mu d\Sigma_\nu
    =  \pm \frac{M \beta \omega}{4 \pi}.
\end{equation*}
Together, the GHY-term and the gauge field contribution yield the Euclidean action
\begin{equation}
\label{eq:emeucact}
    S_E = \pm  5 \frac{M \beta  \omega}{8 \pi} \;.
\end{equation}

Formally equating the partition function calculated from the Euclidean action with the negative 
logarithm of 
the thermal partition function,  $\log(\mathcal{Z}) = -\mathcal{N} S_E = -\beta \Omega$,  yields the grand potential
\begin{equation*}
    \Omega(\beta, \mu) = \frac{\mathcal{N} S_E}{\beta} = 
    \mp 5 \mathcal{N}  \frac{\beta  \mu ^4 \omega}{(8 \pi)^2} \;,
\end{equation*}
which we have written in terms of its natural thermodynamic
variables\footnote{The thermodynamic identities used here and in the 
following have been summarized in Appendix \ref{app_thermo}.} $\beta=1/T$ and $\mu$  using that
\begin{equation*}
    M = - \frac{\beta \mu^4}{8 \pi}, \qquad \mathcal{Q} = -\frac{\mu^3 \beta \omega}{(4\pi)^2}.
\end{equation*} 
We now apply our normalisation condition \eq{EMBoundary}: 
the conserved charge $\mathcal{Q}$ calculated from Gauss' law 
must match the negative $\mu$-derivative of $\Omega$. This fixes
$\mathcal{N} = \mp \frac{1}{5}$ so that the grand potential is determined to 
be
\begin{equation} 
\Omega(\beta, \mu) =  \frac{\beta  \mu ^4 \omega}{(8 \pi)^2} \;.
\end{equation}
The free energy $F(\beta, \mathcal{Q})$ is obtained as the  Legendre transform of the grand potential
\begin{equation}
\label{eq:RNfreeenergy1}
    F(\beta,\mathcal{Q}) = \Omega -  \mu \frac{\partial \Omega}{\partial \mu} = 
    \Omega + \mu \mathcal{Q} = 
    3 \left(-\frac{\pi^2 \mathcal{Q}^4}{4 \beta \omega} \right)^{\frac{1}{3}}\;,
\end{equation} 
where we have used the relation 
\begin{equation*}
    \mu = \left(- \frac{16 \pi^2 \mathcal{Q}}{\omega \beta} \right)^{1/3} \;,
\end{equation*}
to express the free energy in terms of its natural variables $\beta$ and $\mathcal{Q}$. 
From $F$ we can compute the thermodynamic entropy $S$ and check that it
matches the Bekenstein-Hawking entropy $S_{BH}:$ 
\begin{equation}
\label{eq:entEM}
    S = \beta^2 \pardev{F}{\beta} 
    = \left( \frac{\pi^2 \mathcal{Q}^4 \beta^2}{4 \omega} \right)^{\frac{1}{3}}
    = S_{BH}  \;.
\end{equation}
As a further consistency check, we can also verify that the free energy gives us the 
correct chemical potential:
\begin{equation*}
    \pardev{F}{\mathcal{Q}} = \left(- \frac{16 \pi^2 \mathcal{Q}}{\beta \omega} \right)^{1/3} =  \mu.
\end{equation*}
The internal energy $E$, for which we do not have a geometric definition, is 
computed using the free energy:
\begin{equation*}
E =    \pardev{(F \beta)}{\beta} = 
 \left(- \frac{2 \pi^2 \mathcal{Q}^4  }{\beta \omega} \right)^{1/3} = \frac{M\omega}{4\pi} \;.
\end{equation*}
We observe that $E$ is proportional to the parameter $M$, and therefore $E$ is positive.

Using our previous results we can verify that the thermodynamic variables 
$E,T,S,\mu, \mathcal{Q}$ satisfy the Smarr relation
\begin{equation}
E = 2TS + \mu \mathcal{Q}  \;.
\end{equation}
Expressing the internal energy $E$ in terms of its natural variables $S$ and $\mathcal{Q}$ 
we obtain the equation of state
\begin{equation*}
    E(S,\mathcal{Q}) = \frac{\pi \mathcal{Q}^2 }{(S \omega)^{1/2}} \;.
\end{equation*}
The partial derivates of the internal energy are 
\begin{equation*}
    \pardev{E}{S} = - \frac{\pi \mathcal{Q}^2}{2 S^{3/2} \omega^{1/2}} = \frac{1}{\beta} = T, \qquad     \pardev{E}{\mathcal{Q}} = \frac{2 \pi \mathcal{Q}}{(S \omega)^{1/2}} = \mu,
\end{equation*}
where both expressions have been simplified by substituting in $S(\beta, \mathcal{Q})$ using \eq{entEM}. The variation of the internal energy is
\begin{equation*}
    dE = \pardev{E}{S} dS + \pardev{E}{\mathcal{Q}} d\mathcal{Q}
       = T dS + \mu d\mathcal{Q} \;.
\end{equation*}
This relation takes the standard form of the first law of thermodynamics. Note that
this works because we have allowed that the temperature is negative. If we had insisted that the 
temperature is positive, this would have resulted in a non-standard sign for the 
entropy term. 

\section{Planar solutions to the STU model}
\label{sec:pstu}

We are now in a position to turn to our main application, the planar cosmological solutions of
the STU model found in \cite{Gutowski:2019iyo}, for which we will establish thermodynamical relations,
including the first law. The general bosonic Lagrangian for $n$ vector multiplets
coupled to ${\cal N}=2$ supergravity is 
\begin{equation}
\label{Bos_Lag}
 e_4^{-1} \La = -\frac{1}{2 \kappa_4^2}R - \frac{1}{\kappa_4^2} g_{A\bar{B}} \partial_\mu z^A \partial^\mu \bar{z}^B + \frac{1}{4 \kappa_4^2} \I_{IJ} F^I_{\mu \nu} F^{J|\mu \nu} + \frac{1}{4 \kappa_4^2} \cR_{IJ} F^I_{\mu \nu} \star {F}^{J|\mu \nu},
\end{equation}
where compared to \cite{Gutowski:2019iyo} we have restored the four-dimensional gravitational
coupling $\kappa_4$, see for example \cite{Mohaupt:2000mj}. While we used standard 
supergravity conventions where $\kappa_4^2 = 1$ in  \cite{Gutowski:2019iyo}, it will be more
convenient in the following to use relativist's conventions where $G=1$ and $\kappa_4^2 = 8\pi$, in order to avoid non-standard 
numerical factors in thermodynamic relations. The couplings $g_{A\bar{B}}, 
\I_{IJ}$ and $\cR_{IJ}$, where $A,B = 1, \ldots, n$ and $I,J = 0, \ldots, n$
  are functions of the scalar fields $z^A$ which can be expressed in terms of a holomorphic function,
  called the prepotential. The STU model has three vector multiplets, and therefore
  there are three complex scalars $z^A$ and four gauge fields $F^I_{\mu \nu}$, including
  the graviphoton which belongs to the Poincar\'e supergravity multiplet. 
  The Hodge dual gauge fields are denoted $\star F^I_{\mu \nu}$. 
  We refer to 
 \cite{Gutowski:2019iyo} for details and only review the results which are directly relevant
 for the following calculations.

\subsection{Dynamic patch}

The line element in the dynamical patch of the planar symmetric cosmological solution is 
\begin{equation}
\label{eq:STUline}
    ds^2  = - \frac{\Ham(\zeta)}{\cW(\zeta)} d\zeta^2 +  \frac{\cW(\zeta)}{\Ham(\zeta)} d\eta^2 +  G(\zeta)  (dx^2 + dy^2),
\end{equation}
where all functions depend only on the timelike coordinate $\zeta$:
\begin{equation*}
\begin{aligned}
    \cW(\zeta) &= \alpha \zeta - 1, \\
    \Ham_a(\zeta) &= (\beta_a + \gamma_a \zeta) ,\\
    \Ham(\zeta) &= 2 \left(\Ham_0 \Ham_1 \Ham_2 \Ham_3 \right)^{\half}, \\
    G(\zeta) &= \zeta^2  \left[\left(1 + \frac{\beta_0}{\gamma_0 \zeta}\right)\left(1 + \frac{\beta_1}{\gamma_1 \zeta}\right)\left(1 + \frac{\beta_2}{\gamma_2 \zeta}\right)\left(1 + \frac{\beta_3}{\gamma_3 \zeta}\right)\right]^{\half}.
\end{aligned}
\end{equation*}
Compared to \cite{Gutowski:2019iyo} we have performed a rescaling 
$(\bar{x},\bar{y}) \mapsto (x,y)$ of the coordinates of the plane
which changes the corresponding part of the line element as follows
 \begin{equation*}
 \begin{aligned}
          \Ham(\zeta) (d\bar{x}^2 + d\bar{y}^2) &= 2\sqrt{\gamma_0\gamma_1\gamma_2\gamma_3} G(\zeta) (d\bar{x}^2 + d\bar{y}^2), \\ &=  G(\zeta)(dx^2 + dy^2)  \;.
 \end{aligned}
 \end{equation*}
 In our new parametrization the asymptotic form of the planar line element is
 $ds^2_2 = \zeta^2(dx^2 + dy^2)$. 
 
 The integration constants $\beta_a, \gamma_a$ are related to the integration constants found in the solution \cite{Gutowski:2019iyo} via the relations:
\begin{equation*}
    \begin{aligned}
        \beta_a &= \frac{2 K_a}{\alpha} \sinh \bigg(\frac{\alpha h_a}{2 K_a}\bigg), \\
        \gamma_a &= K_a \exp \left(-\frac{\alpha h_a}{2 K_a} \right),
    \end{aligned}
\end{equation*}
where 
\[
K_a = \left(Q_0, P^1, P^2, P^3 \right),
\]
are the four non-zero charges carried by the gauge fields $F^I_{\mu \nu}$. 
To avoid a proliferation of cases, we have chosen $Q_0, P^A$ to be 
positive.\footnote{Otherwise we would need to distinguish several cases, and to carry along $\pm$ signs,
which would be cumbersome without contributing any insights. See
\cite{Errington:2014bta,Dempster:2015,Gutowski:2019iyo} 
for a more detailed discussion.}
While $Q_0$ is an electric charge, $P^A$, $A=1,2,3$ are magnetic charges. 
Explicit formulae for the gauge fields
will be given below. 
The scalar fields are expressed as functions of $\zeta$ through
\begin{equation}
\label{eq:physcal}
  z^1 = -i \left( \frac{\Ham_0 \Ham_1 }{\Ham_2 \Ham_3} \right)^{\half}, \quad z^2 = -i \left( \frac{\Ham_0 \Ham_2 }{\Ham_1 \Ham_3} \right)^{\half}, \quad z^3 = -i \left( \frac{\Ham_0 \Ham_3 }{\Ham_1 \Ham_2} \right)^{\half}.
\end{equation}

Having reviewed the planar cosmological solution of \cite{Gutowski:2019iyo} we now 
apply the same procedure as for planar solutions of Einstein-Maxwell theory. 
The metric \eq{STUline} has a future inner horizon at $\zeta = \zeta_h = \alpha^{-1}$ and 
is asymptotic to a Kasner solution for $\zeta \rightarrow \infty$. 

\paragraph{Temperature:}
Using the Kodama-Hayward formulation, we find that the temperature associated with the future inner horizon is negative and of the form:
\begin{equation}
\label{eq:STUtemp}
\begin{aligned}
        T_H &= -\frac{1}{4 \pi} \partial_\zeta \left(\frac{W(\zeta)}{\Ham (\zeta)} \right) 
        \bigg|_{\zeta=\alpha^{-1}} \\
        &= -\frac{\alpha^{3}}{8 \pi} \left[\left(\alpha  \beta _0+\gamma _0\right) \left(\alpha  \beta
   _1+\gamma _1\right) \left(\alpha  \beta _2+\gamma _2\right) \left(\alpha  \beta
   _3+\gamma _3\right) \right]^{-\half}.
\end{aligned}
\end{equation}
We can simplify this by noting:
\begin{equation*}
    (\alpha \beta_a + \gamma_a) = K_a \exp\left(\frac{\alpha h_a}{2 K_a} \right) = \frac{K_a^2}{\gamma_a} \quad \Rightarrow \quad    T_H = -\frac{\alpha^3}{8 \pi} \frac{\sqrt{\gamma_0 \gamma_1 \gamma_2 \gamma_3}}{Q_0 P^1 P^2 P^3}.
\end{equation*}

\paragraph{Entropy:} Using the Bekenstein-Hawking area law we can compute the 
entropy of the solution:
\begin{equation*}
\begin{aligned}
        S_{BH} &= \frac{G(\zeta_h)}{4} = \frac{1}{4 \alpha^2} \exp\left[\frac{\alpha}{2} \left(\frac{h_0}{Q_0} + \frac{h_1}{P^1} + \frac{h_2}{P^2}+\frac{h_3}{P^3} \right) \right],  \\
        &= \frac{1}{4 \alpha^2} \frac{Q_0 P^1  P^2 P^3}{\gamma_0 \gamma_1 \gamma_2 \gamma_3}.
\end{aligned}
\end{equation*}
Since the planar STU solution has several integration constants, we will suppress the parametric volume
$\omega$ of the planar directions in this section by setting $\omega=1$. This can be interpreted
as either working with densities of divergent extensive quantities, or as compactitfying the 
planar dimensions on a two-torus.

\paragraph{Chemical potentials:}
The solution for the gauge field is \cite{Gutowski:2019iyo}: 
\begin{equation*}
    \begin{aligned}
        F^0_{\zeta \eta} = (\dot{A}^{0})_\eta = -\frac{Q_0}{2(\beta_0 + \gamma_0 \zeta)^2}, \qquad \tilde{F}_{A |\zeta \eta} = (\dot{\tilde{A}}_A)_\eta = \frac{P^A}{2(\beta_A + \gamma_A \zeta)^2}.
    \end{aligned}
\end{equation*}
Here $\tilde{F}_{A|\mu \nu}$ denote the duals of the gauge field $F^A_{\mu \nu}$. Since 
the gauge couplings are field dependent, dualisation is not just Hodge dualisation, but
involves inverting the couplings. We refer to Appendix \ref{sec:cad} for details. As shown there
the precise relation between gauge fields and dual gauge fields is
\begin{equation*}
    \tilde{F}_A = - \star \I_{AB} F^B \quad \Rightarrow \quad F^A = \star \I^{AB} \tilde{F}_B,
\end{equation*}
where $F^A, \tilde{F}_A$ are the two-forms corresponding to the gauge fields, and
where $\star$ is the Hodge-$\star$ operator. Note that the coupling matrix $\I_{IJ}$ 
is invertible, and in our convention is negative definite. The advantage of using
the fields $F^0, \tilde{F}_A$ instead of $F^0, F^A$ is that now all gauge fields
and charges appearing in the solution are electric.\footnote{This is for computational simplicity. In \cite{Hawking:1995ap}, the authors show how magnetic and electric black hole solutions are equivalent in the semi-classical approach applied here.} 
 The corresponding gauge
potentials are found by integration, subject to the standard boundary condition 
$A(\zeta_h) = \tilde{A}(\zeta_h) = 0$:
\begin{equation*}
    (A^{0})_\eta = -\frac{\gamma_0 (\alpha  \zeta -1)}{2 Q_0 \left(\beta _0+\gamma_0  \zeta \right)} ,
\qquad
    (\tilde{A}_{A})_\eta = \frac{\gamma_A (\alpha  \zeta -1)}{2 P^A \left(\beta_A+\gamma_A  \zeta \right)} .
\end{equation*}
We then take the asymptotic limit of the gauge potentials to obtain the chemical potentials
\begin{equation*}
    \mu^0 := \lim_{\zeta \rightarrow \infty} A^0_\eta = -\frac{\alpha}{2Q_0}, 
    \qquad 
    \tilde{\mu}_A := \lim_{\zeta \rightarrow \infty} \tilde{A}_{A |\eta} = \frac{\alpha}{2P^A}. 
\end{equation*}

\paragraph{Electromagnetic charges:}

As with the Einstein-Maxwell solution, the conserved charges are computed using 
Gauss' law. However, we need to take into account that the gauge couplings 
depend on the scalar fields.
The gauge field couplings come from $\I_{IJ}$ and were calculated explicitly in \cite{Gutowski:2019iyo} 
\begin{equation*}
    \I_{IJ} = \text{diag} \left(-stu, -\frac{tu}{s}, -\frac{su}{t}, -\frac{st}{u} \right),
\qquad
    \I^{IJ} = \text{diag} \left(-\frac{1}{stu}, -\frac{s}{tu}, -\frac{t}{su}, -\frac{u}{st} \right),
\end{equation*}
where
\begin{equation*}
    s = -\text{Im}(z^1), \qquad t = -\text{Im}(z^2), \qquad u = -\text{Im}(z^3).
\end{equation*}
Putting in the solution \eq{physcal} for the scalar fields $z^A$ we can write these couplings as:
\begin{equation}
\label{eq:gaugecoup}
\begin{aligned}
        \I_{00} &= -\left(\frac{\Ham_0^3}{\Ham_1 \Ham_2 \Ham_3}\right)^{\half}, \qquad \I_{11} = -\left(\frac{\Ham_0 \Ham_2 \Ham_3}{\Ham_1^3}\right)^{\half}, \\
        \I_{22} &= -\left(\frac{\Ham_0 \Ham_1 \Ham_3}{\Ham_2^3}\right)^{\half}, \qquad \I_{33} = -\left(\frac{\Ham_0 \Ham_1 \Ham_2}{\Ham_3^3}\right)^{\half}.
\end{aligned}
\end{equation}

The charge $\mathcal{Q}_0$ carried by the gauge field $F^0$ is 
\begin{equation}
\label{eq:gaussSTU}
    \mathcal{Q}_0 = \lim_{\zeta \rightarrow \infty} \frac{1}{8 \pi} \int \star (-\I_{00} F^0) ,
\end{equation}

We refer to Appendix \ref{sec:cad} for a derivation of the expressions for the charges.
Evaluating \eq{gaussSTU} we obtain the conserved charge\footnote{Actually
charge density, as we set $\omega=1$.} 
\begin{equation}
\label{eq:stuq}
    \mathcal{Q}_0 = -\frac{1}{16 \pi} \frac{Q_0}{\sqrt{\gamma_0 \gamma_1 \gamma_2 \gamma_3}} .
\end{equation}
We use the normalisation $\epsilon_{\eta \zeta x y} = 1$ for the 
volume form, which is the standard normalisation in the static patch of the solution, 
where $\eta$ is timelike and $\zeta$ spacelike. Note that the Hodge operator contains
a factor of $\zeta^2$, so that when we evaluate the integral in the limit $\zeta\rightarrow \infty$
we read out the coefficient of the leading term in the integrand, which is proportional 
to $1/\zeta^2$. This is the leading behaviour of the field strength $F^0$, while the coupling
$\mathcal{I}_{00}$  approaches a constant. 

As mentioned, we have dualised the magnetic field strengths $F^A$ and instead work with their electric duals $\tilde{F}_A$, but we must remember that when we dualise a gauge potential in the Lagrangian the corresponding coupling is inverted. This means the conserved dual electric charges are
\begin{equation*}
    \tilde{\mathcal{Q}}^A = \lim_{\zeta \rightarrow \infty} \frac{1}{8 \pi} \int \star (-\I^{AA} \tilde{F}_A),
\end{equation*}
which when evaluated on our solution take the values
\begin{equation}    
\label{eq:stup}    
    \tilde{\mathcal{Q}}^A = \frac{1}{16 \pi} \frac{P^A}{\sqrt{\gamma_0 \gamma_1 \gamma_2 \gamma_3}}.
\end{equation}
The dual electric charge $\tilde{\mathcal{Q}}^A$ can be related to the magnetic charge of $F^A$ by 
$\tilde{\mathcal{Q}}^A = - \mathcal{P}^A$. This relationship is expanded upon in Appendix \ref{sec:cad}.

\subsection{Euclidean action}
Employing the triple Wick-rotation
\begin{equation*}
    (\eta, x,y) \rightarrow \pm i (\eta, x, y),
\end{equation*}
the Euclidean line element has (negative) definite signature and is of the form:
\begin{equation}
    ds^2  = - \frac{\Ham(\zeta)}{\cW(\zeta)} d\zeta^2 - \frac{\cW(\zeta)}{\Ham(\zeta)} d\eta^2 - G(\zeta)  (dx^2 + dy^2).
\end{equation}
As we did with the Einstein-Maxwell solution, we evaluate the Euclidean action on-shell, which allows us to write the gauge contributions as boundary terms
\begin{equation*}
\begin{aligned}
        S_E = &\pm \frac{1}{16\pi} \int_{M} \sqrt{g} \left( R + 2 g_{A\bar{B}} \partial_\mu z^A \partial^\mu \bar{z}^B \right) d^4 x 
        \\ &\pm \frac{1}{8\pi} \int_{\partial M} \sqrt{|\gamma|} \; K d^3x 
        \\
        &\pm\frac{1}{16\pi} \int_{\partial M} (\I_{00} F^{\mu \nu | 0})A_\mu^0 d\Sigma_\nu \pm \frac{1}{16\pi} \int_{\partial M} (\I^{AA} \tilde{F}^{\mu \nu}_{A})\tilde{A}_{\mu | A} d\Sigma_\nu.
\end{aligned}
\end{equation*}
We have performed the dualisation procedure such that we work with a purely electric solution.

\paragraph{Cancellation of bulk terms:}
As in the much simpler case of Einstein-Maxwell theory, the bulk term does not 
contribute. This is non-trivial since the Ricci scalar does no longer vanish on-shell,
$R\not=0$. However, the gauge field contribution still is a boundary term, and the scalar contribution 
precisely cancels the gravitational term in the bulk.
The trace of Einstein's equation gives that
\begin{equation*}
    R_{\mu \nu} - \half g_{\mu \nu} R = - 8\pi T_{\mu \nu} \quad  \Rightarrow \quad R = 8\pi T \;.
\end{equation*}
In four dimensions the gauge fields do not contribute to the trace of the energy momentum 
tensor, which therefore is completely given by the scalars:
\begin{equation*}
    T = g^{\mu \nu} T_{\mu \nu} = -\frac{2}{8\pi} g_{A\bar{B}} \left(\partial_\mu z^A  \partial^\mu \bar{z}^B \right) ,
\end{equation*}
which shows that
\begin{equation*}
    -\half R = g_{A\bar{B}} \left(\partial_\mu z^A  \partial^\mu \bar{z}^B \right) ,
\end{equation*}
and therefore the bulk contribution of the solution vanishes. Note that when we
set the scalars constant we recover the electro-vac type solution of Einstein-Maxwell
theory considered in the previous section, 
which is not Ricci flat $R_{\mu \nu} \propto T_{\mu \nu}\not=0$, but has
vanishing Ricci scalar.
 
\paragraph{Calculation of boundary terms:}

With the bulk terms found vanishing, the Euclidean action for the planar solution of the STU model can be found from the boundary terms. Following the same method as for the planar Einstein-Maxwell solution, the GHY-term is calculated to be
\begin{equation*}
    \pm  \frac{1}{8\pi} \int_{\partial M} \sqrt{|\gamma|}K d^3x = \pm  \frac{3}{32\pi} \frac{\alpha \beta}{\sqrt{\gamma_0 \gamma_1 \gamma_2 \gamma_3}}  \;.
\end{equation*} 
The gauge field term is calculated, through simply substituting in the various components and taking the limit of $\zeta \rightarrow \infty$, obtaining
\begin{equation*}
        \pm \frac{1}{16 \pi} \int_{\partial M} (\I_{00} F^{\mu \nu | 0})A_\mu^0 d\Sigma_\nu = \mp \frac{1}{64\pi} \frac{\alpha \beta}{\sqrt{\gamma_0 \gamma_1 \gamma_2 \gamma_3}}
\end{equation*}
and similarly 
\begin{equation*}
        \pm \sum_{A=1}^3 \frac{1}{16 \pi} \int_{\partial M} (\I^{AA} \tilde{F}^{\mu \nu}_{A})\tilde{A}_{\mu | A} d\Sigma_\nu = \mp \frac{3}{64\pi} \frac{\alpha \beta}{\sqrt{\gamma_0 \gamma_1 \gamma_2 \gamma_3}}\;.
\end{equation*}    
Collecting these terms, the Euclidean action is found to be
\begin{equation*}
        S_E = \pm \frac{1}{32\pi} \frac{\alpha \beta}{\sqrt{\gamma_0 \gamma_1 \gamma_2 \gamma_3}} \;.\end{equation*}
As in the Einstein-Maxwell case we admit a multiplicative constant $\mathcal{N}$ in the
relation between the Euclidean action and the grand potential:
\begin{equation*}
    \Omega(\beta, \mu^0, \tilde{\mu}_A)  = \frac{\mathcal{N} S_E}{\beta} = \pm \frac{\mathcal{N}}{32\pi}\frac{\alpha}{\sqrt{\gamma_0 \gamma_1 \gamma_2 \gamma_3}}  \;.
\end{equation*}
The constant $\mathcal{N}$ is
fixed by imposing that one of the thermodynamic relations takes its standard form. We choose
to impose the relation between the $\mu^0$ derivative of $\Omega$ and the charge $\mathcal{Q}_0$:
\begin{equation}
\label{Constraint}
\left(\pardev{\Omega}{\mu^0} \right)_{\beta,\tilde{\mu}_A} \overset{!}{=} -\mathcal{Q}_0 = \frac{1}{16\pi}\frac{Q_0}{\sqrt{\gamma_0 \gamma_1 \gamma_2 \gamma_3}}.
\end{equation}
To impose this condition, we first need to express the grand potential $\Omega$ in terms of its
natural variables. This can be done using the relationship
\begin{equation*}
    \frac{\alpha}{\sqrt{\gamma_0 \gamma_1 \gamma_2 \gamma_3}} = \frac{2\beta}{\pi} \mu^0 \tilde{\mu}_1 \tilde{\mu}_2 \tilde{\mu}_3
\end{equation*}
with the result
\begin{equation}
    \Omega(\beta,\mu^0,\tilde{\mu}_A) = \pm \frac{\mathcal{N}}{16 \pi^2} 
     \beta \mu^0 \tilde{\mu}_1 \tilde{\mu}_2 \tilde{\mu}_3 \;.
\end{equation}
Taking the partial derivate we obtain the conserved charge from the grand potential
\begin{equation*}
        \left(\pardev{\Omega}{\mu^0} \right)_{\beta,\tilde{\mu}_A} = \pm 
        \frac{\mathcal{N}}{16 \pi^2} 
        \beta \tilde{\mu}_1 \tilde{\mu}_2 \tilde{\mu}_3 
= \mp \frac{\mathcal{N}}{16 \pi} \frac{Q_0}{\sqrt{\gamma_0 \gamma_1 \gamma_2 \gamma_3}}  \;.
\end{equation*}
Comparing this with (\refeq{Constraint}) we find that $\mathcal{N}=\mp 1$. 
This determines the grand potential to be
\begin{equation*}
    \Omega(\beta, \mu^0,\tilde{\mu}_A) = - \frac{1}{16 \pi^2} \beta \mu^0 \tilde{\mu}_1 \tilde{\mu}_2 \tilde{\mu}_3 = 
    -\frac{1}{8\pi} \frac{\alpha}{4\sqrt{\gamma_0 \gamma_1 \gamma_2 \gamma_3}}  \;.
\end{equation*}
Note that $\Omega$ has turned out to be 
independent of our choice of sign for the triple Wick-rotation. It is clear
that the other derivatives of $\Omega$ with respect to chemical potentials give the
correct corresponding charges.

To obtain the free energy we must Legendre transform the grand potential:
\begin{equation*}
    \begin{aligned}
        F(\beta, \mathcal{Q}_0, \tilde{\mathcal{Q}}^A) &= \Omega + \mu^0 \mathcal{Q}_0 + \tilde{\mu}_1 \tilde{\mathcal{Q}}^1 + \tilde{\mu}_2 \tilde{\mathcal{Q}}^2 + \tilde{\mu}_3 \tilde{\mathcal{Q}}^3, \\
        &= -\frac{1}{8\pi} \frac{\alpha}{4\sqrt{\gamma_0 \gamma_1 \gamma_2 \gamma_3}} + \frac{1}{8\pi} \frac{\alpha}{\sqrt{\gamma_0 \gamma_1 \gamma_2 \gamma_3}}, 
    \end{aligned}
\end{equation*}
and so the free energy is given by:
\begin{equation}
    F(\beta, \mathcal{Q}_0, \tilde{\mathcal{Q}}^A) = \frac{1}{8\pi} \frac{3\alpha}{4\sqrt{\gamma_0 \gamma_1 \gamma_2 \gamma_3}}.
\end{equation}
To express $F$ in terms of its natural thermodynamical variables we use
\begin{equation*}
    \beta = -\frac{8\pi}{\alpha^3}\frac{Q_0P^1P^2P^3}{\sqrt{\gamma_0 \gamma_1 \gamma_2 \gamma_3}} \quad \Rightarrow \quad \frac{\alpha}{\sqrt{\gamma_0 \gamma_1 \gamma_2 \gamma_3}} = \left(\frac{(16\pi)^5\mathcal{Q}_0 \tilde{\mathcal{Q}}^1 \tilde{\mathcal{Q}}^2 \tilde{\mathcal{Q}}^3}{2\beta}\right)^{\frac{1}{3}},
\end{equation*}
and obtain:
\begin{equation}
    F(\beta, \mathcal{Q}_0, \mathcal{P}^A) = \frac{3}{32\pi}\left(\frac{(16\pi)^5\mathcal{Q}_0 \tilde{\mathcal{Q}}^1 \tilde{\mathcal{Q}}^2 \tilde{\mathcal{Q}}^3}{2\beta}\right)^{\frac{1}{3}}.
\end{equation}

We can now verify that all remaining thermodynamic relations take their standard form. 
First we verify that the Bekenstein-Hawking entropy matches with the thermodynamic definition:
\begin{equation*}
    \begin{aligned}
        S &= \beta^2 \left(\pardev{F}{\beta} \right)_{\mathcal{Q}_0,\tilde{\mathcal{Q}}^A} = \frac{1}{4\alpha^2} \frac{Q_0P^1 P^2 P^3}{\gamma_0 \gamma_1 \gamma_2 \gamma_3}.
    \end{aligned}
\end{equation*}
A further consistency check comes from ensuring that the chemical potentials that were found from the gauge field satisfy  the standard thermodynamic relations for chemical potentials:
\begin{equation*}
    \mu^0 = \left(\pardev{F}{\mathcal{Q}_0} \right)_{\beta,\tilde{\mathcal{Q}}^A}
    = \frac{1}{16 \pi \mathcal{Q}_0} \frac{\alpha}{2 \sqrt{\gamma_0\gamma_1\gamma_2\gamma_3}}
    = -\frac{\alpha}{2 Q_0},
\end{equation*}
and for the dual gauge fields:
\begin{equation*}
    \begin{aligned}
        \tilde{\mu}_A &= \left(\pardev{F}{\tilde{\mathcal{Q}}^A} \right)_{\beta,\mathcal{Q}_0} = \frac{\alpha}{2 P^A},
    \end{aligned}
\end{equation*}
which matches exactly with the chemical potentials found from the asymptotic limit of the vector potentials.

The internal energy of our solution can now be defined by the relation
\begin{equation*}
    \begin{aligned}
        E = \left(\pardev{(\beta F)}{\beta} \right)_{\mathcal{Q}_0,\tilde{\mathcal{Q}}^A} &= \frac{1}{16\pi}\left(\frac{(16\pi)^5\mathcal{Q}_0 \tilde{\mathcal{Q}}^1
         \tilde{\mathcal{Q}}^2 \tilde{\mathcal{Q}}^3}{2\beta}\right)^{\frac{1}{3}}, \\
        &= \frac{1}{16\pi} \frac{\alpha}{\sqrt{\gamma_0 \gamma_1 \gamma_2 \gamma_3}}.
    \end{aligned}
\end{equation*}

Next we express the entropy in terms of its natural thermodynamic variables:
\begin{equation*}
    S(E,\mathcal{Q}_0,\tilde{\mathcal{Q}}^A) = -\frac{16\cdot 4\pi^2 \mathcal{Q}_0 \tilde{\mathcal{Q}}^1 \tilde{\mathcal{Q}}^2 \tilde{\mathcal{Q}}^3}{E^2}.
\end{equation*}
Note that the entropy is positive, due to $\mathcal{Q}_0 <0$ and $\mathcal{Q}^A >0$, 
see \eqref{eq:stuq} and \eqref{eq:stup}, bearing in mind that we we have chosen
$Q_0$ and $P^A$ to be positive.\footnote{Note that the sign of the entropy
does not change if change the signs of charges. We have just chosen certain
charges to be postive or negative in order to avoid carrying around $\pm$
signs or to distinguish several cases.}

We need to verify that the Hawking temperature of our solution satisfies the thermodynamic
relation
\begin{equation*}
\beta = \frac{1}{T_H} = \left(\pardev{S}{E} \right)_{\mathcal{Q}_0, \mathcal{P}^A}.
\end{equation*}
Taking the partial derivate of $S$  with respect to $E$ we find that 
\begin{equation*}
\left(\pardev{S}{E} \right)_{\mathcal{Q}_0 \tilde{\mathcal{Q}}^A} = \frac{16\cdot 8\pi^2 \mathcal{Q}_0 \tilde{\mathcal{Q}}^1 \tilde{\mathcal{Q}}^2 \tilde{\mathcal{Q}}^3}{E^3}.
\end{equation*}
To compare this with the Hawking temperature we restore the original integration constants:
\begin{equation*}
        \left(\pardev{S}{E} \right)_{\mathcal{Q}_0, \tilde{\mathcal{Q}}^A} 
        = -\frac{8\pi Q_0 P^1 P^2 P^3}{\alpha^3 \sqrt{\gamma_0 \gamma_1 \gamma_2 \gamma_3}} = \beta.
\end{equation*}
Thus the Hawking temperature $T_H$, calculated from the geometry of the solution agrees with the thermodynamic 
quantity $T = \partial E/\partial S$.

\paragraph{Smarr relation:}
Evaluating the grand potential we find
\begin{equation*}
        \Omega = E - TS - \mu^0 \mathcal{Q}_0 -\tilde{\mu}_A \tilde{\mathcal{Q}}^A
        = - \frac{\alpha}{32\pi \sqrt{\gamma_0\gamma_1\gamma_2\gamma_3}} = TS \;,
\end{equation*}
which we can rearrange in the form of a 
standard Smarr relation
\begin{equation}
    E = 2T S + \mu^0 \mathcal{Q}_0 +\tilde{\mu}_A \tilde{\mathcal{Q}}^A.
\end{equation}
\paragraph{First law of thermodynamics:} 
We wish to verify the first law:
\begin{equation*}
    dE = T_HdS + \mu^0 d\mathcal{Q}_0 + \tilde{\mu}_1 d\tilde{\mathcal{Q}}^1 + \tilde{\mu}_2 d\tilde{\mathcal{Q}}^2 + \tilde{\mu}_3 d\tilde{\mathcal{Q}}^3.
\end{equation*}
The total differential of $E$ is
\begin{equation*}
    dE = \left(\pardev{E}{S} \right)dS + \left(\pardev{E}{\mathcal{Q}_0} \right) d\mathcal{Q}_0 + \left(\pardev{E}{\tilde{\mathcal{Q}}^1} \right) d\tilde{\mathcal{Q}}^1 + \left(\pardev{E}{\tilde{\mathcal{Q}}^2}\right) d\tilde{\mathcal{Q}}^2 + \left(\pardev{E}{\tilde{\mathcal{Q}}^3} \right) d\tilde{\mathcal{Q}}^3.
\end{equation*}
Having already found that
\begin{equation*}
    \left(\pardev{E}{S} \right) = T_H,
\end{equation*}
we turn our attention to the derivatives with respect to the charges. Using that:
\begin{equation*}
    E^2 = - \frac{16\cdot 4\pi^2 \mathcal{Q}_0 \tilde{\mathcal{Q}}^1 \tilde{\mathcal{Q}}^2 \tilde{\mathcal{Q}}^3}{S},
\end{equation*}
we find
\begin{equation*}
        \left(\pardev{E}{\mathcal{Q}_0} \right)_{S,\tilde{\mathcal{Q}}^A} = -\frac{1}{2E} \frac{16\cdot 4\pi^2 \tilde{\mathcal{Q}}^1 \tilde{\mathcal{Q}}^2 \tilde{\mathcal{Q}}^3}{S} = -\frac{\alpha}{2Q_0} = \mu^0.
\end{equation*}
Taking derivatives with respect to the magnetic charges we verify
\begin{equation*}
    \left(\pardev{E}{\tilde{\mathcal{Q}}^A} \right)_{S,\mathcal{Q}_0} = \frac{\alpha}{2P^A} = \tilde{\mu}_A.
\end{equation*}
Hence we see that the first law of thermodynamics holds.

\section{Comparison to the isolated horizon formalism \label{Sect:Isolated_horionzs}}

An alternative way of formulating the first law is to work entirely on the horizon. This allows one to calculate thermodynamic variables in the static region of the spacetime where the usual definitions of the thermodynamic constants hold. From \cite{Ashtekar:2000hw}, we find that the first law using variables defined on the horizon is:
\begin{equation}
    \label{eq:iso1st}
    \delta E_\Delta = \frac{\kappa \delta a_\Delta}{8\pi G} + \mu_a \delta \mathcal{Q}^a.
 \end{equation}
The ambiguity of the energy in the spacetime is fixed by imposing that the infinitesimal energy (mass) is equal to the RHS \eq{iso1st}. For the remainder of this discussion, we set $G=1$.
We note here that this derivation of the first law of black hole mechanics is still only self-consistent as this does not give a direct way to measure the mass outside of the first law itself.
 The subscript $\Delta$ denotes variables evaluated on the isolated horizon $\Delta$, which for us is the location of our Killing horizon at $\zeta = \alpha^{-1}$. The contracted $a$ index denotes the multiple charges (in our case, we have 4). 

\subsection{Planar solutions of the STU model}

Before we begin, we must make a coordinate change into Eddington-Finkelstein coordinates. From there we identify a Killing vector $\ell$ which we use to find $\kappa$. In a similar way to the previous section, we then determine the electromagnetic terms, but this time evaluated on the horizon rather than for $\zeta \rightarrow \infty$. 

\paragraph{Eddington-Finkelstein coordinates:}
Beginning with the metric from the dynamic region of the spacetime
\begin{equation*}
        ds^2  = - \frac{\Ham(\zeta)}{\cW(\zeta)} d\zeta^2 + \frac{\cW(\zeta)}{\Ham(\zeta)} d\eta^2 + G(\zeta)  (dx^2 + dy^2),
\end{equation*}
we make the coordinate change:
\begin{equation*}
    \eta = u + \bar{\zeta}(\zeta), \qquad d\eta = du + \bar{\zeta}^\prime d\zeta,
\end{equation*}
which we can substitute into the metric to obtain:
\begin{equation*}
    ds^2 = \left(\frac{\cW}{\Ham} (\bar{\zeta}^\prime)^2 - \frac{\Ham}{\cW} \right) d\zeta^2 + \frac{\cW}{\Ham} du^2 + \frac{2\cW}{\Ham} \bar{\zeta}^\prime dud\zeta + G(dx^2 + dy^2).
\end{equation*}
By making the choice
\begin{equation*}
    \bar{\zeta}^\prime = \frac{\Ham}{\cW},
\end{equation*}
we obtain the EF metric which is well defined for $\zeta = \alpha^{-1}$:
\begin{equation}
\label{eq:efmet} 
ds^2 =     \frac{\cW}{\Ham} du^2 + 2dud\bar{\zeta} + G( dx^2 + dy^2).
\end{equation}
This allows us to identify a suitable null normal vector field
\begin{equation*}
    \ell = \pardev{}{u}.
\end{equation*}

\paragraph{Surface gravity and area term:}
We can reuse our calculation for the surface gravity from \eq{STUtemp} to find
\begin{equation*}
    \kappa = -\frac{\alpha}{2} \frac{1}{ \Ham(\alpha^{-1})}.
\end{equation*}
From the metric \eq{efmet} we can read off the infinitesimal change in the area as:
\begin{equation*}
    \delta a_\Delta = \delta(G(\alpha^{-1})) = \delta(c \Ham(\alpha^{-1})) = \Ham(\alpha^{-1}) \delta c + c \delta \Ham(\alpha^{-1}) \ , \quad c = \frac{1}{2 \sqrt{\gamma_0 \gamma_1 \gamma_2 \gamma_3}},
\end{equation*}
where as in the Euclidean action formalism we have set
\begin{equation*}
     \omega = \int dx \wedge dy = 1,
\end{equation*}
for the area of the planar directions. 
 
Putting these together we find that the first term on the RHS of \eq{iso1st} is given by:
\begin{equation*}
    \kappa \delta a_\Delta = -\frac{\alpha}{2} \left( \delta c + \frac{c \delta \Ham(\alpha^{-1})}{\Ham(\alpha^{-1})} \right).
\end{equation*}
We can express:
\begin{equation*}
    \Ham(\alpha^{-1}) = \frac{4 c}{\alpha^2} Q_0 P^1 P^2 P^3,
\end{equation*}
to find:
\begin{equation*}
\begin{aligned}
        \kappa \delta a_\Delta &= -\half \alpha  \delta c - \half \alpha  c \delta \log(c\alpha^{-2} Q_0 P^1 P^2 P^3) \\
        &= -\alpha  \delta c +  c \delta \alpha - \half \alpha  c \delta \log(Q_0 P^1 P^2 P^3).
\end{aligned}
\end{equation*}
\paragraph{Gauge fields and charges:}
From the previous calculation we found that the gauge field strengths are given by the relations:
\begin{equation*}
    \begin{aligned}
        F^0_{\zeta \eta} = -\frac{Q_0}{2(\beta_0 + \gamma_0 \zeta)^2}, \qquad \tilde{F}_{A |\zeta \eta} = \frac{P^A}{2(\beta_A + \gamma_A \zeta)^2}.
    \end{aligned}
\end{equation*}
We need to express the gauge field strength in terms of the EF coordinates and then write down the field strength and the correpsonding gauge couplings on the horizon. Starting with the gauge field strengths, we see that they are all of the form:
\begin{equation*}
    F = f(\zeta) d\zeta \wedge d\eta.
\end{equation*}
Defining a null basis:
\begin{equation*}
        ds^2 = 2 e^+ e^- + \delta_{ij} e^i e^j,
\end{equation*}
\begin{equation*}
        e^+ = du \ , \quad e^- = d\zeta + \frac{\cW}{2\Ham} du \ , \quad e^+ \wedge e^- = - d\zeta \wedge du,
\end{equation*}
and we can easily take the Hodge dual:
\begin{equation*}
    F = - f(\zeta) e^+ \wedge e^- \ , \qquad \star F = f(\zeta) e^1 \wedge e^2 = c \Ham f(\zeta) dx \wedge dy .
\end{equation*}
This allows us to write down the Hodge duals explicitly:
\begin{equation*}
    \star F^0 = - \frac{Q_0}{2(\beta_0 + \gamma_0 \zeta)^2} c\Ham dx\wedge dy \ , \qquad     \star \tilde{F}_A = \frac{P^A}{2(\beta_A + \gamma_A \zeta)^2} c\Ham dx\wedge dy.
\end{equation*}
Evaluated on the horizon, these gauge fields are:
\begin{equation*}
        \star F^0_\Delta = - \frac{Q_0 \alpha^2}{2(\beta_0 \alpha + \gamma_0)^2} c\Ham(\alpha^{-1}) dx\wedge dy \ , \qquad     \star \tilde{F}_{A|\Delta} = \frac{P^A \alpha^2}{2(\beta_A \alpha + \gamma_A)^2} c\Ham(\alpha^{-1}) dx\wedge dy.
\end{equation*}
The last step is to take the gauge couplings \eq{gaugecoup} and evaluate them on the horizon. We find that:
\begin{equation*}
    \I_{00|\Delta} =  -\frac{2(\alpha \beta_0 + \gamma_0)^2}{\alpha^2 \Ham(\alpha^{-1})} \ , \qquad  \qquad \I_{AB|\Delta} = -\delta_{AB} \frac{\alpha^2 \Ham(\alpha^{-1})}{2(\alpha \beta_A + \gamma_A)^2}.
\end{equation*}
We are now in the position to calculate the conserved charges using the integrals\footnote{Note the addition of the extra minus sign as $\mathcal{I}_{IJ} < 0$, see Appendix \ref{sec:cad}.}:
\begin{equation*}
    \mathcal{Q}_{0|\Delta} = -\frac{1}{8 \pi} \int_{\Real^2} \star F^0_\Delta \I_{00| \Delta} \ ,\qquad     \tilde{\mathcal{Q}}^A_\Delta = -\frac{1}{8 \pi} \int_{\Real^2}  \star \tilde{F}_{A|\Delta} \I^{AA}_\Delta ,
\end{equation*} 
which we can calculate by substituting in the above results to find:
\begin{equation*}
    \mathcal{Q}_{0|\Delta} =  \left(-\frac{Q_0 \alpha^2}{2 (\alpha \beta_0 + \gamma_0)^2} c \Ham(\alpha^{-1}) \right) \cdot \left(\frac{2(\alpha \beta_0 + \gamma_0)^2}{\alpha^2 \Ham(\alpha^{-1})} \right) = - \frac{c Q_0}{8 \pi},
\end{equation*}
\begin{equation*}
\tilde{\mathcal{Q}}^A_\Delta = \left(\frac{P^A \alpha^2}{2 (\alpha \beta_A + \gamma_A)^2} c \Ham(\alpha^{-1}) \right) \cdot \left(\frac{2(\alpha \beta_A + \gamma_A)^2}{\alpha^2 \Ham(\alpha^{-1})} \right) = \frac{c P^A}{8 \pi}.
\end{equation*}

\paragraph{Gauge potential and chemical potential:}
In the previous section the gauge fields were found to be:
\begin{equation*}
(A^{0})_\eta = \frac{Q_0}{2\gamma_0(\beta_0 + \gamma_0 \zeta)}
\ , \qquad \qquad     (\tilde{A}_A)_\eta = -\frac{P^A}{2\gamma_A (\beta_A + \gamma_A \zeta)}.
\end{equation*}
Re-expressing these in terms of the new EF coordinates, evaluated on the horizon, we find that:
\begin{equation*}
    A^0_\Delta = \frac{Q_0}{2\gamma_0} \frac{\alpha}{\alpha \beta_0 + \gamma_0} \left(du + \frac{\Ham}{\cW} d\zeta \right),
\end{equation*}
\begin{equation*}
    \tilde{A}_{A|\Delta} = -\frac{P^A}{2\gamma_A} \frac{\alpha}{\alpha \beta_A + \gamma_A} \left(du + \frac{\Ham}{\cW} d\zeta \right).
\end{equation*}
Contracting with the null vector $\ell$ we find the chemical potentials from the identities (this is justified in \cite{Ashtekar:2000hw}):
\begin{equation*}
    \mu^0 = -\iota_\ell A^0 \ , \qquad \tilde{\mu}_A = - \iota_\ell \tilde{A}_A .
\end{equation*}
Simplifying the gauge potential using that
\begin{equation*}
\gamma_a (\alpha \beta_a + \gamma_a) = K_a^2   , 
\end{equation*}
we can write down the chemical potentials:
\begin{equation*}
    \mu^0 = - \frac{\alpha}{2 Q_0} \ , \qquad \tilde{\mu}_A =  \frac{\alpha}{2 P^A}.
\end{equation*}

Now we can write down the second term on the RHS of \eq{iso1st} by combining this with the conserved charge from the previous expression to find:
\begin{equation*}
\begin{aligned}
        \mu_a \delta \mathcal{Q}^a_\Delta &= \frac{\alpha}{16 \pi} \left(\frac{1}{Q_0} \delta( c Q_0) + \frac{1}{P^1} \delta( c P^1) + \frac{1}{P^2} \delta( c P^2) + \frac{1}{P^3} \delta( c P^3) \right) \\
        &= \frac{\alpha }{16 \pi} \left( 4 \delta c  + c \delta \log \left(Q_0 P^1 P^2 P^3 \right) \right) \\
        &= \frac{1}{8\pi} \left( 2 \alpha  \delta c + \half \alpha  c \delta \log \left(Q_0 P^1 P^2 P^3 \right) \right).
\end{aligned}
\end{equation*}

\paragraph{First law of black hole mechanics:}
Using these quantities we are now able to find an expression for the variation of the mass parameter from \eq{iso1st}. Combining our results we find that:
\begin{equation*}
    \begin{aligned}
        \frac{\kappa \delta a_\Delta}{8\pi} + \mu_a \delta \mathcal{Q}^a_\Delta &= {1 \over 8 \pi} \bigg[ -\alpha  \delta c +  c \delta \alpha - \half \alpha  c \delta \log(Q_0 P^1 P^2 P^3)\\
        &+ 2 \alpha  \delta c + \half \alpha  c \delta \log \left(Q_0 P^1 P^2 P^3 \right) \bigg] \\
        &= {1 \over 8 \pi} \big( \alpha \delta c +  c \delta \alpha \big) = \delta \left( {1 \over 8 \pi} \alpha c \right) \\
        &= \delta E_\Delta,
    \end{aligned}
\end{equation*}
and so:
\begin{equation*}
    E_\Delta = {\alpha c\over 8 \pi} = \frac{\alpha}{16 \pi \sqrt{\gamma_0 \gamma_1 \gamma_2 \gamma_3}} \;,
\end{equation*}
which is identical to the calculation from the Euclidean action.

\paragraph{Smarr relation:}
Our last consistency check comes from the Smarr relation we derived in the Euclidean formalism. Taking each piece and summing together we calculate that:
\begin{equation*}
\begin{aligned}
        \mu_a \mathcal{Q}^a_\Delta + {\kappa a_\Delta \over 8 \pi} &=  {1 \over 8 \pi} \bigg[c Q_0 \cdot \frac{\alpha}{2Q_0} +  c P^1 \cdot \frac{\alpha}{2P^1} +  c P^2 \cdot \frac{\alpha}{2P^2} \\ 
        &+  c P^3 \cdot \frac{\alpha}{2P^3} -  c \Ham(\alpha^{-1}) \cdot \frac{\alpha}{2 \Ham(\alpha^{-1})} \bigg]\\
        &= \frac{ \alpha c}{16 \pi} (1 + 1 + 1 + 1 - 1) \\
        &= \frac{3}{16 \pi}  \alpha c = \frac{3 E_\Delta}{2},
\end{aligned}
\end{equation*}
which is identical to the one from the Euclidean action formalism.

\subsection{Planar solutions to Einstein-Maxwell theory}

Einstein-Maxwell theory is a consistent truncation of the STU model where all four gauge fields are 
set equal, while the scalar fields are constant. Therefore 
we can map the solution of the STU model to that of Einstein-Maxwell model by fine tuning the integration constants. The physical scalar fields are given by
\begin{equation*}
    z^A = i \Ham_A \bigg(-\frac{\Ham_0}{\Ham_1 \Ham_2 \Ham_3} \bigg)^{\frac{1}{2}},
\end{equation*}
and we see that they are everywhere constant under the restriction that $\Ham_0 = \Ham_1 = \Ham_2 = \Ham_3$. This means the integration constants must be fine-tuned, such that $Q_0 = P^1 = P^2 = P^3 = K$ and $h_0 = h^1 = h^2 = h^3 = h$. Consequently the four gauge fields take identical values, and
the degrees of freedom contributing to the solution match those of Einstein-Maxwell theory. 
Following this through, the integration constants have the form
\begin{equation*}
    \alpha = 2B, \quad \beta =\frac{2 K}{\alpha} \sinh\left(\frac{\alpha h}{2 K}\right), \quad \gamma = \exp\left(-\frac{\alpha h}{2 K}\right), \qquad \alpha,\beta,\gamma \in (0, \infty)
\end{equation*}
in the Einstein-Maxwell limit, and the line element becomes
\begin{equation*}
    ds^2 = -\frac{2(\beta + \gamma \zeta)^2}{(\alpha \zeta - 1)} d\zeta^2 + \frac{(\alpha \zeta - 1)}{2(\beta + \gamma \zeta)^2} d\eta^2 + \zeta^2 \left(1 + \frac{\beta}{\gamma \zeta} \right)^2 (dx^2 + dy^2).
\end{equation*}
By comparing this with \eq{planarEM2}, we can express the parameters $M,q$ of the Einstein-Maxwell
solution in terms of the constrained integration constants of the STU solution:
\begin{equation*}
    M = \frac{\alpha}{4 \gamma^2}, \qquad q^2 = \frac{\alpha \beta + \gamma }{2\gamma^3} = \frac{K^2}{2 \gamma^4}.
\end{equation*}
For a full discussion of this mapping, see Appendix B of \cite{Gutowski:2019iyo}.

We can now study each piece of the above calculation and see how it changes under this mapping, and see that we recover the results from the Euclidean action formalism in section \ref{sec:pem}. In the 
Einstein-Maxwell limit each of the relavant pieces simplifies into the form
\begin{equation*}
    \kappa = - \frac{\alpha^3 \gamma^2}{4 K^4} = - \frac{4 M^3}{q^4}, \qquad a_\Delta  = \frac{K^4}{\gamma^4 \alpha^2} = \frac{q^4}{4 M^2},
\end{equation*}
\begin{equation*}
        \mathcal{Q}_{0|\Delta} = - \tilde{\mathcal{Q}}^A_\Delta = - \frac{K}{16 \pi \gamma^2} = - \frac{\sqrt{2} q}{16 \pi},
\end{equation*}        
\begin{equation*}
     \mu^0 = - \tilde{\mu}_A = -\frac{\alpha}{2 K} = - \frac{2 M}{\sqrt{2} q},
\end{equation*}
where we have used that $c^{-1} = 2\gamma^2$.

Looking at the variation of the appropriate terms:
\begin{equation*}
    \kappa \delta a_\Delta = - \frac{4 M^3}{q^4} \left( \frac{q^3}{M^2} \delta q - \frac{q^4}{2M^3} \delta M \right) = - 4 \cdot \left( \frac Mq \delta q - \half \delta M \right),
\end{equation*}
\begin{equation*}
    \mu_a \delta \mathcal{Q}^a = 4 \cdot \left(\frac{2 M}{\sqrt{2} q} \frac{\sqrt{2} }{16 \pi} \delta q  \right) = \frac{M}{2 \pi q} \delta q ,
\end{equation*}
we obtain an expression for the variation of the energy by imposing the first law
\begin{equation*}
    \delta E_\Delta := \frac{\kappa \delta a_\Delta}{8 \pi} + \mu_a \delta \mathcal{Q}^a = \frac{\delta M}{4 \pi}.
\end{equation*}
When this is integrated up, we obtain an expression for the internal energy
\begin{equation*}
    E_\Delta = \frac{M}{4 \pi},
\end{equation*}
which matches exactly with the internal energy derived from the free energy, via the Euclidean action formalism in section \ref{sec:pem}.

\section{Planar solutions to Einstein-anti Maxwell theory}
\label{sec:antimax}

In this section, we report on a `dual version' of the planar Einstein-Maxwell solution,
where the static and dynamic regions are exchanged such that the first law and other
thermodynamic relations can be derived using a conventional Wick-rotation. The price
to be paid for this is to flip the sign of the Maxwell term. This theory is sometimes referred to as \emph{Einstein-anti-Maxwell}, and in general fields with a flipped sign kinetic terms are referred to as ``phantom" fields \cite{Sabra:2015vca,Taam:2015sia}. 
Fields with negative kinetic energy have been discussed in the context 
of cosmology, because some data suggest that the current expansion of the universe
is over-exponential, leading to a `big-rip' cosmological singularity in finite time. While naively the
negative kinetic energy renders the theory unstable, $p$-form gauge fields with inverted
kinetic terms appear in the type-II$^*$ string theories which are related to type-II string theories
by timelike T-duality transformations. In these cases, the theory is made consistent through 
the presence of massive string modes and the related higher gauge symmetries 
\cite{Hull:1998vg}. Gauge fields with flipped kinetic terms also appear in ``Fake-Supergravity" 
theories.

We will now show that the Einstein-anti-Maxwell theory admits a planar solution which 
can be viewed as the `dual' partner of 
to the planar cosmological solution of Einstein-Maxwell theory. This solution realizes the
same thermodynamical system as studied in Section \ref{sec:pem}, in the sense that both 
solutions have the same Euclidean action, and therefore the same grand potential and other 
thermodynamic potentials. More precisely, the range of some of thermodynamic quantities (temperature,
energy) will turn out to be different, suggesting that the two solutions represent
different `phases' of the same system. We will discuss 
the interpretation of these observations in Section \ref{Sect:Disc+Outlook}.

%

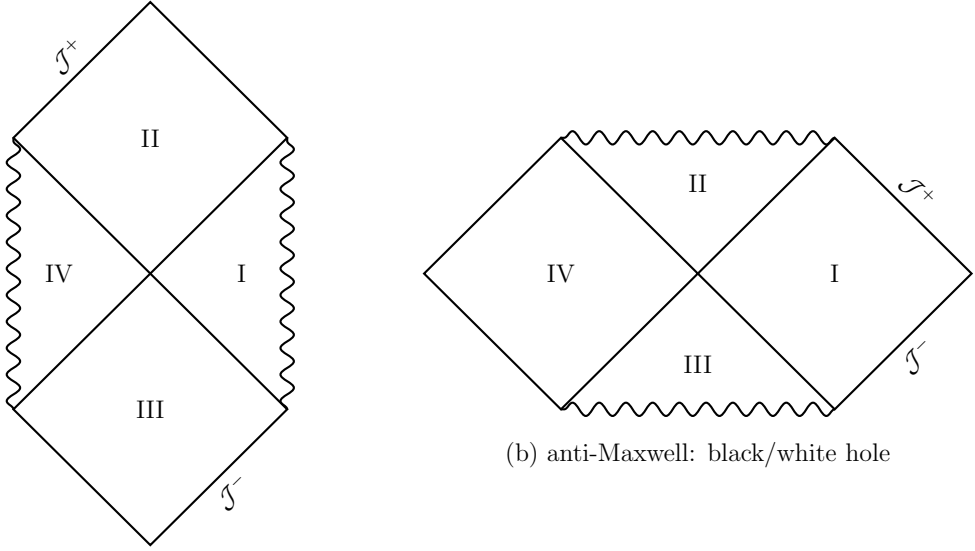
\begin{figure}[h!]
\centering
\begin{tikzpicture}[scale=0.6, every node/.style={scale=0.8}]

\node (I) at (-6,6) {II};
\node (II) at (-6,0) {III};
\node (III) at (-8,3) {IV};
\node (IV) at (-4,3) {I};
\node (label) at (-6,-4) {(a) Maxwell: cosmological};

\path 
 (I) +(90:3) coordinate[label=90:] (IItop)
 +(-90:3) coordinate(IIbot)
 +(0:3) coordinate[label=360:] (IIright)
 +(180:3) coordinate[label=180:] (IIleft)
 ;
\draw[thick] (IIleft) -- (IItop)  node[midway, above, sloped] {$\mathcal{J}^+$} -- (IIright)  -- (IIbot) -- (IIleft) -- cycle;

\path 
 (II) +(90:3) coordinate (Itop)
 +(-90:3) coordinate (Ibot)
 +(180:3) coordinate (Ileft)
 +(0:3) coordinate (Iright)
 ;
\draw[thick] (Ileft) -- (Itop) -- (Iright) -- (Ibot) node[midway, below, sloped] {$\mathcal{J}^-$} -- (Ileft) -- cycle;

\draw[decorate,decoration=snake,draw=black, thick] (Ileft) -- (IIleft);

\draw[decorate,decoration=snake, draw=black, thick] (Iright) -- (IIright);

\node (Ia) at (6,5) {II};
\node (IIa) at (6,1) {III};
\node (IIIa) at (3,3) {IV};
\node (IVa) at (9,3) {I};
\node (labela) at (6,-1) {(b) anti-Maxwell: black/white hole};

\path 
 (IIIa) +(90:3) coordinate[label=90:] (IIrighta)
 +(-90:3) coordinate(IIlefta)
 +(0:3) coordinate[label=360:] (IItopa)
 +(180:3) coordinate[label=180:] (IIbota)
 ;
\draw[thick] (IIlefta) -- (IItopa) -- (IIrighta) -- (IIbota) -- (IIlefta) -- cycle;

\path 
 (IVa) +(90:3) coordinate (Irighta)
 +(-90:3) coordinate (Ilefta)
 +(180:3) coordinate (Ibota)
 +(0:3) coordinate (Itopa)
 ;
\draw[thick] (Ilefta) -- (Itopa) node[midway, below, sloped] {$\mathcal{J}^-$}  -- (Irighta) node[midway, above, sloped] {$\mathcal{J}^+$}  -- (Ibota) -- (Ilefta) -- cycle;

\draw[decorate,decoration=snake,draw=black, thick] (Ilefta) -- (IIlefta);

\draw[decorate,decoration=snake, draw=black, thick] (Irighta) -- (IIrighta);
\end{tikzpicture}
\caption{Comparison of the conformal diagrams of cosmological and black hole solutions. Left side: Planar cosmological solution of Einstein-Maxwell theory. Right side: Planar black bole solution of
Einstein-anti-Maxwell theory, same as for the (spherical) Schwarzschild solution of pure Einstein theory.
\label{Fig:compar}}
\end{figure}

\subsection{The solution}

We start with an action which simultaneously describes both theories, 
\begin{equation*}
S = \int d^4 x \sqrt{-g}\left( - \frac{1}{2\kappa_4^2} R + \frac{\varepsilon}{4g^2} F^2\right),    
\end{equation*}
where $\varepsilon = \pm 1$ and $g^2=4\pi$ is the gauge coupling. Introducing $g$ 
is convenient because it allows us to relate both theories by analytic continuation 
of the coupling constant $g \rightarrow ig$. Alternatively we could relate them
by analytic continuation of the gauge fields $F$, but we prefer to keep $F$ 
real valued in both theories. This said, we revert to our standard conventions
where $G=1, \kappa^2_4 = 8\pi$ and $g^2=4\pi$. 

Solving Einstein's equations with a static, planar symmetric ansatz yields a line element of the form:
\begin{equation}
    ds^2 = -f(r) dt^2 + f(r)^{-1} dr^2 + r^2 d\vec{X}^2,  \quad f(r) = \frac{2 c}{r} +  \frac{\varepsilon q^2}{r^2}.
\end{equation}
For spherically symmetric solutions, the value of the integration constant $c$ is set by comparing the result in a weak field limit to Newtonian results. In planar symmetric theories, this is not possible as there is no asymptotically flat region. Instead we choose the sign of $c$ by imposing 
the existence of a Killing horizon, which implements cosmic censorship 
 by placing the singularity at $r=0$ behind a horizon. With this in mind, we can write the line element as
\begin{equation}
    ds^2 = -f(r) dt^2 + f(r)^{-1} dr^2 + r^2 d\vec{X}^2,  \quad f(r) = \varepsilon \left(\frac{2M}{r} -  \frac{q^2}{r^2} \right),
\end{equation}
where the integration constant $M$ is always positive. In this form, it is easy to see that the sign of $f(r)$ is set by $\varepsilon$. Namely, when $\varepsilon = -1$, the asymptotic region is dynamic and the static patch for the solution is a finite region of the spacetime, bounded by
\begin{equation*}
0 < r < \frac{q^2}{2 M}.
\end{equation*}
Conversely, for $\varepsilon = 1$ the static region is found for coordinate values of 
\begin{equation*}
r > \frac{q^2}{2 M},
\end{equation*}
and we see that for Einstein-anti-Maxwell, the asymptotic region of the spacetime is static. 
Asymptotically this metric is the vacuum Taub solution \cite{Taub:1951}.
This allows the Wick-rotation of the timelike coordinate $t$ to produce a (positive) definite, real line element. Unlike the Einstein-Maxwell solution, we have the standard relation between 
quantum mechanics and statistical mechanics, which identifies 
the saddle point approximation of the gravitational functional integral with a thermodynamic potential.

In the following, we set $\varepsilon=1$ and calculate the Euclidean action and thermodynamic
potentials. The line element in the static region is
\begin{equation*}
    ds^2 = -f(r)^2 dt^2 + \frac{dr^2}{f(r)} + r^2 d\vec{X}^2, \qquad f(r) = \frac{2M}{r} - \frac{q^2}{r^2}, \quad r_h = \frac{q^2}{2M}.
\end{equation*}

\paragraph{Chemical potential:}

The gauge field is  
\begin{equation}
    F = \left( - \frac{q}{r} \right) dt \wedge dr, \qquad     A = \left(- \frac{q}{r} + \frac{q}{r_h} \right) dt,
\end{equation}
and by taking the asymptotic limit of the gauge potential, we obtain the chemical potential 
\begin{equation*}
    \mu = \lim_{r \rightarrow \infty} A_t = \frac{2M}{q} \;.
\end{equation*}

\paragraph{Conserved charge:}

The sign flip of the gauge field coupling leads to a sign flip in the conserved charge, 
as we explain in Appendix \ref{sec:cad}. Therefore 
\begin{equation*}
    \mathcal{Q} = -\frac{1}{4 \pi} \int_{\partial \Sigma} \star F = - \frac{q}{4 \pi}.
\end{equation*}
Note that we have set $\omega=1$ for simplicity.

\paragraph{Entropy \& Temperature:}
The Einstein-anti-Maxwell solution has an exterior region with a timelike Killing vector which allows the surface gravity to be calculated by the standard method from the Killing vector field.
\begin{equation*}
    \kappa = \frac{4M^3}{q^4} \quad \Rightarrow \quad \beta = \frac{\pi q^4}{2M^3}.
\end{equation*}
We remark that using the Kodama-Hayward expression \eq{THK}, we obtain an identical result. We show in Appendix \ref{app:classification} that the horizon 
separating the static exterior from  a dynamic interior 
is a future outer horizon. In Figure \ref{Fig:compar} these
are regions I and II (or regions IV and II) in the conformal diagram on the right hand side.
This being a future outer horizon, we have $\kappa \propto T_H$ and the temperature
is positive, but has the same magnitude as in the Einstein-Maxwell solution.

The Bekenstein-Hawking area law gives 
\begin{equation*}
    S_{BH} = \frac{r_h^2 }{4} = \frac{q^4 }{16 M^2}.
\end{equation*}

\subsection{Euclidean action}
After the Wick-rotation $t \rightarrow -i t$ the Euclidean action is given by \eq{euclact}, with the addition of a sign flip for the gauge field contribution 
\begin{equation*}
\begin{aligned}
            S_E 
        &= \frac{1}{16 \pi} \int_M \sqrt{g} R  d^4 x  \\
        &- \frac{1}{8 \pi} \int_{\partial M} \sqrt{|\gamma|} K  d^3x + \frac{1}{8 \pi} \int_{\partial M} F^{\mu \nu} A_{\mu} d\Sigma_{\nu}.
\end{aligned}
\end{equation*}
The bulk term does not contribute, since $R=0$. 
The GHY-term is
\begin{equation*}
    -\frac{1}{8\pi} \int_{\partial M} \sqrt{\gamma} K = -\frac{3M \beta}{8\pi} .
\end{equation*} 
The gauge field contribution is identical to the one of the Einstein-Maxwell solution
\begin{equation*}
    \frac{1}{8\pi} \int_{\partial M} F^{\mu \nu}A_\mu d\Sigma_\nu = - \frac{2 M \beta}{8 \pi} ,
\end{equation*}
and when these two pieces are taken together, the Euclidean action is found to be
\begin{equation*}
    S_E = - \frac{5\beta M}{8 \pi} \;,
\end{equation*}
which is the same the Euclidean action of the triple Wick-rotated Einstein-Maxwell action
 \eq{emeucact}.

As the charge is kept fixed, we associate the grand canonical thermodynamic partition function with the saddle-point approximation of the gravitational partition function:
\begin{equation*}
    \log \mathcal{Z} = -{\cal N} S_E = - \beta \Omega \quad \Rightarrow \quad \Omega(\beta, \mu) = - \frac{5 \mathcal{N} \beta \mu^4 }{(8 \pi)^2}\;,
\end{equation*}
where we have expressed $\Omega$ in terms of its natural thermodynamical variables. 
The normalisation constant $\mathcal{N}$ is fixed by imposing the relation
\begin{equation*}
    \pardev{\Omega}{\mu} = -\mathcal{Q} \;.
\end{equation*}
Since
\[
\pardev{\Omega}{\mu} = - 20 \mathcal{N}  \frac{\mu^3 \beta}{(8\pi)^2} \;,\;\;\;
- \mathcal{Q} = \frac{q}{4\pi} = \frac{\mu^3 \beta}{(4 \pi)^2} \;, \;\;\;
\mu = \left( - \frac{(4\pi)^2 \mathcal{Q}}{\beta} \right)^{1/3},
\] 
this fixes $\mathcal{N}=-\frac{1}{5}$, so that the grand potential for the 
planar Einstein anti-Maxwell solution is
\[
\Omega(\beta, \mu) = \frac{\beta \mu^4}{(8\pi)^2} \;.
\]
The free energy is then obtained by a Legendre transformation:
\begin{equation*}
    F(\beta, \mathcal{Q}) = \Omega + \mu \mathcal{Q} =
    - 3 \left( \frac{\mathcal{Q}^4 \pi^2}{4 \beta} \right)^{\frac{1}{3}}.
\end{equation*}
By taking partial derivatives we can verify that the following two thermodynamic
relations take their standard forms:
\begin{equation*}
    \pardev{F}{\mathcal{Q}} = \left(- \frac{(4\pi)^2 \mathcal{Q}}{\beta} \right)^{\frac{1}{3}} = \mu, \qquad     S = \beta^2 \pardev{F}{\beta} = \left(\frac{\mathcal{Q}^4 \beta^2 \pi^2}{4} \right)^{\frac{1}{3}} = S_{BH}.
\end{equation*}
Therefore we are confident in defining the internal energy as
\begin{equation*}
    E = \pardev{(F \beta)}{\beta}  = -\frac{M}{4 \pi} <0 \;.
\end{equation*}
We note that $E$ is negative, which reflects that in the Einstein-anti Maxwell theory the
vector field has negative kinetic energy. 
By expressing $E$ in terms of its natural thermodynamic variables we obtain 
the following equation of state:
\begin{equation*}
    E(S, \mathcal{Q}) = - \frac{\pi \mathcal{Q}^2}{\sqrt{S}} \;.
\end{equation*}
Finally, we compute the total differential of the internal energy, 
\begin{equation*}
    dE = \pardev{E}{S} dS + \pardev{E}{\mathcal{Q}} \mathcal{Q} = \beta^{-1} dS + \mu d\mathcal{Q},
\end{equation*}
and find that the first law is satisfied. It is interesting to note that the Euclidean action and
grand potential, as well as other thermodynamic relations, are the same as for the planar
solutions of Einstein-Maxwell theory, except for the range of some of the variables. For
the Einstein-Maxwell solution temperature is negative and internal energy is positive, while for 
the Einstein-anti Maxwell solution temperature is positive and internal energy is negative. 
While both solutions exhibit features indicating instabilities (negative temperature and negative
internal energy, respectively), they obey all formal relations of thermodynamics and have the
same underlying Euclidean action. 


\section{Discussion and outlook\label{Sect:Disc+Outlook}}
\label{sec:discussion}

In this paper, we have developed a modified version of the Euclidean approach to
horizon thermodynamics, which can be applied to a class of cosmological 
spacetimes whose causal structure is related to black hole solutions by
exchanging the role of exterior and interior. By applying a triple Wick-rotation 
we obtain a finite Euclidean on-shell action which defines a grand thermodynamical
potential, from which all thermodynamic quantities, the Smarr relation and the 
first law can be derived. For planar solutions of 
Einstein-Maxwell theory and of the STU model, which are asymptotic
to Kasner cosmological solutions, the formalism
allows the definition of a positive mass-like quantity. 
The results obtained using the triple Wick-rotation are consistent with those
from the isolated horizon formalism, which is another check of its validity.
Both approaches are complementary. The isolated horizon formalism is
quasi-local, and does not require knowledge of the global spacetime geometry. 
But it therefore misses out on finding an underlying Euclidean action which 
defines the grand potential and to obtain a mass-like quantity from the 
thermodynamic formalism. In the isolated horizon formalism, the mass is instead
determined locally by imposing that the first law holds. The Euclidean action 
is also required to make the connection between planar solutions of Einstein-Maxwell
and Einstein-anti-Maxwell theory.

For the thermodynamic formalism, we used the future inner horizons of the
maximally extended solutions. This is natural because these horizons 
can be crossed by causal geodesics from the outside to the inside, which 
is the same situation as for black holes. For future inner horizons
the surface gravity and temperature are both negative, when 
computed according to \cite{Hayward:1997jp,Binetruy:2014ela,Helou:2015zma},
and we have shown that the first law takes its standard form.
It is natural to ask what happens if we use the past horizons instead, 
where causal geodesics cross from the interior to the exterior. This 
is analogous to asking about the thermodynamics of white holes. 
For our cosmological solutions the past horizons are past inner horizons. 
Since the surface gravity is still negative, the expression $\kappa dA$ and
hence  the `first law of horizon mechanics'
retains its standard form. However, the temperature is now positive and the
`first law of thermodynamics' takes a non-standard form where the sign
of the temperature/entropy term is flipped,  $TdS \rightarrow -T dS$.
This depends, of course, on accepting both 
the definition of the surface gravity by Kodama-Hayward and the sign
of the Hawking temperature being determined by the Parikh-Wilczek
tunneling method. We leave the investigation of this observation to future
work. Another question which deserves further attention is the interpretation 
of the negative temperature and, for planar black holes, negative 
energy. There is also the question how the thermodynamics 
defined using the triple Wick rotation relates to an underlying microscopic
description, given that the `Hamiltonian' we use is actually related to spatial
translations. Finally, one should ask whether one can formulate the thermodynamics
of horizons for more general non-static solutions using a variational approach as in \cite{An:2016fzu}, 
modified to imposing initial rather than boundary conditions.

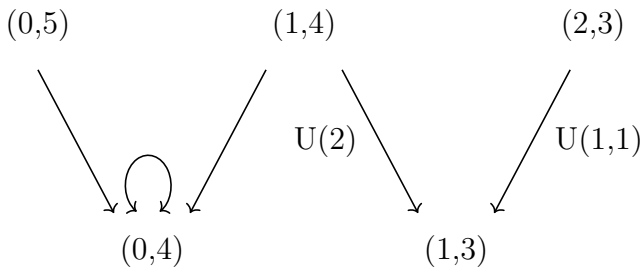
\begin{figure}[!h]
\centering

\begin{tikzpicture}[scale=2]


\node at (1,2.25) {(0,5)};
\node at (2.75,2.25) {(1,4)};
\node at (4.65,2.25) {(2,3)};


\node at (1.75,0.75) {(0,4)};
\node at (3.75,0.75) {(1,3)};

\draw[semithick,<-]  (1.5,1) -- (1,1.95) ;
\draw[semithick,->]  (2.5,1.95) -- (2,1) ;
\draw[semithick,<-]  (3.5,1) -- (3,1.95) node [midway, left=0.15cm] {U(2)};
\draw[semithick,->]  (4.5,1.95) -- (4,1) node [midway, right=0.15cm] {U(1,1)};

\draw [<->, semithick, rotate=0] (1.8,1.01) arc [start angle=-60, end angle=240, x radius=0.15cm, y radius=0.2cm];

\end{tikzpicture}

\caption{This diagram summarises the relations between five-dimensional and four-dimensional 
vector multiplet theories with spacetime signature $(t,s)$, that is, $t$ timelike and $s$ spacelike 
dimensions \cite{Cortes:2019mfa}. The two four-dimensional theories in a given signature differ by relative 
signs between terms in their Lagrangians. In Euclidean signature, these signs can be changed by 
a suitable field redefinition, and the Euclidean theory is unique. In Minkowski signature there
are two non-isomorphic supersymmetry algebra which are distinguished by their 
R-symmetry groups $U(2)$ and $U(1,1)$, respectively. Therefore the corresponding 
vector multiplets theories cannot be related by a field redefinition.
\label{Fig:dim_red}}
\end{figure}

Besides formulating thermodynamics using a non-stationary spacetime patch, 
we also have uncovered 
a curious relation or `duality' between
cosmological and black hole solutions, induced by flipping the 
sign of the Maxwell term, which exchanges interior and exterior,
spacelike and timelike singularities, and which relates solutions with negative
temperature to solutions with negative energy. We think 
that a promising  way to better understand these features
is the embedding into string theory, to which we turn now.
The realisation of a duality between two distinct Lorentzian solutions, via the equivalence of their Euclidean actions, has an interesting relation to recent results
in ${\cal N}=2$ supersymmetry.
In \cite{Cortes:2019mfa} four-dimensional 
${\cal N}=2$ supersymmetry algebras have been classified for all possible 
signatures $(t,s)$, where $t$ is the number to timelike and $s$ the number of
spacelike dimensions. It was found that while the ${\cal N}=2$ supersymmetry
algebra is unique in Euclidean signature $(0,4)$, there are two inequivalent algebras 
in Minkowski signature $(1,3)$, namely the standard algebra with compact R-symmetry 
group $U(2)$ and a twisted version with R-symmetry group $U(1,1)$. The 
corresponding vector multiplet theories are distinguished by relative signs
between various terms in the Lagrangian, including a relative sign between
Maxwell and scalar terms. Already in \cite{Sabra:2017xvx} it has been shown
that a non-standard ${\cal N}=2$ supergravity theory coupled to vector multiplets 
with inverted signs for all Maxwell-like terms results from dimensional 
reduction of five-dimensional supergravity coupled to vector multiplets 
with signature $(2,3)$. This theory reduces to  Einstein-anti-Maxwell theory 
upon truncating out the matter fields and the gravitini. We remark that 
while vector multiplet theories in signature $(0,4)$ can likewise be obtained
in two ways from five dimensions, the resulting relative signs can be 
removed by a suitable field redefinition, since the underlying Euclidean 
supersymmetry algebra is unique up to isomorphism \cite{Cortes:2019mfa}.
The situation is summarized in Figure \ref{Fig:dim_red}.

The relative sign flips between the Minkowski signature theories are of the same type as those between type-II and type-II$^*$ string theory, which 
are related to each other by timelike T-duality \cite{Hull:1998vg}. Moreover,
${\cal N}=2$ supergravity with vector (and hyper) multiplets arises by compactification 
of type-II string theory on Calabi-Yau threefolds. In a future publication we will present
the details of the embedding of the twisted ${\cal N}=2$ supergravity theory into 
type-II$^*$ theory and show that the STU and anti-STU model (which generalizes
the Einstein-anti-Maxwell theory considered in this paper) are related by T-duality
\cite{Third}. We expect that this will shed more light onto the thermodynamics of planar
solutions and their microscopic interpretation in terms of string theory. We remark that
when combining timelike and spacelike T-duality with S-duality, it is possible to
change spacetime signature in type-II string theory, which provides a second way,
besides analytical continuation, of relating theories in Euclidean and in Minkowski 
signature \cite{Hull:1998ym}. Solutions in neutral and in general signature
have recently found attention in the literature, see for example
\cite{Klemm:2015mga,Gutowski:2019hnl,Sabra:2020gio}.

Both from the point of view of thermodynamics and from the one 
of T-duality, certain spacetime geometries naturally form pairs which share the same
underlying Euclidean description. If one takes the Euclidean functional integral 
as fundamental and allows both the spacetime 
and the field space to be complex-valued, this will correspond to pairs of 
complex saddle points of the functional integral which represent dual
Minkowksi signature solutions. At this point it is not clear whether the 
two dual solutions are actually `the same', that is, gauge equivalent 
under a chain of string duality transformations, or just have `the same
thermodynamics.' In either case
one could also look for relations to solutions in neutral signature. 
It will be interesting to further investigate these
intriguing relations between geometry, thermodynamics and dualities.

\section*{Acknowledgements}
Giacomo would like to thank Alessandro Torrielli, for the enlightening discussions about the physical interpretation of a triple Wick-rotation.

\newpage
\appendix

\section{Conventions}
\label{sec:Conventions}

We use the same parametrization of conventional signs as in \cite{Freedman:2012zz}.
When studying general relativity this involves three conventional sign
choices $s_i=\pm 1$, $i=1,2,3$. The first is the overall sign of the Minkowski metric
\begin{equation*}
	\eta_{a b} = s_1 \text{diag}(-+++),
\end{equation*}
and decides whether we work with a ``mostly-plus'' or ``mostly-minus'' signature. 
The second sign choice comes from the definition of the Riemann tensor:
\begin{equation*}
	\tensor{R}{^\mu_{\nu \rho \sigma}} = s_2(\partial_\rho \tensor{\Gamma}{^\mu_{\nu \sigma}} - \partial_\sigma \tensor{\Gamma}{^\mu_{\nu \rho}} +  \tensor{\Gamma}{^\tau_{\nu \sigma}} \tensor{\Gamma}{^\mu_{\tau \rho}} -  \tensor{\Gamma}{^\tau_{\nu \rho}} \tensor{\Gamma}{^\mu_{\tau \sigma}} )\;,
\end{equation*} 
and the third sign from the Einstein equations
\begin{equation*}
	s_3 \left( R_{\mu \nu} - \half g_{\mu \nu} R \right)=  \kappa_4^2 T_{\mu \nu}\;,
\end{equation*}
where it is understood that $T_{00}$ is always positive (for normal matter). The signs $s_2,s_3$
enter into the definitions of the Ricci tensor and Ricci scalar:
\begin{equation*}
	s_2 s_3 R_{\mu \nu} =  \tensor{R}{^\rho_{\mu \rho \nu}}, \qquad R = g^{\mu \nu} R_{\mu \nu}.
\end{equation*}
These three signs enter into a Lagrangian for gravity, vector and scalar fields as:
\begin{equation*}
	\La = \left(s_1 s_3 \frac{R}{2 \kappa_4^2} - s_1 \frac{1}{\kappa_4^2} \partial_\mu \phi \partial^\mu \phi  - \frac{1}{4 g^2} F_{\mu \nu} F^{\mu \nu} \right).
\end{equation*}
The conventions used in a particular paper can usually be reconstructed using that 
the kinetic terms are positive. This depends of cause on knowing that the overall sign of the Lagrangian 
has been fixed accordingly, and that we are not dealing with a non-standard theory 
with flipped kinetic terrms. 
We also need to assume that 
the energy momentum Tensor is defined such that $T_{00}$ is positive and 
and therefore:
\begin{equation*}
	T_{\mu \nu} = -s_1 \frac{2}{\sqrt{-g}} \frac{ \delta(
	\La_m \sqrt{-g})}{\delta g^{\mu \nu}},
\end{equation*}
where $\La_m$ is the matter contribution to the Lagrangian. 

In this work we use the same sign conventions as in 
\cite{Gutowski:2019iyo} which in turn where taken over from 
\cite{Dempster:2015}. This is a parametrization where the Einstein-Hilbert and
scalar term enter with a minus sign:
\begin{equation*}
	\La = \left(-\frac{R}{2 \kappa_4^2} - \frac{1}{\kappa_4^2} \partial_\mu \phi \partial^\mu \phi  - \frac{1}{4 g^2} F_{\mu \nu} F^{\mu \nu} \right)\;.
\end{equation*} 
From this we can read off 
\begin{equation*}
	s_1 = 1, \qquad s_3 = -1.
\end{equation*}
Defining the Ricci tensor such that  $s_2 s_3=1$, consistency determines the
overall sign of the Riemann tensor as 
$s_2 = -1$, 
\begin{equation*}
	\tensor{R}{^\mu_{\nu \rho \sigma}} = -(\partial_\rho \tensor{\Gamma}{^\mu_{\nu \sigma}} - \partial_\sigma \tensor{\Gamma}{^\mu_{\nu \rho}} +  \tensor{\Gamma}{^\tau_{\nu \sigma}} \tensor{\Gamma}{^\mu_{\tau \rho}} -  \tensor{\Gamma}{^\tau_{\nu \rho}} \tensor{\Gamma}{^\mu_{\tau \sigma}} ).
\end{equation*} 
It follows that Einstein's equations are:
\begin{equation*}
	 R_{\mu \nu} - \half g_{\mu \nu} R =  -\kappa_4^2 T_{\mu \nu}.
\end{equation*}

With these conventions, a spacelike surface of positive curvature has sign$(R) = s_1 s_3 = -1$, such that a positively curved space has a negative Ricci scalar. From the perspective of the (anti) de Sitter solutions, we take the action to be of the form
\begin{equation*}
	S = - \frac{1}{16\pi} \int_{M} (R - 2\Lambda) \sqrt{-g} d^4 x
\end{equation*}
such that when solving the equations of motion, the Ricci scalar is proportional to the cosmological constant. This means that for the de Sitter solution, we have $\Lambda < 0$ and for the anti-de Sitter solution, we have that $\Lambda > 0$. This follows from our choice of an action with  $s_3 = -1$. While this
is different from the conventions used in most research papers 
on (anti-)de Sitter solutions, it allows us to be consistent with previous work for 
solutions without a cosmological constant, which also is the main focus in the present paper.

Following these conventions through to the Euclidean action, we find that these signs appear as:
\begin{equation}
\label{eq:totact}	
\begin{aligned}
		S = \frac{s_1 s_3}{2\kappa_4^2} \int_M \sqrt{g} R d^4 x + \frac{s_1 s'_4 \epsilon}{\kappa_4^2} \int_{\partial M} \sqrt{|\gamma|} K d^3x ,
\end{aligned}
\end{equation}
where the fourth sign $s'_4$, which arises from the definition of the second fundamental form, 
is discussed in the next appendix. Note that $s'_4$ is distinct from $s_4$ in \cite{Freedman:2012zz},
which is related to the spin connection.  Since we only consider bosonic fields, 
this sign $s_4$ is irrelevant for us.  We also do not need to fix the fifth parameter $s_5$ of \cite{Freedman:2012zz}, 
which determines the overall sign of the $\varepsilon$-tensor, because the numerical
value of the $\varepsilon$-tensor is not relevant for our calculations.

\section{Extrinsic curvature \label{app:extrinsic_curvature}}
\label{sec:extrinsiccurvature}

A well-posed variational problem for the Einstein-Hilbert action requires the inclusion of 
boundary terms which `live' on the boundary $\partial M =\Sigma$, the exception being when
$M$ is closed (compact, without boundary). 
For spacetimes of infinite volume the boundary is defined as the limit of 
a family of hypersurfaces; for example, the boundary of $\mathbb{R}^4$ can be
defined as the limit of a family of three-spheres $S^3_R$ of radius $R$, 
where $R\rightarrow \infty$. In such situations, the boundary terms must be
added to the Einstein-Hilbert action to deal with variations of field configurations which do 
not fall off fast at infinity. 

The boundary term involves the (trace of the) extrinsic curvature of the 
boundary, regarded as an embedded submanifold. In this appendix we review
how the extrinsic curvature $K_{\mu \nu}$ and its trace $K$ can be computed.
We use a parameter $\epsilon=\pm 1$ to encode whether the boundary $\Sigma$ is
timelike, $\epsilon=-1$ or spacelike, $\epsilon=1$. We denote the outer
unit normal vector field of $\Sigma$  by $n$, so that $g(n,n)=n_\mu n^\mu = \epsilon$.

The first fundamental form, $\gamma$, is constructed out of the spacetime metric $g$ and the unit normal $n$. This tensor is transversal to $\Sigma$, that is $\gamma(N,\cdot)= 0$ for all vectors $N$
which are normal to $\Sigma$. When evaluated on vectors tangent to $\Sigma$, the tensor $\gamma$ 
agrees with the pull-back metric  $(\iota^* g)$ on $\Sigma$ which is induced
by the embedding $\iota: \Sigma \rightarrow M$. Its 
mixed components $\gamma^\mu_{\,\nu}= \delta^\mu_\nu - \epsilon n^\mu n_\nu$ can be used
to project tensors on $M$ onto tensors on $\Sigma$. 
For example, the projection of a vector $X$ at a point
$p\in \Sigma$ onto a vector $X_\parallel$ tangent to $\Sigma$ takes the form
$X^\mu \mapsto X_{\parallel}^ \mu = \gamma^\mu_{\,\nu}X^\nu$ in local coordinates.

The second fundamental form $K$, or the \emph{extrinsic curvature tensor}, measures the failure of a normal vector to remain normal to $\Sigma$ under parallel transport with the Levi-Civita connection 
$\nabla$ of $(M,g)$  along  curves on $\Sigma$. To define the second fundamental form, 
consider the parallel transport of a normal vector $N$ along a curve $C$ on $\Sigma$.
Then $\nabla_X N =X^\mu \nabla_\mu X^\nu= 0$, where $X$ is the tangent vector field of $C$.  If $N$ remains
a normal vector field to $\Sigma$ under parallel transport on $\Sigma$, then $g(N,Y)=0$
for all points on $C$ and all tangent vectors $Y$ to $\Sigma$ along $C$. In other words
we can measure the failure of $N$ remaining normal to $\Sigma$ by studying the 
variation 
\begin{equation*}
	X\cdot g(N , Y) = X^\mu \nabla_\mu (Y^\nu N_\nu) = N_\nu X^\mu \nabla_\mu Y^\nu
\end{equation*}
of $g(N,Y)$ along $C$. To define the extrinsic curvature, we use the normal unit vector
field $n$ on $\Sigma$, which we can extend around each point $p\in M$ to a unit vector field
on a  neighbourhood in $M$. The extrinsic curvature
is defined, using the projections introduced above, by 
\begin{equation}
\label{eq:excurdef}
	K(X,Y) = -s_4 n_\mu \left(\nabla_{X_\parallel} Y_{\parallel} \right)^\mu,
\end{equation}
for vector fields $X,Y$ on $M$. It can be shown that this definition is independent of how 
$n$ is extended away from $\Sigma$. The overall sign of $K$ is coventional. 
 In our work, we use that $s_4 = 1$, but note that many other authors choose $s_4 = -1$ (see for example eq. (5) in \cite{York1986}). Using that $n_\rho Y^\rho_\parallel = 0$ we can evaluate $K(X,Y)$:
\begin{equation}
			K(X,Y) 
			= - n_\rho X_\parallel^\sigma \nabla_\sigma Y_\parallel^\rho 
			= X_\parallel^\sigma Y_\parallel^\rho \nabla_\rho n_\sigma 
		= (\gamma^\sigma_\mu \gamma^\rho_\nu  \nabla_\rho n_\sigma) X^\mu Y^\nu 
		 = K_{\mu \nu} X^\mu Y^\nu\;.
	\end{equation}
This leads to the expressions
\begin{equation}
\label{eq:extcur0}
	K_{\mu \nu} =  \gamma^\rho_\mu \gamma^\sigma_\nu \nabla_\rho n_\sigma 
	= \nabla_\mu n_\nu - \epsilon n^\rho n_\mu \nabla_\rho n_\nu
	=  \gamma^\rho_\mu \nabla_\rho n_\nu
\end{equation}
for the extrinsic curvature, where we used $n^\rho \nabla_\mu n_\rho = \frac{1}{2} \nabla_\mu (n^\rho n_\rho)=0$. The boundary can locally be described as the level set of a function $f$.
Then $N_\mu = \partial_\mu f$ is a normal vector field, and the corresponding 
unit normal vector field $n_\mu$ is hypersurface orthogonal and satisfies the 
Frobenus integrability condition
\begin{equation}
n_{[\mu} \nabla_\nu n_{\rho]} = 0 \;.
\end{equation}
Contracting this relation with $n^\rho$ it is straightforward to obtain a relation which allows to show
 that $K_{\mu\nu}$ 
is symmetric, $K_{\mu \nu} = K_{\nu \mu}$.

An alternative definition commonly used in the literature is  \cite{York1986}
\begin{equation}
\label{eq:extcur2}
	K_{\mu \nu} = \half \La_n \gamma_{\mu \nu},
\end{equation}
where $\La_n$ is the Lie derivative with respect to $n$. 
We can verify that this agrees with our definition by writing out the Lie derivative
in terms of covariant derivatives, and using  $\gamma_{\mu \nu}= g_{\mu \nu} - \epsilon
n_\mu n_\nu$:
\begin{equation*}
\begin{aligned}
		\half \La_n \gamma_{\mu \nu} 		&=  \half (n^{c}\nabla_\rho (g_{\mu \nu} - \epsilon n_\mu n_\nu) + (g_{\mu \rho} - \epsilon n_\mu n_\rho) \nabla_\nu n^\rho + (g_{\rho \sigma} - \epsilon n_\rho n_\nu) \nabla_\mu n^\rho) \\
				&= \half (\nabla_\mu n_\nu + \nabla_\nu n_\mu - \epsilon n^\rho \nabla_\rho (n_\mu n_\nu)) 
				= \nabla_\mu n_\nu - \epsilon n_\mu n^\rho \nabla_\rho n_\nu \;,
\end{aligned}
\end{equation*}
were we used the contracted Frobenius integrability property of $n^\mu$. 
In our calculations, we need the trace of the extrinsic curvature,
\begin{equation}
\label{eq:trext}	
	K = g^{\mu \nu} K_{\mu \nu}.
\end{equation}
The trace can be easily calculated from the expression
\begin{equation}
\begin{aligned}
		K &= g^{\mu \nu} K_{\mu \nu} = \half g^{\mu \nu}(\nabla_\mu n_\nu + \nabla_\nu n_\mu - \epsilon n^\rho \nabla_\rho (n_\mu n_\nu)) \\
		&= \half (\nabla_\mu n^{\mu} + \nabla^\mu n_\mu - {\epsilon n^\rho \nabla_\rho (g^{\mu \nu} n_\mu) n_\nu} ) 
		= \nabla_\mu n^\mu \;. \label{K_as_divergence}
\end{aligned}
\end{equation}
We have defined the first and second fundamental form as transversal tensors on $M$, that is
as tensors which vanish when contracted with normal vectors to $\Sigma$. 
Alternatively, they can be defined as tensors on $\Sigma$, see for example \cite{Poisson}. The resulting expressions are related by
\[
\gamma_{mn} = (\iota^*g)_{mn} = e^\mu_m e^\nu_n \gamma_{\mu \nu} \left( = 
e^\mu_m e^\nu_n g_{\mu \nu}  \right)\;,\;\;\;
K_{mn} = e^\mu_m e^\nu_n K_{\mu \nu} \;,
\]
where the vector fields $e_m=(e^\mu_m)$, $m=1,2,3$ define an orthonormal coordinate frame on $\Sigma$. 

\section{Charges and Hodge dualisation \label{sec:cad}}

The thermodynamic formalism for the STU-model makes use of an electric-magnetic
duality frame where the magnetic charges $\mathcal{P}^A$, $A=1,2,3$ have been 
replaced by electric charges $\tilde{\mathcal{Q}}_A$, so that all charges excited in 
our solution are electric. In this appendix we give some details of the dualization 
procedure and derive the expressions for the charges used in the main part of
this article. 

To explain the idea we use a theory with a single Abelian vector field
$A_\mu$ with field strength $F_{\mu \nu} = \partial_\mu A_\nu - \partial_\nu A_\mu$
and a curved spacetime Maxwell type action
\begin{equation}
S[A] = \int \left( - \frac{1}{4g^2} F_{\mu \nu} F^{\mu \nu} \right) e \; d^4 x \;.
\end{equation}
For Maxwell theory $g$ is a constant, but to cover the case of $\mathcal{N}=2$
vector multiplets we promote $g$ to background field which depends on the 
spacetime coordinates. This background coupling can be used as a proxy for
the scalar field dependent couplings $\mathcal{I}_{IJ}$, as we will see below. 
We promote the Bianchi identity $\epsilon^{\mu \nu \rho \sigma} \partial_\nu F_{\rho \sigma}=0$ 
to a field equation by introducing the Lagrange multiplier vector field $\tilde{A}_\mu$:
\begin{equation}
\label{Action_w_Multiplier}
S[F,\tilde{A}] =  \int \left( 
- \frac{1}{4g^2} F_{\mu \nu} F^{\mu \nu} 
- \frac{1}{2} \varepsilon^{\mu \nu \rho \sigma} \partial_\mu \tilde{A}_\nu F_{\rho \sigma}
\right) e\; d^4  x\;.
\end{equation}
Note that $\varepsilon^{\mu \nu \rho \sigma}$ denotes the Levi-Civita tensor, not 
the permutation symbol $\epsilon^{\mu \nu \rho\sigma} = e \varepsilon^{\mu \nu \rho \sigma} = 
\delta^{\mu \nu \rho \sigma}_{0123}$, which is a tensor density. This parametrization is convenient because 
the metric determinant $e$ appears as an overall factor. While variation with respect to $\tilde{A}_\mu$
imposes the Bianchi identity, variation with respect to $F_{\mu \nu}$ produces its algebraic
equation of motion
\begin{equation}
\frac{1}{g^2} F^{\mu \nu} = - \varepsilon^{\mu \nu \rho \sigma} \partial_\rho \tilde{A}_\sigma
= - \frac{1}{2}  \varepsilon^{\mu \nu \rho \sigma} \tilde{F}_{\rho \sigma} \;,
\end{equation}
where we defined the dual field strength tensor
\begin{equation}
\tilde{F}_{\rho \sigma} = \partial_\rho \tilde{A}_\sigma - \partial_\sigma \tilde{A}_\rho \;.
\end{equation}
Substituting this back into \eqref{Action_w_Multiplier} we obtain the dual action
\begin{equation}
S[\tilde{A}] = \int \left( -\frac{g^2}{4} \tilde{F}_{\mu \nu} \tilde{F}^{\mu \nu} \right) e \, d^4 x 
= \int \left( -\frac{1}{4\tilde{g}^2} \tilde{F}_{\mu \nu} \tilde{F}^{\mu \nu} \right) e\, d^4 x \;,
\end{equation}
where we defined the dual coupling 
\begin{equation}
\tilde{g} = \frac{1}{g} \;.
\end{equation}
The relation between the dual and the
orginal field strength is
\begin{equation}
\tilde{F}_{\mu \nu} = \frac{1}{g^2} \star F_{\mu \nu} \;,
\end{equation}
where 
\begin{equation}
\star F_{\mu \nu} = \frac{1}{2} \varepsilon_{\mu \nu \rho \sigma} F^{\rho \sigma} =
\frac{1}{2} e \epsilon_{\mu \nu \rho \sigma} F^{\rho \sigma}
\end{equation}
is the Hodge-dual field strength. Thus the duality exchanges electric and 
magnetic fields and inverts the coupling. It also exchanges Euler-Lagrange
equations and Bianchi identities:
\begin{eqnarray}
\partial^\mu \left( \frac{e}{g^2} F_{\mu \nu}\right) =0 &\Leftrightarrow &
\epsilon^{\mu\nu \rho \sigma}  \partial_\nu \tilde{F}_{\rho\sigma} = 0  \;, \\
\epsilon^{\mu\nu \rho \sigma}  \partial_\nu {F}_{\rho\sigma}  = 0 &\Leftrightarrow &
\partial^\mu \left( \frac{e}{\tilde{g}^2} \tilde{F}_{\mu \nu} \right)= 0  \;.
\end{eqnarray}
Note that we can write both field equations as Bianchi identities:
\begin{equation}
d \tilde{F}=0 \;,\;\;\;dF = 0  \;.
\end{equation}
This tells us how to define conserved charges. 
In a theory with field dependent coupling $g$, the electric and magnetic charge
are defined by 
\begin{equation}
\label{charges}
\mathcal{Q} = \frac{1}{8 \pi} \int_X \tilde{F}  = \frac{1}{8\pi} \int_X \frac{1}{g^2} \star F \;,\;\;\;
\mathcal{P} = \frac{1}{8\pi} \int_X F \;,
\end{equation}
where we have chosen the normalization to be same as in the main part of the
paper. For Maxwell theory, where $g=\mbox{const}$, the electric charge is
given by integral of the Hodge-dual two-form $\star {F}$, leading to the 
standard expression for the Maxwell electric charge. For point-like charges 
the two-surface $X$ has the topology of a sphere. For solutions with planar symmetry,
we take $X$ to be a plane. The resulting integral is divergent, but defines 
a finite charge (density) upon formally dividing by the volume of $X$, or by 
compactifying $X$ into a two-torus. The equations of motions and Bianchi identities, 
which are valid outside charges, tell us that 
both $F$ and $\tilde{F}$ are closed. 
This allows one to deform the integration surfaces $X$
continuously, as long as one avoids moving them through the charges, which
for our solutions are located at the singularities. Often it is
convenient to evaluate the charges in a limit where $X$ is pushed to infinity, 
and this is in particular how charges are computed in the main part of this paper.

The dualization procedure can be used to replace magnetic charges by electric charges.
This can be convenient since in a fixed duality frame electric charges are Noether charges
and can couple minimally to the gauge field, whereas magnetic charges are topological
and do not have local couplings to the gauge field. For black hole thermodynamics we find
it convenient to replace magnetic charges by electric charges in the main part of the paper. 
The dual charges are found by replacing $F$ by $\tilde{F}$. Using that 
\[
\tilde{\tilde{F}} =  \frac{1}{{\tilde{g}}^2} \star \tilde{F}= \frac{1}{\tilde{g}^2 g^2} \star \star F=-F \;,
\]
we find
\begin{eqnarray}
\tilde{\mathcal{Q}}& =&  \frac{1}{8\pi} \int_X \tilde{\tilde{F}} = - \frac{1}{8\pi} \int_X  F = - \mathcal{P} \;, \\
\tilde{\mathcal{P}} &=& \frac{1}{8\pi} \int_X \tilde{F} = \mathcal{Q} \;.
\end{eqnarray}
Note that the transformation $(\mathcal{Q}, \mathcal{P}) \rightarrow (-\mathcal{P}, \mathcal{Q})$ 
is symplectic. 

The relation between the Einstein-Maxwell theory and the Einstein-anti-Maxwell theory
where the sign of the Maxwell term has been flipped, can be interpreted as an
analytic continuation of the coupling: $g\rightarrow ig$, $g^{-2} \rightarrow - g^{-2}$. 
From \eqref{charges} it is clear that this flips the signs of electric charges, 
$\mathcal{Q} \rightarrow - \mathcal{Q}$. The same conclusion is reached 
when including electric sources and defining the electric charge using Gauss law,
\begin{equation*}
    \mathcal{Q} = - \frac{1}{8\pi} \int_{\Sigma} \star j,
\end{equation*}
where $\Sigma$ is a hypersurface such that $X = \partial \Sigma$, and where 
$j^\mu$ is the charge density. Maxwell's equations relate the gauge field to the charge density:
\begin{equation*}
    d\left( \frac{1}{g^2}  \star F  \right)= - \star j.
\end{equation*}
Integrating over a hypersurface $\Sigma$ and applying the Gauss-Stokes 
theorem we obtain:
\begin{equation*}
    \mathcal{Q} = \frac{1}{8\pi}  \int_{\Sigma} d \left( \frac{1}{g^2}\star F\right) 
    = \frac{1}{8\pi}  \int_{\partial \Sigma}\frac{1}{g^2} \star F.
\end{equation*} 

As we consider sources as external, the analytical continuation of the coupling $g$ 
changes the sign of $\mathcal{Q}$.

Electric-magnetic duality can be extended to theories with multiplet vector fields, 
including $\mathcal{N}=2$ vector multiplets with bosonic Lagrangian \eqref{Bos_Lag}.
In these theories electric-magnetic duality becomes part of a larger group of
symplectic transformations, which acts continuously on gauge fields, but is broken
to a discrete subgroup once charge quantization is taken into account. 
We refer to \cite{LopesCardoso:2019mlj} for a general discussion. In this paper we
consider the STU-model, and for our solutions we only need the consistent truncation
where the coupling matrices are restricted by 
$\mathcal{R}_{IJ}=0$. Since the remaining coupling matrix $\mathcal{I}_{IJ}$ is diagonal,
the vector field part of the Lagrangian \eqref{Bos_Lag} reduces to 
\begin{equation}
 e_4^{-1} \La =  
 + \frac{1}{4 \kappa_4^2} \left(  \I_{00} F^0_{\mu \nu} F^{0|\mu \nu} 
+ \sum_{A=1}^3  \I_{AA} F^A_{\mu \nu} F^{A|\mu \nu} \right) + \cdots
\end{equation}
This amounts to four copies of the type of vector field Lagrangian we have considered
before. Our solution carries charges $(\mathcal{Q}_0, \mathcal{P}^A)$, which we 
can map to the purely electric charges $(\mathcal{Q}_0, \tilde{\mathcal{Q}}_A)$, where 
$\tilde{\mathcal{Q}}_A=-\mathcal{P}^A$. 

For reference we bring the expressions for the charges to the form used for explicit computations
in the main part of the paper:
\begin{eqnarray}
\mathcal{Q}_0 &=& \frac{1}{8\pi} \int_X \tilde{F}_0 = \frac{1}{8\pi} \int_X 
\star \left( - \mathcal{I}_{00}  F^0\right)\;, \\
\tilde{\mathcal{Q}}_A &=&- \mathcal{P}^A = - \frac{1}{8\pi} \int_X F^A =
\frac{1}{8\pi} \int_X \star \left( - \mathcal{I}^{AA} \tilde{F}_A \right) \;.
\end{eqnarray}

\section{Kruskal extensions and classification of trapping horizons}
\label{app:classification}

In this section, we perform Kruskal-like coordinate transformations which allow us to construct the maximal analytic extensions of all metrics considered within this paper. This allows us to understand their global causal structure and to identify the type of all horizons using the classification of trapping horizons 
reviewed in Section \ref{sec:surfacetemp}. Solutions fall into two categories, which are distinguished
by the causal relation between their interior and exterior regions. For all solutions, we call regions
{\em exterior} if transverse/radial null geodesics reach a horizon in one direction, but can 
be extended to infinite affine parameter in the other direction. 
In terms of our standard 
transverse/radial coordinate, the asymptotic region is at 
 $r \rightarrow \infty$. In contrast, regions are called {\em interior} regions if transverse/radial null geodesics
 terminate at a curvature singularity in one direction and reach a horizon in the other. 
We refer to solutions with a static exterior region as \emph{black hole solutions}, and those with a time-dependent exteriors as \emph{cosmological solutions}. 
When using the Kodama-Hayward method \cite{Kodama:1980, Hayward:1997jp} for computing the surface gravity,
it is positive for black hole solutions, but negative for cosmological solutions. 

The static line elements considered in the main part are used to define ingoing and outgoing
null geodesics and to fix the global time orientation of the maximally extended spacetime. 
This is important since the extension contains two isometric static regions, where the line
element takes same form in terms of coordinates $t,r$, but where the timelike Killing vector
field $\partial_t$ is future-directed in one region, but past-directed in the other (where future-directed
is defined globally by picking one of the patches to fix the time orientation). Starting from a
`standard static patch,' which fixes the definition of ingoing/outgoing and determines
the direction of time, we define Kruskal coordinates and obtain a maximally extended
spacetime containing three additional regions. By computing the expansions of
null geodesic congruences for each regions, we can identify the types of the horizons 
separating them. For thermodynamics we consider future horizons, where the exterior region can causally 
influence the interior, but not vice versa. The horizons between such regions are 
\emph{future outer horizons} for black hole solutions and 
\emph{future inner horizons} for cosmological solutions. 
For the thermodynamic formalism based on the Euclidean action that we use 
in the main part, we assume that temperature and surface gravity are related
according to \cite{Binetruy:2014ela, Helou:2015zma}, that is they are proportional.
Then black holes have positive temperature, while (contracting) cosmologies have negative temperature. 

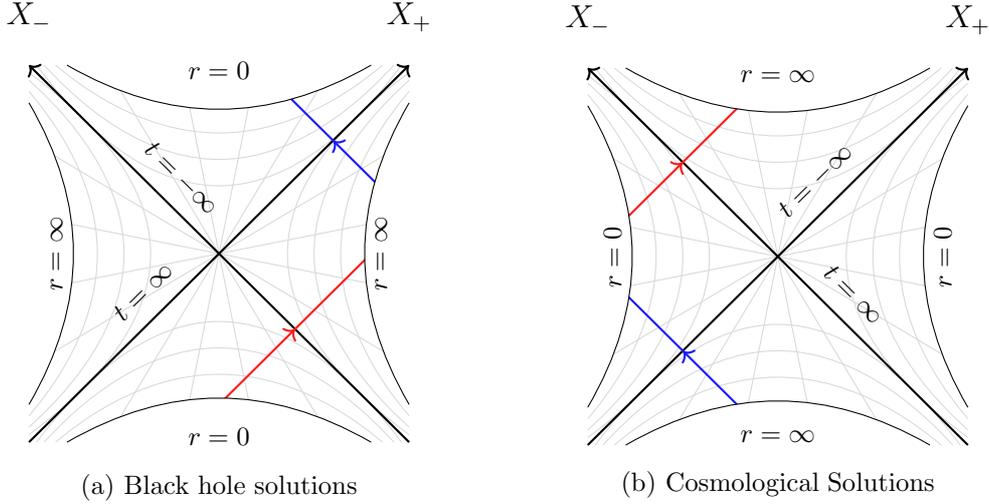
\begin{figure}
\centering
\begin{subfigure}{.5\textwidth}
  \centering
    \begin{tikzpicture}
    
    \draw[black!15!white] (0,1) -- (5,4);
    \draw[black!15!white] (0,2) -- (5,3);
    \draw[black!15!white] (0,3) -- (5,2);
    \draw[black!15!white] (0,4) -- (5,1);
    \draw[black!15!white] (1,0) -- (4,5);
    \draw[black!15!white] (2,0) -- (3,5);
    \draw[black!15!white] (3,0) -- (2,5);
    \draw[black!15!white] (4,0) -- (1,5);

    \draw[black!15!white] (2.5,4.1) parabola (4.65,5);
    \draw[black!15!white] (2.5,4.1) parabola (0.35,5);    
    \draw[black!15!white] (2.5,3.75) parabola (4.8,5);
    \draw[black!15!white] (2.5,3.75) parabola (0.2,5);
    \draw[black!15!white] (2.5,3.4) parabola (4.9,5);
    \draw[black!15!white] (2.5,3.4) parabola (0.1,5);

    \draw[black!15!white] (2.5,0.9) parabola (4.65,0);
    \draw[black!15!white] (2.5,0.9) parabola (0.35,0);    
    \draw[black!15!white] (2.5,1.25) parabola (4.8,0);
    \draw[black!15!white] (2.5,1.25) parabola (0.2,0);
    \draw[black!15!white] (2.5,1.6) parabola (4.9,0);
    \draw[black!15!white] (2.5,1.6) parabola (0.1,0);
    
    \begin{scope}[shift={(5,0)}, rotate=90]
    \draw[black!15!white] (2.5,4.1) parabola (4.65,5);
    \draw[black!15!white] (2.5,4.1) parabola (0.35,5);    
    \draw[black!15!white] (2.5,3.75) parabola (4.8,5);
    \draw[black!15!white] (2.5,3.75) parabola (0.2,5);
    \draw[black!15!white] (2.5,3.4) parabola (4.9,5);
    \draw[black!15!white] (2.5,3.4) parabola (0.1,5);
    \end{scope}
    
    \begin{scope}[shift={(0,5)}, rotate=-90]
    \draw[black!15!white] (2.5,4.1) parabola (4.65,5);
    \draw[black!15!white] (2.5,4.1) parabola (0.35,5);    
    \draw[black!15!white] (2.5,3.75) parabola (4.8,5);
    \draw[black!15!white] (2.5,3.75) parabola (0.2,5);
    \draw[black!15!white] (2.5,3.4) parabola (4.9,5);
    \draw[black!15!white] (2.5,3.4) parabola (0.1,5);
    \end{scope}

    \draw[thick, ->] (0,0) -- node[at end, above=0.3cm]{$X_+$} (5,5);
    \draw[thick, ->] (5,0) -- node[at end, above=0.3cm]{$X_-$} (0,5);
    
    \draw[thick, red!90!white, ->-]  (2,0) -- (5,3);
    \draw[thick, blue!90!white, ->-]  (5,3) -- (3,5);
    
    \draw[thick, white] (0,0.5) edge  (0,4.5)
         edge[bend right,fill=white, draw=black, thin] node[rotate=90, midway,above, black, font=\fontsize{10}{0}] {$r=\infty$} (0,4.5);
         
    \draw[thick, white] (5,0.5) edge  (5,4.5)
         edge[bend left,fill=white, draw=black, thin] node[rotate=90, midway,below, black, font=\fontsize{10}{0}] {$r=\infty$} (5,4.5);
         
    \draw[thick, white] (0.5,0) edge  (4.5,0)
         edge[bend left,fill=white, draw=black, thin] node[midway,below=0.25cm, black, font=\fontsize{10}{0}] {$r=0$} (4.5,0);
    
    \draw[thick, white] (0.5,5) edge  (4.5,5)
         edge[bend right,fill=white, draw=black, thin] node[midway,above=0.25cm, black, font=\fontsize{10}{0}] {$r=0$} (4.5,5);
         
    \node[rotate=45, font=\fontsize{10}{0}] (a) at (1.5,2) {$t = \infty$};
    \node[rotate=-45, font=\fontsize{10}{0}] (a) at (2,3.5) {$t = -\infty$}; 
        
    \end{tikzpicture}
   \caption{Black hole solutions}
  \label{fig:bhkrus1}
\end{subfigure}%
\begin{subfigure}{.5\textwidth}
  \centering
    \begin{tikzpicture}
    \draw[black!15!white] (0,1) -- (5,4);
    \draw[black!15!white] (0,2) -- (5,3);
    \draw[black!15!white] (0,3) -- (5,2);
    \draw[black!15!white] (0,4) -- (5,1);
    \draw[black!15!white] (1,0) -- (4,5);
    \draw[black!15!white] (2,0) -- (3,5);
    \draw[black!15!white] (3,0) -- (2,5);
    \draw[black!15!white] (4,0) -- (1,5);

    \draw[black!15!white] (2.5,4.1) parabola (4.65,5);
    \draw[black!15!white] (2.5,4.1) parabola (0.35,5);    
    \draw[black!15!white] (2.5,3.75) parabola (4.8,5);
    \draw[black!15!white] (2.5,3.75) parabola (0.2,5);
    \draw[black!15!white] (2.5,3.4) parabola (4.9,5);
    \draw[black!15!white] (2.5,3.4) parabola (0.1,5);

    \draw[black!15!white] (2.5,0.9) parabola (4.65,0);
    \draw[black!15!white] (2.5,0.9) parabola (0.35,0);    
    \draw[black!15!white] (2.5,1.25) parabola (4.8,0);
    \draw[black!15!white] (2.5,1.25) parabola (0.2,0);
    \draw[black!15!white] (2.5,1.6) parabola (4.9,0);
    \draw[black!15!white] (2.5,1.6) parabola (0.1,0);
    
    \begin{scope}[shift={(5,0)}, rotate=90]
    \draw[black!15!white] (2.5,4.1) parabola (4.65,5);
    \draw[black!15!white] (2.5,4.1) parabola (0.35,5);    
    \draw[black!15!white] (2.5,3.75) parabola (4.8,5);
    \draw[black!15!white] (2.5,3.75) parabola (0.2,5);
    \draw[black!15!white] (2.5,3.4) parabola (4.9,5);
    \draw[black!15!white] (2.5,3.4) parabola (0.1,5);
    \end{scope}
    
    \begin{scope}[shift={(0,5)}, rotate=-90]
    \draw[black!15!white] (2.5,4.1) parabola (4.65,5);
    \draw[black!15!white] (2.5,4.1) parabola (0.35,5);    
    \draw[black!15!white] (2.5,3.75) parabola (4.8,5);
    \draw[black!15!white] (2.5,3.75) parabola (0.2,5);
    \draw[black!15!white] (2.5,3.4) parabola (4.9,5);
    \draw[black!15!white] (2.5,3.4) parabola (0.1,5);
    \end{scope}
    
    \draw[thick, ->] (0,0) -- node[at end, above=0.3cm]{$X_+$} (5,5);
    \draw[thick, ->] (5,0) -- node[at end, above=0.3cm]{$X_-$} (0,5);
    
    \draw[thick, red!90!white, ->-] (0,2.5) -- (2.5,5);
    \draw[thick, blue!90!white, ->-] (2.5,0) -- (0,2.5);
    
    \draw[thick, white] (0,0.5) edge  (0,4.5)
         edge[bend right,fill=white, draw=black, thin] node[rotate=90, midway,above, black,  font=\fontsize{10}{0}] {$r=0$} (0,4.5);
         
    \draw[thick, white] (5,0.5) edge  (5,4.5)
         edge[bend left,fill=white, draw=black, thin] node[rotate=90, midway,below, black,  font=\fontsize{10}{0}] {$r=0$} (5,4.5);
         
    \draw[thick, white] (0.5,0) edge  (4.5,0)
         edge[bend left,fill=white, draw=black, thin] node[midway,below=0.25cm, black,  font=\fontsize{10}{0}] {$r=\infty$} (4.5,0);
    
    \draw[thick, white] (0.5,5) edge  (4.5,5)
         edge[bend right,fill=white, draw=black, thin] node[midway,above=0.25cm, black,  font=\fontsize{10}{0}] {$r=\infty$} (4.5,5);
         
    \node[rotate=-45, font=\fontsize{10}{0}] (a) at (3.5,2) {$t = \infty$};
    \node[rotate=45, font=\fontsize{10}{0}] (a) at (3,3.5) {$t = -\infty$}; 
    \end{tikzpicture}
    \caption{Cosmological Solutions}
  \label{fig:coskrus}
\end{subfigure}
\caption{Kruskal diagrams for black hole and cosmological solutions. Surfaces of constant $r$ are hyperbola and surfaces of constant $t$ are straight lines. Also included are the ingoing (blue) and outgoing (red) null geodesics which are future-pointing.}
\label{fig:kruskalcomp1}
\end{figure}

\begin{figure}[!h]
\centering
\begin{subfigure}{.5\textwidth}
  \centering
  \begin{tikzpicture}
    \fill[green!15!white] (0,0) -- (2.5,2.5) -- (-2.5,2.5);
    
    \fill[green!15!white] (0,0) -- (2.5,-2.5) -- (-2.5,-2.5);
    
    \begin{scope}
    \draw[decoration={snake, amplitude =.7mm, segment length = 2.5mm}, fill=white] 
    decorate {(-2.2, 2.5) to [bend right]  (2.2, 2.5)}
    to (-1.8, 2.5) -- cycle;
    \draw[white, line width=0.25cm] (-2.5, 2.6) --  (2.5, 2.6);
    \end{scope}
    
    \begin{scope}[rotate=180]
    \draw[decoration={snake, amplitude =.7mm, segment length = 3mm}, fill=white] 
    decorate {(-2.2, 2.5) to [bend right]  (2.2, 2.5)}
    to (-1.8, 2.5) -- cycle;
    \draw[white, line width=0.25cm] (-2.5, 2.6) --  (2.5, 2.6);
    \end{scope}
    
    \node (a) at (0, 2.4) {\footnotesize{$r = 0$}};
    \node (a) at (0, -2.4) {\footnotesize{$r = 0$}};
    
    \draw[thick, ->] (-2.5,-2.5) -- node[at end, above=0.3cm]{$X_+$} (2.5,2.5);
    \draw[thick, ->] (2.5,-2.5) -- node[at end, above=0.3cm]{$X_-$} (-2.5,2.5);
    
    \draw[thick, ->] (-3,-3) -- node[midway, left=0.3cm]{$T$} (-3,3);
    \draw[thick, ->] (-3,-3) -- node[midway, below=0.3cm]{$R$} (3,-3);
    
    \draw[thick, ->] (-2.2,1.2) to [bend left] (-2.2,-1.2);
    \draw[thick, ->] (2.2,-1.2) to [bend left] (2.2,1.2);    
    
    \node (a) at (1.1,0) {I};
    \node (a) at (-1.1,0) {IV};
    \node (a) at (0,1.1) {II};
    \node (a) at (0,-1.1) {III};
    
    \end{tikzpicture}
   \caption{Black hole solutions}
  \label{fig:sub1}
\end{subfigure}%
\begin{subfigure}{.5\textwidth}
  \centering
  \begin{tikzpicture}
    \fill[green!15!white] (0,0) -- (2.5,2.5) -- (2.5,-2.5);
    
    \fill[green!15!white] (0,0) -- (-2.5,2.5) -- (-2.5,-2.5);
    
    \draw[thick, ->] (-2.5,-2.5) -- node[at end, above=0.3cm]{$X_+$} (2.5,2.5);
    \draw[thick, ->] (2.5,-2.5) -- node[at end, above=0.3cm]{$X_-$} (-2.5,2.5);

	\begin{scope} [rotate=90]   
    \begin{scope}
    \draw[decoration={snake, amplitude =.7mm, segment length = 2.5mm}, fill=white] 
    decorate {(-2.2, 2.5) to [bend right]  (2.2, 2.5)}
    to (-1.8, 2.5) -- cycle;
    \draw[white, line width=0.25cm] (-2.5, 2.6) --  (2.5, 2.6);
    \end{scope}
    
    \begin{scope}[rotate=180]
    \draw[decoration={snake, amplitude =.7mm, segment length = 3mm}, fill=white] 
    decorate {(-2.2, 2.5) to [bend right]  (2.2, 2.5)}
    to (-1.8, 2.5) -- cycle;
    \draw[white, line width=0.25cm] (-2.5, 2.6) --  (2.5, 2.6);
    \end{scope}
    \end{scope}
    
    \node[rotate=90] (a) at (2.4, 0) {\footnotesize{$r = 0$}};
    \node[rotate=-90] (a) at (-2.4, 0) {\footnotesize{$r = 0$}};
    
    \draw[thick, ->] (-3,-3) -- node[midway, left=0.3cm]{$T$} (-3,3);
    \draw[thick, ->] (-3,-3) -- node[midway, below=0.3cm]{$R$} (3,-3);

    \draw[thick, ->] (-1.7,-1.2) to [bend right] (-1.7,1.2);
    \draw[thick, ->] (1.7,1.2) to [bend right] (1.7,-1.2);
    
    \node (a) at (0.8,0) {I};
    \node (a) at (-0.8,0) {IV};
    \node (a) at (0,1.1) {II};
    \node (a) at (0,-1.1) {III};
    
    \end{tikzpicture}
    \caption{Cosmological Solutions}
  \label{fig:sub2}
\end{subfigure}
\caption{Comparison of the Kruskal diagrams for black hole and cosmological solutions. Shaded regions correspond to the interior regions, curved lines show the direction of the Killing vector field in the static patches of the spacetime.}
\label{fig:comparison}
\end{figure}
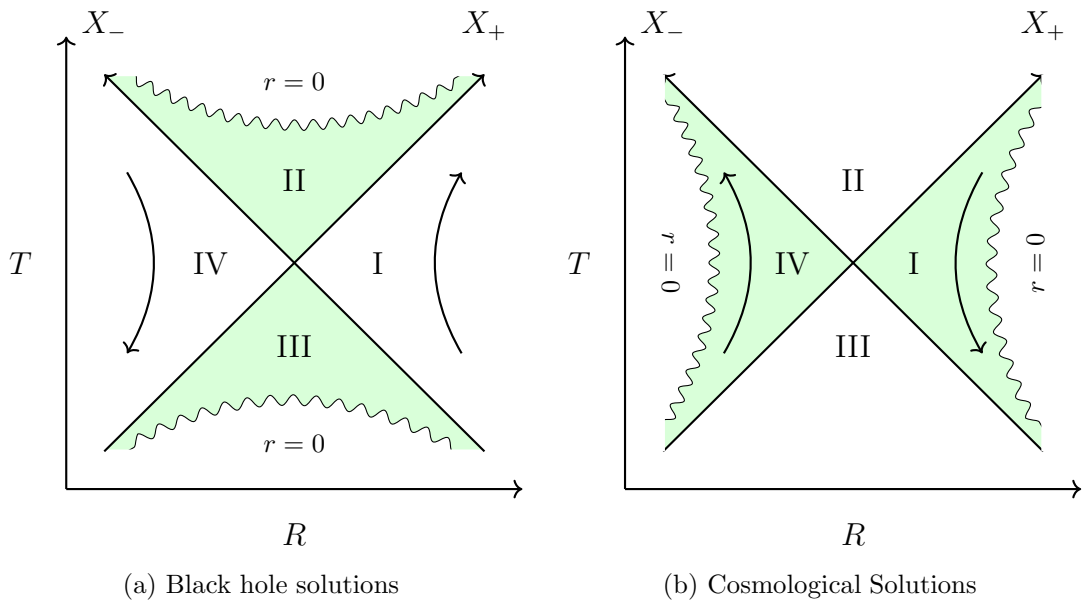

\subsection{Black hole solutions}

For the following discussion, we will be considering solutions with a line element given by
\begin{equation}
\label{eq:krusbh}
    ds^2 = -f(r) dt^2 + f(r)^{-1} dr^2 + r^2 d\vec{X}^2 \;,
\end{equation}
where $d\vec{X}^2 = d\Omega^2$ for spherically symmetric solutions, and $d\vec{X}^2 = dx^2+ dy^2$ for planar symmetric solutions. The function $f(r)$ is assumed to have a simple zero at $r=r_h>0$, and to be 
positive in the exterior region $r_h<r<\infty$, $-\infty < t < \infty$. This implies that the surface gravity of the horizon is positive:
\begin{equation*}
    \kappa = \half \partial_r f(r) \bigg|_{r = r_h} > 0.
\end{equation*}
The solution is static in the exterior region, but the Killing vector field $\partial_t$ becomes
spacelike for $r<r_h$. This situation is known from the event horizons of black holes. 
The two explicit choices for $f$ that are relevant for us are 
\begin{enumerate}
    \item The Schwarzschild solution
    \begin{equation*}
        f(r) = 1 - \frac{2M}{r}, \qquad \kappa = \frac{1}{4M}  \;.
    \end{equation*}
    \item The planar anti-Einstein-Maxwell solution
    \begin{equation*}
        f(r) = \frac{2M}{r} - \frac{q^2}{r^2}, \qquad \kappa =  \frac{4M^3}{q^4}  \;.
    \end{equation*}
\end{enumerate}
While we do not explicitly work with the Schwarzschild solution in this paper, we include
it so that one compare computations with the 
well-known results \cite{Wald:106274}. For the remainder of the discussion, we will keep $f(r)$ general and sometimes use its Taylor expansion near the horizon:
\begin{equation}
\label{eq:taylorbh}
        f(r) = 2 \kappa (r - r_h) + \Op((r - r_h)^2) \;.
\end{equation}

\subsubsection{Defining Kruskal coordinates}

As a first step to extending the solution beyond the horizon\footnote{It will turn out that
the naive extension of the line element to $r<r_h$ covers part of the maximal extension.
This is of course well known for standard solutions, such as Schwarzschild.}
we replace the transversal/radial coordinate $r$ by the tortoise coordinate 
\begin{equation*}
  r_\star = \int f(r)^{-1} dr, \qquad -\infty < r_\star < \infty \;.
\end{equation*}
Then we can define the future-pointing null coordinates
\begin{equation*}
    x_+ = t + r_\star, \qquad x_- = t - r_\star, \qquad -\infty < x_\pm < \infty.
\end{equation*}
In these coordinates,  the metric is given by
\begin{equation*}
  ds^2 = - f(r) dx_+ dx_- + r^2 d\vec{X}^2,
\end{equation*}
where we have used that
\begin{equation*}
    dx_+ = dt + f(r)^{-1} dr, \qquad dx_- = dt - f(r)^{-1} dr \;.
\end{equation*}
It is understood that $r$ is now a function of $x_\pm$, which we only 
need to define implicitly,
\begin{equation*}
    r = r[r_\star(x_+, x_-)] \;.
\end{equation*}
Outgoing null congruences consist of null rays 
with $x_- = $ constant and propagate in the 
positive $ x_+$ direction, while ingoing null congruences are defined by $x_+ =$ constant  and
move in the positive $x_-$ direction.

To extend our solution, we define Kruskal coordinates by 
\begin{equation*}
\begin{aligned}
    X_- &= - e^{-x_- \kappa}, & \qquad -\infty &< X_- < 0 \;\; &&\Leftrightarrow \;\; -\infty < x_- < \infty  \;,\\ 
    X_+ &= + e^{x_+ \kappa }, & \qquad 0 &< X_+ < \infty \;\;&&\Leftrightarrow \;\; -\infty < x_+ < \infty. \\ 
\end{aligned}
\end{equation*}
The Kruskal extension is obtained by dropping the constraints $X_-<0, X_+>0$, resulting 
in the four regions of Figure \ref{fig:comparison}. 
In order for this extension to be well defined, we must make sure that the metric
is non-degenerate at the horzions $r=r_h$, where $f(r)$ has its zero. It is for this purpose that factors of the
surface gravity $\kappa$ have been included in the definition of $X_\pm$. 

The line element in Kruskal coordinates is given by
\begin{equation}
\label{eq:krus1bh}
     ds^2 = - \frac{1}{\kappa^2} f(r) e^{-2\kappa r_\star} dX_+ dX_- + r^2 d\vec{X}^2\;,
\end{equation}
where $r(X_+,X_-)$ is determined implicitly by  
\begin{equation*}
    X_+ X_- = - e^{2 \kappa r_\star} \;.
\end{equation*}
We can also implicitly define the coordinate $t(X_+,X_-)$ from
\begin{equation*}
  \frac{X_+}{X_-} = -e^{2 \kappa t} \;,
\end{equation*}
To show that \eq{krus1bh} is regular on the horizon, we  look at the Taylor expansion \eq{taylorbh}, and integrate to obtain $r_\star$, 
\begin{equation*}
    -2 \kappa r_\star = \int \left( - \frac{2 \kappa }{2\kappa(r - r_h)}  +  \Op(r - r_h) \right) dr = -\log(r - r_h) + \Op((r - r_h)^2).
\end{equation*}
We see that the zeros are cancelled, 
\begin{equation*}
    -f(r) e^{-2\kappa r_\star} = - 2\kappa (r - r_h) e^{-\log(r - r_h)} = -2\kappa  + \Op((r - r_h)^2), 
\end{equation*}
and thus, on the horizon we find the line element
\begin{equation*}
     ds^2 = -\frac{2}{\kappa} dX_+ dX_- + r_h^2 d\Omega^2.
\end{equation*}
As a result, one can consider the metric \eq{krus1bh} with the coordinates extended to
\begin{equation*}
    - \infty < X_\pm < \infty \;,
\end{equation*}
subject to the constraint that $r(X_+,X_-) > 0$. This produces the four regions of the diagram in Figure \ref{fig:comparison}.

With our choice of signs in \eqref{eq:krus1bh}, the static region we started with is identified
with region I. The Kruskal extension contains a second static region, region IV. 
In Region IV the null Kruskal coordinates take values $X_- > 0$ and $X_+ < 0$. One can introduce null coordinates $x_\pm$ by
\begin{equation*}
\begin{aligned}
    X_- &= + e^{-x_- \kappa}, & \qquad 0 &< X_- < \infty \;\; &&\leftrightarrow \;\; \infty > x_- > - \infty  \;.\\ 
    X_+ &= - e^{x_+ \kappa}, & \qquad -\infty &< X_+ < 0 \;\;&&\leftrightarrow \;\; \infty > x_+ > - \infty \;.\\ 
\end{aligned}
\end{equation*}
Observe that $x_+,x_-$ are directed opposite to $X_+,X_-$ in Region IV. If we go back from $x_\pm$, to $t, r_\star$ and further to $t, r$, using the same relations as in Region I, the metric assumes the same local form \eq{krus1bh} as in Region I. But globally, there is a difference compared to Region I, as 
$t, r$ point the opposite way: $t$ downwards, $r$ leftwards. As a result, ingoing lightfronts move in positive $X_+ =$ negative $x_+$ direction. Outgoing lightfronts move in the positive $X_- =$ negative $x_-$ direction. The association of $X_-, X_+$ with in/out-moving lightfronts is reversed compared to Region I.

The global spacetime is time-orientable and time-reversal symmetric. We should not conclude that time is flowing backwards in Region IV. If we choose a global time orientation that points in the same direction as $t$ in Region I, then physical time in Region IV is measured by $-t$. We refer to Region I 
as the `standard' static region, and the definition of Kruskal coordinates as the `standard 
embedding' of the static region into its Kruskal extension. 
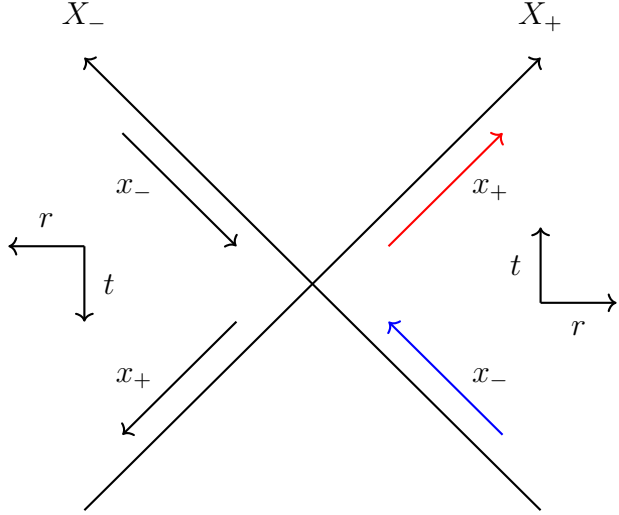
\begin{figure}[!h]
\centering
\begin{tikzpicture}

    \draw[thick, ->] (6,0) -- node[at end, above=0.2cm] {$X_-$} (0,6);
    \draw[thick, ->] (0,0) -- node[at end, above=0.2cm] {$X_+$} (6,6);
    
    \begin{scope}[shift={(-1,0)}]
    \draw[thick, ->] (7,2.75) -- node[midway, left=0.1cm] {$t$} (7,3.75);
    \draw[thick, ->] (7,2.75) -- node[midway, below=0.1cm] {$r$} (8,2.75);
    \end{scope}
    
    \begin{scope}[shift={(1,0)}]    
    \draw[thick, ->] (-1,3.5) -- node[midway, right=0.1cm] {$t$} (-1,2.5);
    \draw[thick, ->] (-1,3.5) -- node[midway, above=0.1cm] {$r$} (-2,3.5);
    \end{scope}

    \draw[thick, ->] (0.5, 5) -- node[midway, left=0.2cm] {$x_-$} (2, 3.5);
    \draw[thick, ->] (2, 2.5) -- node[midway, left=0.2cm] {$x_+$} (0.5, 1);
        
    \draw[thick, ->, red] (4, 3.5) -- node[midway, right=0.2cm, black] {$x_+$}  (5.5, 5);
    \draw[thick, ->, blue] (5.5, 1) -- node[midway, right=0.2cm, black] {$x_-$}(4, 2.5);
\end{tikzpicture}
\caption{Flow of coordinates in the static regions of the Kruskal diagram for black hole solutions. The red arrow denotes future directed outgoing null geodesics, the blue arrow denotes future directed ingoing null geodesics}
\end{figure}

\subsubsection{Calculation of expansions}

With the global causal structure understood, we are now in a position to write down future-directed ingoing and outgoing null geodesics and then calculate their expansions. Before this, we perform some intermediate calculations. 

For $X_\pm$ in region I, we can calculate $dr$ from
\begin{equation*}
    X_+ dX_- +  X_- dX_+ = - 2 \kappa e^{2\kappa r_\star} dr_\star = 2 \kappa X_+ X_- dr_\star \;,
\end{equation*}
which allows us to write
\begin{equation}
\label{eq:drcalc}
    dr = \frac{f(r)}{X_+ X_-} \frac{1}{2 \kappa} \left( X_+ dX_- + X_- dX_+ \right).
\end{equation}
Similarly, we can write down
\begin{equation*}
    dt = - \frac{1}{X_+ X_-} \frac{1}{2 \kappa} \left( X_+ dX_- - X_- dX_+ \right) \;.
\end{equation*}
In the original coordinates, the Killing vector field is given by $\xi  = \pardev{}{t}$, and using the metric tensor we can write down the co-vector $\xi^\flat = -f(r)dt$.\footnote{We use
`musical' notation to distinguish between vectors and the corresponding co-vectors (one-forms).} Using the results from above, we can rewrite the Killing co-vector field in terms of our new Kruskal-like coordinates
\begin{equation*}
    \xi^\flat = - f(r) dt = \frac{f(r)}{X_+ X_-} \frac{1}{2 \kappa} \left( X_+ dX_- - X_- dX_+ \right)\;.
\end{equation*}
The corresponding vector field is
\begin{equation*}
    \xi = \kappa \left(- X_- \pardev{}{X_-} + X_+ \pardev{}{X_+} \right) \;.
\end{equation*}
To calculate the expansions we choose the following normal null co-vector fields for
our two null congruences:
\begin{equation}
\label{eq:geodesics}
    N_{+}^\flat = -\frac{1}{\kappa} dX_-, \qquad N_{-}^\flat = -\frac{1}{\kappa} dX_+
\end{equation}
or, as vector fields,
\begin{equation*}
    N_+ = -2 \kappa \frac{ X_+ X_- }{f(r)} \left(\pardev{}{X_+} \right), \qquad N_- = -2 \kappa \frac{ X_+ X_- }{f(r)} \left(\pardev{}{X_-} \right).
\end{equation*}
The normalisation and overall sign has been set in \eq{geodesics} such that 
\begin{equation*}
    \xi \cdot N_+ = X_-, \qquad \xi \cdot N_- = -X_+\;,
\end{equation*}
which ensure that the normal null vectors $N_\pm$ are future directed, $\xi \cdot N_\pm < 0$, 
in region I where the Killing vector flows in the direction of global time. Their expansions are calculated 
\begin{equation*}
    \theta_\pm = \nabla_\mu N^{\mu}_\pm = \frac{1}{\sqrt{-g}} \partial_\mu \left(\sqrt{-g} N^\mu_\pm \right), \qquad \sqrt{-g} = -  \frac{f(r) \sqrt{h}}{X_+ X_-} \frac{r^2}{2 \kappa^2}\;,
\end{equation*}
where we use that $h = \det \left( d\vec{X}^2 \right)$ for either the two-sphere or the two-plane depending on the symmetry of the solution. This results in
\begin{equation*}
    \theta_\pm = - \frac{4 \kappa}{r} \frac{X_+ X_-}{f(r)} \pardev{r}{X_{\pm}} \;.
\end{equation*}
To calculate the sign, we use \eq{drcalc} to find that
\begin{equation*}
    \pardev{r}{X_\pm} = \frac{f(r)}{X_+ X_-} \frac{X_\mp}{2 \kappa},
\end{equation*}
which gives the result
\begin{equation*}
    \theta_+ = -\frac{2 X_-}{r}, \qquad \theta_- = -\frac{2 X_+}{r} \;.
\end{equation*}
Thus the expansions change signs across horizons, and the resulting pattern
already completely determines the nature of each horizon. For completeness,
we calculate the Lie derivatives of expansions vanishing  at the horizon, though we stress that their signs
are completely determined by the sign change of the corresponding expansion. 
We find that
\begin{equation*}
    \La_{N_-} \theta_+ = - 2 \kappa \frac{ X_+ X_- }{f(r)}  \partial_- \left( - \frac{2 X_-}{r} \right) = \frac{4 \kappa}{r} \left( \frac{X_+ X_-}{f(r)} \right) - \frac{2 X_+ X_-}{r^2},
\end{equation*}
which evaluated on the horizon, gives
\begin{equation*}
    \La_{N_-} \theta_+ \bigg|_{r = r_h} = \frac{4 \kappa}{r_h} \cdot \left(- \frac{1}{2\kappa} \right) = -\frac{2}{r_h}\;.
\end{equation*}
By symmetry, we can see that
\begin{equation*}
    \La_{N_+} \theta_- \bigg|_{r = r_h} = -\frac{2}{r_h} \;.
\end{equation*}
We can now look at the two horizons in region I, where we have $X_- < 0$ and $X_+ > 0$. For the horizon given by $X_+ = 0$, we have:
\begin{equation*}
    \theta_+ > 0, \quad \theta_- = 0, \quad \La_{N_+} \theta_- < 0 \;,
\end{equation*}
which is a \emph{past outer horizon}. For the horizon set by $X_- = 0$, we have that 
\begin{equation*}
    \theta_+ = 0, \quad \theta_- < 0, \quad \La_{N_-} \theta_+ < 0 \;,
\end{equation*}
which is a \emph{future outer horizon}.

Since the causal and expansion properties of the Kruskal diagram do not depend on details
of the function $f$, the interpretation is the same as for the Kruskal-Schwarzschild solution,
for all members of this class. 
The future outer horizon is the event horizon of a black hole, and the past outer horizon is the horizon for the white hole region. For thermodynamics we use
the horizon where causal geodesic cross from the exterior to the interior, which is the
future outer horizon between Regions I and II. This horizon has positive temperature,
$T_H \propto \kappa > 0$.

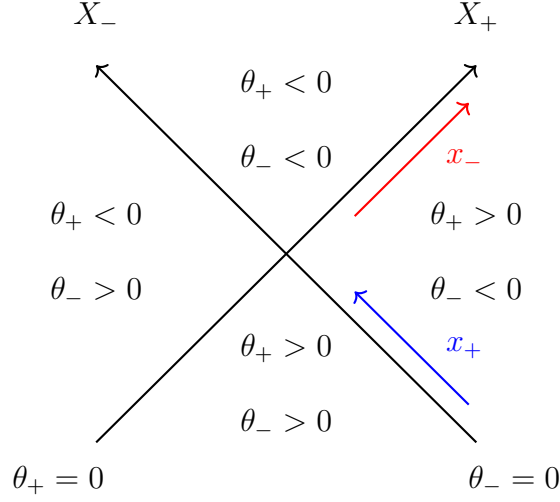
\begin{figure}[!h]
\centering
\begin{tikzpicture}

\draw[thick, ->] (-2.5,-2.5) -- node[at end, above=0.3cm]{$X_+$} (2.5,2.5);
\draw[thick, ->] (2.5,-2.5) -- node[at end, above=0.3cm]{$X_-$} (-2.5,2.5);

\node (a) at (2.5,0.5) {$\theta_+ > 0$};
\node (a) at (2.5,-0.5) {$\theta_- < 0$};

\node (a) at (-2.5,0.5) {$\theta_+ < 0$};
\node (a) at (-2.5,-0.5) {$\theta_- > 0$};

\node (a) at (0,2.25) {$\theta_+ < 0$};
\node (a) at (0,1.25) {$\theta_- < 0$};

\node (a) at (0,-1.25) {$\theta_+ > 0$};
\node (a) at (0,-2.25) {$\theta_- > 0$};

\node (0) at (-3, -3) {$\theta_+ = 0$};
\node (0) at (3, -3) {$\theta_- = 0$};

\draw[thick, ->, blue] (2.4,-2) -- node[midway, right=0.3cm] {$x_+$} (0.9,-0.5);
\draw[thick, ->, red] (0.9,0.5) -- node[midway, right=0.3cm] {$x_-$} (2.4, 2);

\end{tikzpicture}
\caption{Signs of the expansions $\theta_\pm$ in the four quadrants of the Kruskal diagram for a our black hole solutions for which $\kappa  > 0$.}
\end{figure}

\subsection{Cosmological solutions}

We now turn our attention to the cosmological solutions in this paper.
The line element takes the same form 
\begin{equation*}
    ds^2 = -f(r) dt^2 + f(r)^{-1} dr^2 + r^2 d\vec{X}^2 
\end{equation*}
as before, but now  $f(r)$ is positive for $r_{\mathrm{sing}} < r < r_h$, where $r_{\mathrm{sing}}$ is the position of
the singularity. The metric is static in this region, so that compared to the previous 
class the roles of exterior and interior are exchanged, that is, the interior region is static. 
We assume that $f(r)$ has a simple zero and therefore changes sign at $r=r_h$. 
Since the Killing vector field $\partial_t$ becomes spacelike for $r>r_h$, the outside
region is dynamical. We assume that $f(r)$ is negative for $r_h < r < \infty$, with 
$r\rightarrow \infty$ at infinite distance. Thus the horizon at $r=r_h$ is a cosmological 
horizon. Under the conditions we have imposed on $f$ the surface gravity is \emph{negative}
\begin{equation*}
    \kappa = \half \partial_r f(r) \bigg|_{r = r_h} < 0.
\end{equation*}
Explicit choices for $f(r)$, which we consider in this paper, are
\begin{enumerate}
    \item De Sitter solution
    \begin{equation*}
        f(r) = 1 - \frac{r^2}{L^2}, \qquad \kappa = - \frac{1}{L}  \;.
    \end{equation*}
    \item Planar Einstein-Maxwell solution
    \begin{equation*}
        f(r) = \left( - \frac{2M}{r} + \frac{q^2}{r^2} \right), \qquad \kappa = - \frac{4M^3}{q^4}  \;.
    \end{equation*}
    \item Planar STU solution
    \begin{equation*}
        f(r) = \frac{1 - \alpha r}{2\sqrt{\Ham_0 \Ham_1 \Ham_2 \Ham_3}}, \qquad \Ham_a = \beta_a + \gamma_a r, \qquad \kappa = - \frac{\alpha^3}{4} \frac{\sqrt{\gamma_0 \gamma_1 \gamma_2 \gamma_3}}{Q_0 P^1 P^2 P^3} \;.
    \end{equation*}
\end{enumerate}
For the remainder of the discussion, we will keep $f(r)$ general and sometimes use its Taylor expansion near the horizon:
\begin{equation*}
        f(r) = 2 \kappa (r - r_h) + \Op((r - r_h)^2) \;.
\end{equation*}
\subsubsection{Defining Kruskal coordinates}
As in the black hole case we start by defining a
Tortoise coordinate 
\begin{equation*}
  r_\star = \int f(r)^{-1} dr, \qquad 0 < r_\star < \infty \;.
\end{equation*}
Then we introduce future-pointing null coordinates
\begin{equation*}
    x_+ = t + r_\star, \qquad x_- = t - r_\star, \qquad -\infty < x_\pm < \infty \;,\;\;\;
    r(x_+,x_-) > r_{\mathrm{sing}} \;.
\end{equation*}
The  line element takes the form
\begin{equation*}
  ds^2 = - f(r) dx_+ dx_- + r^2 d\vec{X}^2\;,
\end{equation*}
where $r$ is an implicitly defined function of $x_\pm$. 
Note that $x_+$ is future- and outward-pointing while $x_-$ is future-  and 
inward-pointing in the interior region. This is the same assignment as in black hole solutions considered previously. With coordinates fixed in this way, we can clearly see what is the difference compared to the static patch of the black hole solutions. Since $r$ points in the opposite direction, the roles of interior and exterior are exchanged, where interior means $r  <r_h$. While $x_+$ points outwards in both cases, it points away form the horizon for the black hole solutions, but towards the horizon for the cosmological ones. 
This makes it natural to define Kruskal coordinates such that the standard static region, 
which we use to fix the overall time orientation, is Region IV, rather than Region I.

We start with the static line element, rewritten using null coordinates $x_\pm$,
where $x_+$ is future-pointing and outward-pointing, while $x_-$ is future-pointing and inward-pointing, relative to the local coordinates $t, r$. This fixes the definitions of the expansion $\theta_\pm$, 
and the direction of physical time. Next, we define global null Kruskal coordinates $X_\pm$ such that they point in the same direction as $x_\pm$. 
\begin{equation*}
\begin{aligned}
    X_+ &= - e^{\kappa x_+} \qquad & - \infty &< X_+ < 0 \; \;&&\Leftrightarrow \; \; -\infty < x_+ < \infty  \;.\\
    X_- &= e^{- \kappa x_-} \qquad & 0 &< X_- < \infty \; \;&&\Leftrightarrow \; \; -\infty < x_- < \infty  \;,
\end{aligned}
\end{equation*}
where the factors of the surface gravity have been included to make manifest that the 
metric is regular at $r=r_h$ where $f(r)$ has its zero. We have also used 
that $\kappa < 0$. The standard static patch is Region IV, and it is illustrative to 
compare the black hole case and the cosmological case in Figures 
\ref{fig:kruskalcomp1} and \ref{fig:comparison}.

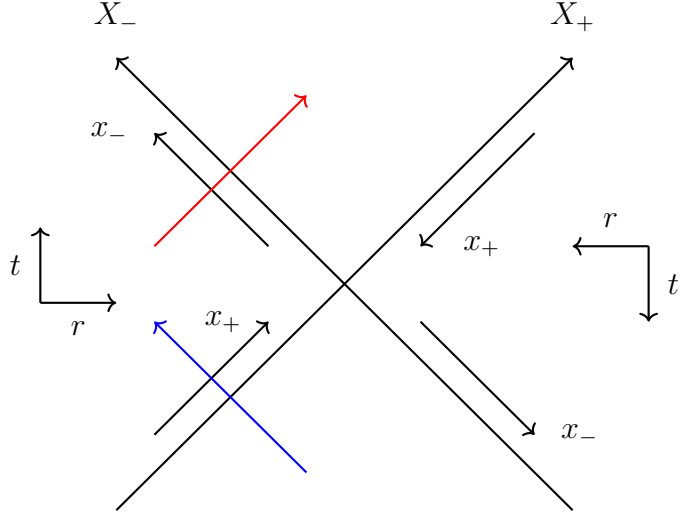
\begin{figure}[!h]
\centering
\begin{tikzpicture}

    \draw[thick, ->] (6,0) -- node[at end, above=0.2cm] {$X_-$} (0,6);
    \draw[thick, ->] (0,0) -- node[at end, above=0.2cm] {$X_+$} (6,6);
    
    \begin{scope}[shift={(0.5,0)}]
    \draw[thick, ->] (-1.5,2.75) -- node[midway, left=0.1cm] {$t$} (-1.5,3.75);
    \draw[thick, ->] (-1.5,2.75) -- node[midway, below=0.1cm] {$r$} (-0.5,2.75);
    \end{scope}
    
    \begin{scope}[shift={(-0.5,0)}]    
    \draw[thick, ->] (7.5,3.5) -- node[midway, right=0.1cm] {$t$} (7.5,2.5);
    \draw[thick, ->] (7.5,3.5) -- node[midway, above=0.1cm] {$r$} (6.5,3.5);
    \end{scope}

    \draw[thick, ->] (2, 3.5) -- node[at end, left=0.2cm] {$x_-$} (0.5, 5);
    \draw[thick, ->] (0.5, 1) -- node[at end, left=0.2cm] {$x_+$} (2, 2.5);
        
    \draw[thick, ->, black]  (5.5, 5) -- node[at end, left=-1.2cm, black] {$x_+$} (4, 3.5);
    \draw[thick, ->, black] (4, 2.5) -- node[at end, right=0.2cm, black] {$x_-$}(5.5, 1);
    
    \draw[thick, red, ->] (0.5, 3.5) -- (2.5, 5.5);
    \draw[thick, blue, ->] (2.5, 0.5) -- (0.5, 2.5);
\end{tikzpicture}
\caption{Flow of coordinates in the static regions of the Kruskal diagram for cosmological solutions. The red arrow denotes future directed outgoing null geodesics, the blue arrow denotes future directed ingoing null geodesics}
\label{fig:orientationcos}
\end{figure}

The line element in Kruskal coordinates is given by
\begin{equation*}
     ds^2 = - \frac{f(r) e^{-2\kappa r_\star}}{\kappa^2} dX_+ dX_- + r^2 d\vec{X}^2 \;.
\end{equation*}
We can show this is regular on the horizon using the same method as previously. By expanding at the horizon, we find
\begin{equation*}
    -2 \kappa r_\star = \int \left( - \frac{2 \kappa  }{2\kappa(r - r_h)} +  \Op(r - r_h) \right) dr  = -\log(r_h - r)  +  \Op((r - r_h)^2)\;,
\end{equation*}
where we note that the $\log(r_h - r)$ is different from the black hole case, as in the static patch we have $r < r_h$. Putting this together, we obtain
\begin{equation*}
    -f(r) e^{-2\kappa r_\star} = - 2\kappa (r - r_h) e^{-\log(r_h - r)} = 2\kappa + \Op((r - r_h)^2) \;,
\end{equation*}
such that on the horizon, the line element is given by
\begin{equation*}
     ds^2 = \frac{2}{\kappa} dX_+ dX_- + r_h^2 d\vec{X}^2 \;.
\end{equation*}
We see that we can extend the Kruskal null coordinates such that
\begin{equation*}
    -\infty < X_\pm < \infty \;,
\end{equation*}
subject to the constraint that $r(X_+,X_-) > r_{\mathrm{sing}}$, where we implicitly write
\begin{equation*}
        X_+ X_- = - e^{2 \kappa r_\star}, \qquad \frac{X_+}{X_-} = -e^{2 \kappa t} \;.
\end{equation*}
The direction of various coordinates in the respective regions is shown in Figure \ref{fig:orientationcos}.

\subsubsection{Calculating the expansions}
As before, we precalculate a few useful relations for our calculations, which we find are identical to the results for the black hole solutions, namely
\begin{equation*}
    dr = \frac{f(r)}{X_+ X_-} \frac{1}{2 \kappa} \left( X_+ dX_- + X_- dX_+ \right) \;,
\end{equation*}
\begin{equation*}
    dt = - \frac{1}{X_+ X_-} \frac{1}{2 \kappa} \left( X_+ dX_- - X_- dX_+ \right) \;.
\end{equation*}

Again as in the calculation for black hole solutions, we start with the Killing vector field in our original coordinates: $\xi  = \pardev{}{t}$. Using the metric tensor write down the co-vector field $\xi^\flat = -f(r) dt$, which we can rewrite with the results above to write down the Killing co-vector field in terms of the Kruskal-like coordinates
\begin{equation*}
    \xi^\flat = - f(r) dt = \frac{f(r)}{X_+ X_-} \frac{1}{2 \kappa} \left( X_+ dX_- - X_- dX_+ \right) \;,
\end{equation*}
which the corresponding vector given by 
\begin{equation*}
    \xi = \kappa \left(- X_- \pardev{}{X_-} + X_+ \pardev{}{X_+} \right) \;.
\end{equation*}
We now write down the geodesics which are future-pointing within region IV, where the normalisation of the co-vectors is handpicked to ensure this property:
\begin{equation}
\label{eq:cosnullvec}
    N^\flat_{+} = \frac{1}{\kappa} dX_-, \qquad N^\flat_{-} = \frac{1}{\kappa} dX_+ \;,
\end{equation}
or as vectors
\begin{equation*}
    N_+ = 2 \kappa \frac{ X_+ X_- }{f(r)} \left(\pardev{}{X_+} \right), \qquad N_- = 2 \kappa \frac{ X_+ X_- }{f(r)} \left(\pardev{}{X_-} \right) \;.
\end{equation*}
Double checking that the normals are future-pointing, we look at the inner product of these with the Killing vector field
\begin{equation*}
    \xi \cdot N_+ = -X_-, \qquad \xi \cdot N_- = X_+ \;,
\end{equation*}
which we see obeys $\xi \cdot N_\pm < 0$ in region IV. The expansions are calculated in the same manner, and we find that the relative sign imposed in the normalisation by considering region IV, rather than region I introduces a relative sign in the expansions compared to the black hole case,
\begin{equation*}
    \theta_+ = \frac{2 X_-}{r}, \qquad \theta_- = \frac{2 X_+}{r}.
\end{equation*}
While this already determines the types of all horizons, 
we also calculate the Lie derivative at the horizon explicitly. We find that
\begin{equation*}
    \La_{N_-} \theta_+ = 2 \kappa \frac{ X_+ X_- }{f(r)}  \partial_- \left( \frac{2 X_-}{r} \right) = \frac{4 \kappa}{r} \left( \frac{X_+ X_-}{f(r)} \right) - \frac{2 X_+ X_-}{r^2},
\end{equation*}
and on the horizon, we find
\begin{equation*}
    \La_{N_-} \theta_+ \bigg|_{r = r_h} = \frac{4 \kappa}{r_h} \cdot \frac{1}{2\kappa} = \frac{2}{r_h}, \qquad     \La_{N_+} \theta_- \bigg|_{r = r_h} = \frac{2}{r_h}.
\end{equation*}

\begin{figure}[!h]
\centering
\begin{tikzpicture}

\draw[thick, ->] (-2.5,-2.5) -- node[at end, above=0.3cm]{$X_+$} (2.5,2.5);
\draw[thick, ->] (2.5,-2.5) -- node[at end, above=0.3cm]{$X_-$} (-2.5,2.5);

\node (a) at (2.5,0.5) {$\theta_+ < 0$};
\node (a) at (2.5,-0.5) {$\theta_- > 0$};

\node (a) at (-2.5,0.5) {$\theta_+ > 0$};
\node (a) at (-2.5,-0.5) {$\theta_- < 0$};

\node (a) at (0,2.25) {$\theta_+ > 0$};
\node (a) at (0,1.25) {$\theta_- > 0$};

\node (a) at (0,-1.25) {$\theta_+ < 0$};
\node (a) at (0,-2.25) {$\theta_- < 0$};

\node (0) at (-3, -3) {$\theta_+ = 0$};
\node (0) at (3, -3) {$\theta_- = 0$};

\draw[thick, ->, blue] (-0.8,0.5) -- node[midway, left=0.3cm] {$x_-$} (-2.3,2);
\draw[thick, ->, red] (-2.3,-2)-- node[midway, left=0.3cm] {$x_+$} (-0.8, -0.5);

\end{tikzpicture}
\caption{Signs of the expansions $\theta_\pm$ in the four quadrants of the Kruskal diagram for a cosmological solution where $\kappa < 0$.}
\label{fig:flipped}
\end{figure}
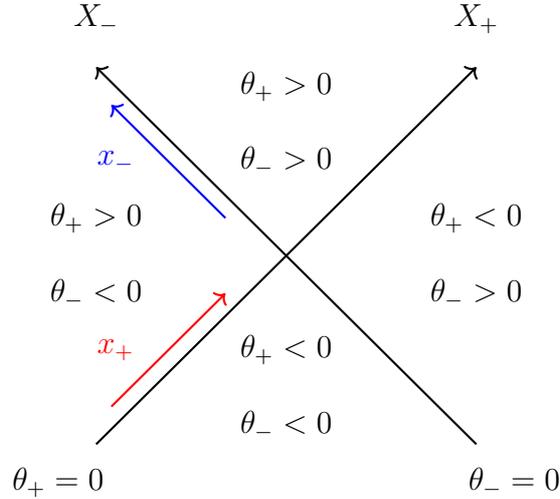

Let us look at the left part of the Kruskal diagram that is regions III, IV and II. The physics of
the sequence III, I, IV is equivalent, but parametrized differently since $\partial_t$ is 
past-pointing in region I. 
In region IV of the Kruskal diagram, we have $X_- > 0$ and $X_+ < 0$. On the horizon given by $X_+ = 0$, we have:
\begin{equation*}
    \theta_+ > 0, \quad \theta_- = 0, \quad \La_{N_+} \theta_- > 0
\end{equation*}
which shows that this is a \emph{past inner horizon}. For the horizon set by $X_- = 0$, we have that 
\begin{equation*}
    \theta_+ = 0, \quad \theta_- < 0, \quad \La_{N_-} \theta_+ > 0
\end{equation*}
which is a \emph{future inner horizon}. The expansions for all four regions is illustrated in Figure \ref{fig:flipped}. 

When considering future-directed causal geodesics which pass through a horizon from the exterior to the interior, we must consider the future inner horizon between region III and region IV. For a future inner horizon, we have that $T_H \propto \kappa < 0$, which aligns with the signs of the temperatures we consider throughout this paper while employing the triple Wick rotation for our cosmological solutions. Overall the sequence III, IV (or I), II, describes a cosmic bounce,
since the solution is a contracting cosmology in III and an expanding cosmology in II. 
We expand on the global interpretation of the cosmological solution in \ref{app:asymptotics}.

\subsection{Eddington-Finkelstein coordinates}

For completeness we also derive expressions for advanced and retarded
Eddington-Finkelstein coordinates. We start with the line element
\begin{equation}
\label{Static_patch}
ds^2 = -f(r) dt^2 + f(r)^{-1} dr^2 + r^2 d\vec{X}^2 \;,
\end{equation}
where we assume that there is some interval for $r$ such that
$f(r)>0$, so that the line element is static, with Killing vector
field $\partial_t$. We also assume that $f(r)$ has a simple zero
at $r=r_h$, so that there is a second region where the Killing
vector field becomes spacelike. At this point we do not specify
whether the static region is $r>r_h$ or $r<r_h$, so that we can
cover black holes and cosmological solutions simultaneously. 

We first define advanced Eddington Finkelstein coordinates
$(x_+, r, \ldots)$, where we omitted two further coordinates, which
are $\theta,\phi$ for spherical and $x,y$ for planar symmetry. 
The null coordinate $x_+$ is defined as before,
\begin{equation}
dx_+ = dt + f(r)^{-1} dr = dt + dr_\star \;.
\end{equation}
The line element takes the form 
\begin{equation}
ds^2 = -f(r) dx_+^2 + 2 dx_+ dr + r^2 d\vec{X}^2 \;,
\end{equation}
which is manifestly regular at $r=r_h$, so that we can cover
both the static and the non-static domain. Now we consider
radial/transversal null geodesics, which must satisfy
\begin{equation}
0 = -f(r) dx_+^2 + 2 dx_+ dr \Rightarrow dx_+ = 0 \text{  or  } \frac{dr}{dx^+} = \frac{f}{2} \;.
\end{equation}
Using $\lambda=x^+$ as an affine parameter,
the normal vector field 
for the second null congruence is 
\begin{equation}
U_+= \left( \frac{dx^\mu}{d\lambda}  \right)= (U_+^+, U_+^r,\ldots ) = \left(1,\frac{f}{2}, \ldots \right)  \;.
\end{equation}
The corresponding co-vector has the form\footnote{We note here that the normal vector fields $U_\pm$ differs in their overall normalisation from $N_\pm$ defined in  the previous section \eq{cosnullvec}. The normalisation has been chosen to avoid carrying around an irrelevant numerical factor.}
\begin{equation}
U_{+}^\flat= (U_{+|+}, U_{+|r}, \ldots) = \left( - \frac{f}{2}, 1, 0 , 0 \right)  \;.
\end{equation}


The normal vector field $U_-$ for the congruence $dx^+=0$ is a constant vector field. 
We normalize it such that $U_+\cdot U_- = -2$:
\begin{equation}
U_-= (U_-^+, U_-^r, \ldots ) = (0,-2,\ldots) \;,\;\;\;
U_{-}^\flat = (U_{-|+}, U_{-|r} , \ldots ) = (-2,0,\ldots) \;.
\end{equation}
Both $U_\pm$ can be checked to be future-pointing. We could now proceed and
compute the expansions $\theta_\pm$ but one nice feature of Eddington-Finkelstein 
coordinates is that they allow one to `visualize' the expansion properties of geodesics,
and the causal structure. To do this we introduce the advanced time
coordinate 
\begin{equation}
\bar{t} := x_+ - r  = t + r_\star -r  \Leftrightarrow x_+ = \bar{t} + r = t + r_\star \;,
\end{equation}
and transform the null congruences $U_\pm$ from coordinates
$(x_+,r,,\ldots)$ to coordinates $\bar{t}, r,\ldots)$. The result is 
\begin{eqnarray*}
(U_+^{\bar{t}}, U_+^r, \ldots ) &=& \left( 1 - \frac{f}{2}, \frac{f}{2}, \ldots \right) \;, \\
(U_-^{\bar{t}}, U_-^r, \ldots ) &=& \left( 2,-2 , \ldots  \right) \;.
\end{eqnarray*}
Now it is manifest that $U_-$ is future and inward-pointing for all $r$, 
while $U_+$ is future-pointing, but will switch between outward-pointing and
inward-pointing at $r=r_h$. Further details depend on whether $f$ is
positive for $r>r_h$ or for $r<r_h$. Since the black hole case $r>r_h$ is
familiar, we focus on the cosmological case, $r<r_h$. Then $U_+$ is
pointing outwards on the inside $r<r_h$, but pointing inwards on the outside
$r>r_h$. At the horizon, $r=r_h$, $U_+$ points `upwards', that is light rays
are stuck at the horizon. Drawing the lightcones associated with $U_\pm$ 
we see that future-pointing null and timelike geodesics can cross from the
outside to the inside, but not the other way. This shows that $r<r_h$ corresponds
to the standard static region IV in the cosmological Kruskal diagram, while $r>r_h$ corresponds
to region III (and not to region II). In particular, the limit $r\rightarrow \infty$ 
corresponds to past timelike infinity and the coordinate $r$ is timelike and past-pointing
for $r>r_h$.

Outside the horizons $r=r_h$ we can use the expression \eqref{Static_patch}
in all regions of the Kruskal diagram. In the main part of the paper our convention is to 
relabel coordinates $r\leftrightarrow t$ in 
non-static patches, so that the timelike coordinate is always denoted $t$. 
We also multiply $f$ by a minus sign in non-static regions, so that $f$
remains positive. Let us here be more explicit and define $\tilde{f}(x) = - f(x)$. 
If we extend our solution from the static region IV to the non-static region III,
 then $t\rightarrow \infty$ corresponds to 
past timelike infinity, and $t$ is past-pointing. If we prefer to use a time coordinate
in region III which is future pointing, we should relabel $t\rightarrow -t$, so that the line
element in region III is\footnote{Note that $f$ and $\tilde{f}$ are, in general, not even functions.}
\begin{equation}
\label{Region_IV_EF}
ds^2 = - \frac{dt^2}{\tilde{f}(-t)} + \tilde{f}(-t) dr^2 + t^2 d\vec{X}^2
\;,\;\; -\infty < t < -r_h \;.
\end{equation}

Let us now briefly discuss retarded Eddington-Finkelstein coordinates
$(x_-,r,\ldots)$, where, in the static patch,  we replace $t$ by the ingoing future-pointing
null coordinate $x_-$,
\begin{equation}
dx_- = dt - f(r)^{-1} dr  = dt - dr_\star \;.
\end{equation}
The line element is
\begin{equation}
ds^2 = -f(r) dx_-^2 - 2 dx_- dr + r^2 d\vec{X}^2
\end{equation}
with radial/transversal null geodesics
\begin{equation}
dx_- = 0 \;,\;\;\;
\frac{dr}{dx_-} = - \frac{1}{2} f(r) \;.
\end{equation}
The future-pointing null normal vector fields of the null congruences are
\begin{eqnarray}
U_+&=& (U_+^-, U_+^r, \ldots ) = (0,2, \ldots ) \;,\\
U_- &=& (U_-^-, U_-^r, \ldots) = \left(1, -\frac{f}{2}, \ldots \right)  \;,
\end{eqnarray} 
which satisfy $U_+\cdot U_- = -2$. Introducing the new `retarded' time coordinate
$\bar{t} = x_- + r$, and transforming from coordinates $(x_-, r, \ldots)$ to 
coordinates $(\bar{t}, r, \ldots)$ they transform into 
\begin{eqnarray}
U_+ &=& (U_+^{\bar{t}}, U_+^r, \ldots) = (2,2, \ldots) \;,\\
U_- &=&  (U_-^{\bar{t}}, U_-^r, \ldots) = \left(1-\frac{f}{2} ,-\frac{f}{2}, \ldots \right)  \;.
\end{eqnarray}
This shows that $U_+$ is future and outward-pointing for all $r$ while
$U_-$ is future-pointing but changes between pointing inwards and pointing outwards
at $r=r_h$. For black holes, where $f>0$ for $r>r_h$, this provides the 
extension into the white hole part of the Kruskal extension (region III). For cosmological
solutions we see that $U_-$ is pointing inwards for $r<r_h$, which we identify
with the standard static region IV, and outwards for $r>r_h$. Future-pointing null and timelike
geodesics can only cross from the inside to the outside, which shows that
the outside region is region II in the cosmological Kruskal diagram. 
Relabeling coordinates $r\leftrightarrow t$, the metric in region II
is 
\begin{equation}
\label{Region_II}
ds^2 = - \frac{dt^2}{\tilde{f}(t)} + \tilde{f}(t) dr^2 + t^2 d\vec{X}^2  \;,\;\;\;
r_h < t < \infty \;,
\end{equation} 
where $t$ is a future-pointing timelike coordinate, and where $t\rightarrow \infty$ 
corresponds to future timelike infinity. Comparing to \eqref{Region_IV_EF}
we see that the regions II and III are related by time reflection.

\subsection{Asymptotic limits of planar Reissner-Nordstr\"om-like solutions\label{app:asymptotics}}

Evaluating \eqref{Region_II} for the planar Reissner-Nordstr\"om solution
in the asymptotic limit $t\rightarrow \infty$ we obtain
\begin{equation}
ds^2_{II, (+\infty)} = - \frac{t}{2M} dt^2 + \frac{2M}{t} dr^2 + t^2 d\vec{X}^2 \;,
\end{equation}
which is the `positive mass' version of the planar (type D) A-III vacuum solution
of pure Einstein gravity.\footnote{See \cite{Griffiths:2009dfa} for background 
and original literature.} Introducing a new time coordinate $t \propto \tau^{2/3}$
and absorbing numerical factors by rescaling $r,x,y$ this becomes
\begin{equation}
ds^2_{II, (+\infty)} = -d\tau^2 + \tau^{-2/3} dr^2 + \tau^{4/3} d\vec{X}^2 \;,
\end{equation}
which belongs to the class of type D Kasner solutions. These are the simplest
homogeneous but anisotropic vacuum cosmological solution of pure 
Einstein gravity. The A-III/Kasner solution is defined for $0<t,\tau <\infty$ and
describes a universe starting in a big bang at $t=\tau=0$ and then expanding
in the $(x,y)$-directions while contracting in the transverse direction $r$. 
Its time-reversed version, which is the asymptotic solution for Region III, 
\begin{eqnarray}
ds^2_{III, (-\infty)} &=& \frac{t}{2M} dt^2 - \frac{2M}{t} dr^2 + t^2 d\vec{X}^2 \;,\;\;\;
-\infty < t < 0 \;, \\
&=&  -d\tau^2 + \tau^{-2/3} dr^2 + \tau^{4/3} d\vec{X}^2\;, \;\;\;\;-\infty < \tau < 0\;,
\end{eqnarray}
describes
a universe which contracts in the $(x,y)$ directions, expands transversally, 
and end in a big crunch at $t=\tau=0$. The planar Reissner-Nordstr\"om solution that 
has been obtained by adding non-trivial Maxwell fields describes a bouncing
cosmology which interpolates between a contracting and an expanding Kasner cosmology. This removes
the spacelike big crunch and big bang singularities at $t=\tau=0$ 
and replaces them by
an intermediate region containing two 
timelike singularities which are shielded behind event horizons. These singularities
can be interpreted as sources, and by embedding Einstein-Maxwell theory 
into the STU-supergravity and subsequently into string theory, these 
source can be identified as certain brane configurations \cite{Gutowski:2019iyo}.

The asymptotic solution in Regions I and IV for the planar solution of
the Einstein-anti-Maxwell theory is the `negative mass' , static version 
of the planar (type D) A-III vacuum solution of pure Einstein gravity,
which was first described by Taub \cite{Taub:1951}. 
It was interpreted as the geometry outside an infinite static
plane or domain wall, but the observation that neutral particles are
repelled indicated a negative mass and made the physical interpretation
problematic. In the planar solution to the Einstein-anti-Maxwell theory
the naked timelike 
singularity is replaced by a spacelike singularity shielded
by a horizon, so that the deformed solution describes 
a planar black hole.
The negative mass issue becomes clearer in this description, since
it is related to the negative kinetic energy of the Maxwell field in 
this flipped sign version of Einstein-Maxwell theory. It is an interesting
question whether the embedding of this theory into type-II* string theory
will allow one to give a viable physical interpretation of this solution.

\section{Grand canonical ensemble (with fixed volume) \label{app_thermo}}

In textbook thermodynamics, the internal energy $E$ (often denoted $U$) 
in the grand canonical ensemble depends on the extensive variables entropy 
$S$, volume $V$ and particle number $N$. In relativistic thermodynamics the
particle number is not conserved, and therefore it is replaced by conserved charges. Let
us consider the case of a single conserved charge $\mathcal{Q}$. We take the 
volume (which in black hole thermodynamics corresponds to angular momentum, or
for planar solutions, linear momentum)
to be  fixed, so that the internal energy only depends on entropy and charge,
$E=E(S,\mathcal{Q})$. The free energy $F(T,\mathcal{Q}) = E-TS$ and the 
grand potential $\Omega(T,\mu) =E-TS - \mu \mathcal{Q}$ are related to $E(S,\mathcal{Q})$ by 
Legendre transformations which exchange the extensive variables $S,\mathcal{Q}$ with
the intensive variables temperature $T=1/\beta$ and chemical potential $\mu$. Various
partial derivatives can be read off from the total differentials
\begin{equation}
dE = TdS + \mu d\mathcal{Q}\;, \;\;
dF = - S dT + \mu d \mathcal{Q} \;,\;\;
d\Omega = - S dT - \mathcal{Q} d\mu \;.
\end{equation}
In particular, we obtain the following relations used in the main text:
\[
\mathcal{Q} = - \frac{\partial \Omega}{\partial \mu} \;,\;
\mu = \frac{\partial F}{\partial \mathcal{Q}} \;,\;\;
\beta = \frac{1}{T} = \frac{\partial S}{\partial E} \;,\;\;
S = - \frac{\partial F}{\partial T} = \beta^2 \frac{\partial F}{\partial \beta}
\]
and
\[
\frac{\partial (\beta F)}{\partial \beta} = F + TS = E \;.
\]

\providecommand{\href}[2]{#2}\begingroup\raggedright\endgroup


\end{document}